\documentclass[12pt,preprint]{aastex}
\begin{document}
\title{The Kinematic State of the Local Volume}

\author{Alan B. Whiting}
\affil{Cerro Tololo Interamerican Observatory}
\email{whiting@ctio.noao.edu}

\begin{abstract}
The kinematics of galaxies with 10 megaparsecs (10 Mpc) of the Milky Way is
investigated using published distances and radial velocities.  With respect to
the average Hubble flow (isotropic or simple anisotropic), there is {\em no}
systematic relation between peculiar velocity dispersion and absolute
magnitude over a range of 10 magnitudes; neither is there any apparent
variation with galaxy type or between field and cluster members.  There are
several possible explanations for the lack of variation, though all have difficulties:
either there is no relationship between light and mass on these scales, or the peculiar velocities
are not produced by gravitational interaction, or the background dynamical picture is wrong in 
some systematic way.  The extremely
cold local flow of 40-60 km s$^{-1}$ dispersion reported by some authors is shown to be an
artifact of sparse data, a velocity dispersion of over 100 km s$^{-1}$
being closer to the actual value.  Galaxies with a high (positive) radial velocity
have been selected against in studies of this volume, biasing numerical results. 
\end{abstract}

\keywords{galaxies: kinematics and dynamics---cosmological parameters---
cosmology: observations---dark matter}

\section{Mass and Motion Models}

Under the physically well-motivated assumption that the only important
force on the largest astronomical scales and for most of the history of the
universe is gravity, there should be an intimate relationship
between the cosmic velocity field and the cosmic density field.  Predicting,
calculating and observing the details of this relationship have been the
major occupations of cosmology for most of the past century.

The first-order description of the kinematics of the universe takes the form
of a homogeneous, isotropic expansion.  To investigate this it is best to
examine the largest spatial scales, where the density field is closest to
the homogeneous ideal. 
In that case the observational goal is to determine the Hubble constant
($H_0$) and the major problem is the determination of reliable
distances.  A subsidary problem appears in practice, seen for example in \citet{VB79},
in the fact that the distribution of galaxies with the most reliable
distances (the closer ones) is not homogeneous and their motions are subsequently not
uniform.

With improving observational and calculational techniques this subsidary
problem has been turned into an area of research in its own right.
\citet{CW00} provide an overview; specific techniques and results are found in,
for instance, \citet{BTF99} and \citet{DEK99}.  One of the objects of this work
is the estimations of the biasing parameter,
which measures the relationship between actual density fluctuations
and observed galaxies.  To do this convincingly, however, requires a volume
large enough that density contrasts remain in or near the linear region,
hence this is the study of large-scale structure.  Unfortunately, distance
indicators for galaxies so far away are not very accurate and are subject to
various systematic biases; one must
take large samples and be careful about manipulating statistics.

The situation reverses itself in the Local Volume, the region within about 10
megaparsecs (Mpc).  Here we are several rungs down on the
cosmic distance ladder and distance indicators are based on resolved stars,
allowing accuracies better than 7\% under good conditions.  With redshifts good
to a few km s$^{-1}$ from HI observations, the data are (potentially) excellent.  
Balancing this are the complications of nonlinear dynamics (sufficiently nonlinear
that pertubative techniques, such as the Zeldovich approximation and expansion
to a few orders, cannot be used) and the
fact that peculiar velocities are comparable to the Hubble flow.

Dynamically, this is almost unknown territory.  It lies between the well-studied
galaxy and galaxy group scales on one hand, and that of large-scale structure on
the other.  Neither are there any galaxy clusters in the Local Volume, Virgo
being just outside.  The potential scientific returns of examining this unexplored
region, especially when excellent data are available, are great.  For instance,
dark matter on galaxy scales seems to be concentric with luminous matter, but far more
extended; on the linear-structure scale, both seem to be distributed the same
way.  The biasing parameter $b$, which measures the concentration of luminous
matter relative to dark matter, is found to be in the range 0.5-1.0 on large scales
(\citet{DEK99}; \citet{BTF99}); within a spiral galaxy it must range from
zero beyond the visible disk to some number possibly greater than unity at the
center.  In between these two scales, what happens?

Without a dynamical model of obvious applicability to the Local Volume, we
must begin by studying the kinematics.  The procedure in what follows is to calculate
overall average motions, and then examine deviations from them for systematic
effects.  In effect, we are looking for some kind of non-randomness among peculiar
(radial) velocities.  

Two models will be used for the ``overall average motions.''  Both include the average
velocity of all galaxies in a sample, showing up as the solar reflex velocity;
the first model then adds 
a uniform expansion (the local reflection of the universal Hubble flow).  The second is
suggested by the general distribution of galaxies within the Volume.
The Supergalactic Plane is well-defined in this region, clear in \citet{TF87} as well
as with updated data in \citet{KMa96} and \citet{LV00}.  There should be
therefore an anisotropy in the overall flow, expansion normal to the
Supergalactic Plane being slower than in the Plane.  In fact \citet{VB79}
found something of the sort, as have \citet{KM01} and \citet{KMa96}.
In addition, the tidal effect of the Virgo Cluster should be visible as
an elongation (high eigenvalue) in the direction of the Cluster and a
contraction (low eigenvalues) normal to it.  Since Virgo is almost in the
Plane, at roughly Supergalactic Longitude $100\arcdeg$, we expect the highest
eigenvalue in that direction, the lowest at the Supergalactic poles, and
an intermediate eigenvalue in the Plane at longitude about $10\arcdeg$.
This relationship between the distribution of galaxies in the Supergalactic
Plane and velocities can be seen either as dynamic (the concentration of mass causing
anisotropic motions) or as kinematic (the concentration of galaxies having been
produced by anisotropic motions), or more accurately both.

\section{The Overall Hubble Flow}

\subsection{Calculations}

The mathematical tools used here are similar to some of those developed in
\citet{LB88}, though details and the use to which they are put differ.

Suppose that the velocity field in the Local Volume is
smooth, continuous and differentiable.
Under these conditions we may expand this field in terms of the vector
distance from us, ${\bf r}$, as a Taylor series:
\begin{equation}
{\bf v} = {\bf v}_0 + {\bf r} \cdot \frac{\partial {\bf v}}{\partial
{\bf r}} + \frac{1}{2}{\bf r} \cdot \frac{\partial^2 {\bf v}}{\partial {\bf r}^2}
\cdot {\bf r}+ \cdots
\end{equation}
If we truncate this series at the term linear in distance and recognize
that only the radial velocities are observable, we find that
\begin{equation}
v_{\rm obs} = {\bf v} \cdot \hat{\bf r} = {\bf v}_0 \cdot \hat{\bf r}
+ {\bf r} \cdot {\bf H} \cdot \hat{\bf r}
\label{eq:Hubble}
\end{equation}
where ${\bf H}$ is the symbol for the first-order partial-derivative tensor
and $\hat{\bf r}$ is the unit vector in the ${\bf r}$ direction.
If the tensor is isotropic, ${\bf H}$ reduces to a scalar; if the
region over which it is determined is representative of the universe as
a whole, it is the Hubble constant.  It is therefore convenient to call
${\bf H}$ the Hubble tensor\footnote{However, it is worth emphasizing that the
determination of the Hubble tensor over the Local Volume has no necessary
cosmic implication, since this region is too small to be
a fair sample of the universe.}.  The Hubble tensor, quantifying
anisotropic motion, is the next more complicated description of
cosmic motion after a simple uniform expansion.

To determine the components of the Hubble tensor from a set of data, a
least-squares method is the most straightforward.  We take as a measure
of goodness of fit the average square of the difference between
the predicted radial velocity and the observed radial velocity\footnote{
Compare \citet{KM01}, in which a slightly different measure is used.},
\begin{equation}
\sigma^2 = \frac{1}{N} 
\sum_i \sigma_i^2 = \frac{1}{N} \sum_i \left( v_{\rm obs} - \left( {\bf v}_0 \cdot
\hat{\bf r} + \hat{\bf r} \cdot {\bf H} \cdot {\bf r} \right) \right)^2
\label{eq:dispersion}
\end{equation}
Taking the derivatives of this with respect to the three components of
${\bf v}_0$ and the six independent components of ${\bf H}$ and setting
them equal to zero gives nine linear equations to be solved for the nine
unknowns, a straightforward if tedious calculation\footnote{In practice,
this was done by the Mathematica program.}.  An isotropic
solution is determined similarly, using four equations in four unknowns.

\subsection{Data}

Distance and radial velocity data for galaxies within 10 Mpc were gathered
from the literature and are summarized in Table~\ref{table:Data}.
In gathering the data
much use has been made of the NASA/IPAC Extragalactic Database (NED)\footnote{
NED is operated by the Jet Propulsion Laboratory, California Institute of 
Technology, under contract with the National Aeronautics and Space Administration.}.
The column headings are: (1) Designation; for those galaxies which
have been catalogued several times, only one was chosen, to maintain
readability of the table\footnote{Unfortunately, most galaxies in this
sample have at least three popular designations, and a given author will
mention one while another will only use a different one.  The designation
used here is generally that found in the distance reference; others
may conveniently be found by referring to NED.};
(2) Apparent $B$ magnitude, from NED; (3) Morphological type, from NED;
(4) Supergalactic longitude, in degrees; (5)
Supergalactic latitude, in degrees; (6) Radial velocity, in km s$^{-1}$;
(7) the source for the radial velocity; (8) Distance, in Mpc; (9)
the source for the distance; (10) the method used to derive the distance.
The references corresponding to the source codes are listed at the
end of the table.  An asterisk denotes data which are not known.

Galaxies known to be part of the Local Group are not included, to avoid
a possible dynamical bias at the small end of the distance scale.
Of course, distances derived from a velocity model
were of no use for this purpose.  Those available were based on
Cepheids (``ceph'' in the table), brightness of the tip of the Red Giant
Branch (TRGB; in one case, the tip of the Asymptotic Giant Branch,
TAGB), surface brightness fluctuations (SBF), a geometric
method involving water masers (geo), or the
brightness of the brightest stars (stars).  The former methods 
have quoted accuracies of $\pm$ 0.2 magnitudes or less in distance modulus.
The last method was found by \citet{KT94} to have an accuracy of 0.3 to
0.45 magnitudes, depending on exactly how it was applied.  However, in
several cases (see, for example, \citet{CSH00}, and compare \citet{KMa96}
with \citet{AT00} for the case of DDO 109 = UGC 9240)
it has been found to be in error by a factor of two or more. 
In what follows calculations will be performed separately on
both on the full data set of 98 galaxies (so as to take
advantage of the larger number of objects) and on the set of 35 galaxies
with more reliable distances. 

Although
the data come from many sources and some difficulties could be anticipated
from that fact, in practice there were no problems along those lines.
Each of the TRGB distances used the same calibration \citep{LFM93}, as did the
Cepheid distances within the stated errors; and all used as a zero-point
the Large Magellanic Cloud at a distance modulus of 18.50.  The SBF method
itself was calibrated using TRGB and Cepheid data, most of which found
its way into this data set.  The internal consistency of the better-quality
distances is therefore reliable\footnote{This agreeable state of affairs
probably won't last long, as someone is sure to come up with a ``better''
calibration of some distance method soon.}.

As noted, the brightest-star distances are treated with less confidence here,
and the better-quality data will be examined separately when possible.
In the case of NGC 4258 the Cepheid and geometric distances do not
agree within their stated errors (7.98 and 7.2 Mpc, respectively).
The distance shown is the average.

It is worth pointing out that the data amount to a minority of the
galaxies estimated to lie in this volume.  Including the brightest-star
galaxies possibly as much as a quarter of the population is represented;
restricting ourselves to the better data, less than a tenth\footnote{So even
if one were to ignore nonlinear effects, a POTENT-like treatment \citep{DEK99} of the Local
Volume is impossible: the locations of most of the luminous masses are just not
known.}.  In addition, these
are not all the largest or brightest, nor uniformly distributed.  It is
worthwhile, then, to look into possible effects of this uneven sampling.

The average motion of all the galaxies in each sample will show up in the
solar reflex velocity.  Subsidiary ``bulk flows,'' motions of one part of
the Volume with respect to the whole, will appear as peculiar velocities
systematically different in that part---but only if there are galaxies 
there to show it.  Similarly, flows on a smaller scale (or a larger) than is sampled will
go unnoticed.

More subtle would be effects on overall parameters due to badly-distributed
data.  To examine those, suppose that we are attempting to fit a model of
anisotropic flow (as above) to an actual flow.  The ``real'' flow is
described by a flow with noise, in part due to distance errors (which obey
a distribution function $\delta r$) and in part due
to real peculiar velocities (which obey a vector distribution function ${\bf e}$).
The quantity we attempt to minimize during least-squares fitting is then
\begin{eqnarray*}
\sigma^2 &= &\frac{1}{N} \sum \left( v_{obs} - v_{model} \right)^2 \\
 & = & \frac{1}{N} \sum_i \left( {\bf v}_{0a} \cdot {\bf \hat{r}}_i +
{\bf \hat{r}}_i \cdot {\bf H}_a \cdot {\bf \hat{r}}_i \left(r_i + \delta r_i \right)
+ {\bf e}_i \cdot {\bf \hat{r}}_i - {\bf v}_{0m} \cdot {\bf \hat{r}}_i
- {\bf \hat{r}}_i \cdot {\bf H}_m \cdot {\bf \hat{r}}_i r_i \right)^2
\end{eqnarray*}
where the subscript $a$ stands for ``actual'' and $m$ for ``model,'' and the
vector ${\bf r}$ is written as a product of magnitude and direction, $r {\bf \hat{r}}$.
The various terms may be regrouped as follows:
\begin{eqnarray*}
\sigma^2 & = & \frac{1}{N} \sum_i \left[ 
\left( \left( {\bf v}_{0a} -{\bf v}_{0m} \right) \cdot {\bf \hat{r}}_i +
{\bf \hat{r}}_i \cdot \left( {\bf H}_a - {\bf H}_m \right) \cdot {\bf \hat{r}}_i r_i
\right)^2 \right. \\
 & & + 2 \left( {\bf v}_{0a} -{\bf v}_{0m} \right) \cdot {\bf \hat{r}}_i \:
{\bf \hat{r}}_i \cdot {\bf H}_a \cdot {\bf \hat{r}}_i \delta r_i \\
& & + 2 {\bf \hat{r}}_i \cdot \left( {\bf H}_a - {\bf H}_m \right) \cdot {\bf \hat{r}}_i 
r_i \: {\bf \hat{r}}_i \cdot {\bf H}_a \cdot {\bf \hat{r}}_i \delta r_i \\
& & +2 {\bf \hat{r}}_i \cdot \left( {\bf H}_a - {\bf H}_m \right) \cdot {\bf \hat{r}}_i r_i
\: {\bf e}_i \cdot {\bf \hat{r}}_i \\
& & +2 \left( {\bf v}_{0a} -{\bf v}_{0m} \right) \cdot {\bf \hat{r}}_i \: {\bf e}_i 
\cdot {\bf \hat{r}}_i \\
& & \left. + \left( {\bf \hat{r}}_i \cdot {\bf H}_a \cdot {\bf \hat{r}}_i \delta r_i
+ {\bf e}_i \cdot {\bf \hat{r}}_i \right)^2 \right]
\end{eqnarray*}
The first line, the squared term, is the difference between the model and the actual
situation in the absence of errors or noise; its minimum is zero, when the actual and
model parameters are identical.  The last line is the collected ``noise'' and does 
not contain the model parameters at all; its only effect on the process is to set
a positive minimum to the quantity $\sigma^2$.  The cross terms require some attention.

If we assume that the real peculiar velocity distribution ${\bf e}$ does not depend on
direction or distance, the two cross terms containing it will average to zero.  Also,
if the distance errors $\delta r_i$ do not depend on direction (that is, our measures
are no more uncertain in one part of the sky than another), the first cross term vanishes
also.  The second cross term is harder to deal with, since in general we do expect a
correlation between the distance $r_i$ and the distance error $\delta r_i$:
the distance error should be larger as the distance itself increases, so that galaxies
farther away will contribute terms of larger magnitude to this sum.  However, as long as the
error distribution function itself averages to zero, this cross term also vanishes.

So if peculiar velocities and distance errors have no systematic correlation with direction
then the actual distribution of galaxies (the particular ${\bf r}_i$ from which kinematic
model parameters are derived) is irrelevant.  The lack of galaxies at high Supergalactic
latitudes, for instance, and in some directions in the Supergalactic Plane should not bias
the values for the solar reflex velocity and Hubble tensor.  Put another way, if all the
galaxies in the sample take part in the same motion with random deviations, the motion
can be reconstructed from any part of it (though with varying amounts of uncertainty).

A peculiar velocity distribution which does show some systematic variation with position,
of course, will change the derived overall parameters.  In the simplest case it would be
a sort of ``bulk flow'' on a smaller scale than the total sample, and should be visible
in plots of peculiar velocity against position (which will be shown below).  In the case of
a mass concentration like a cluster, a locally increased velocity dispersion should be
visible on the same plots.  Of course, in volumes containing few or no sample galaxies such flows
or mass concentrations will not be visible.

There remains the possibility of inducing biases by the choice of the galaxy sample.  This
is potentially serious, since the available data were gathered from many different sources
who had obtained them for many different reasons, and there is no guarantee of even
coverage in any parameter.  An example of subtle biases of this sort is found in \citet{RFR73}.
There, a study of galaxies in a narrow magnitude range showed a significant velocity
anisotropy, most evident in two concentrations at 4950 km s$^{-1}$ and 6400 km s$^{-1}$.
Further analysis \citep{FJ76} showed that the narrow magnitude range actually isolated different parts
of the galaxy luminosity function in each case, so in fact different kinds of galaxy were
being compared; a further attempt to limit the sample by redshift only exacerbated the
problem \citep{JJC91}.  

The present study is not subject to quite the same kind of bias.  The distances used here
are not dependent on a certain selection technique for the galaxies, and especially for the
higher-accuracy sample are quite good.  The range in absolute magnitude is very wide and
there is no great concentration in any luminosity interval (to anticipate, see Figures~\ref{mplot98}
and \ref{mplot35}); neither is there a concentration at any particular distance within the
10 Mpc limit (Figure~\ref{distplot}).  It appears that the hetrogeneous nature of the sample
has acted as a sort of random selection, at least in those variables.

There still remains the possibility of a slightly different bias, perhaps stemming from uneven spatial
coverage or some effect which causes kinematically important galaxies to be underrepresented
in published data.  Especially given the fact that a minority of galaxies known to be within
this chosen region is represented, it is well to keep this in mind.  Unfortunately, it is impossible
to correct for a bias whose form (and even existence) is unknown.

\begin{deluxetable}{lrrrrrrlrll}
\rotate
\tablewidth{0pc}
\tablecaption{Local Volume Galaxy Data \label{table:Data}}
\tablehead{
\colhead{Designation} & \colhead{Mag} & \colhead{Type}
 & \colhead{L} & \colhead{B} & \colhead{RV}
 & \colhead{RV Source} & \colhead{Dist} & \colhead{Dist
 Source} & \colhead{Method} }
\startdata
A0554+0728 & 18.4 & Im & 353.90 & -62.79 & 411 & KM96 & 5.50 & KM96 & stars \\
Antlia & 16.2 & dE & 139.93 & -44.80 & 361 & RC3 & 1.32 & A97 & TRGB \\
BK3N & 17.1 & Im & 41.10 & 0.40 & -40 & KMa96 & 2.79 & KMa96 & stars \\
DDO13 & 14.4 & Im & 315.26 & -6.14 & 631 & RC3 & 9.04 & S96 & stars \\
DDO50 & 11.1 & Im & 33.30 & -2.40 & 158 & RC3 & 3.05 & H98 & ceph \\
DDO53 & 14.5 & Im & 35.90 & -6.10 & 19 & RC3 & 3.08 & KMa96 & stars \\
DDO63 & 13.0 & IAB & 38.76 & 1.34 & 136 & RC3 & 6.95 & KT94 & stars \\
DDO66 & 14.3 & Im & 41.30 & 0.70 & 46 & RC3 & 3.41 & KMa96 & stars \\
DDO70 & 11.8 & Im & 95.53 & -39.62 & 301 & RC3 & 1.33 & S97 & TRGB/ceph \\
DDO71 & 18.0 & Im & 43.51 & -0.58 & -126 & NED & 3.50 & KKD00 & TRGB \\
DDO75 & 11.9 & IBm & 109.25 & -40.67 & 324 & RC3 & 1.39 & P94 & ceph \\
DDO82 & 13.5 & Sm: & 41.69 & 3.85 & 40 & KMa96 & 4.48 & KMa96 & stars \\
DDO155 & 14.7 & ImV & 103.13 & 4.66 & 214 & RC3 & 2.24 & T95 & ceph \\
DDO165 & 12.8 & Im & 49.60 & 15.60 & 37 & RC3 & 4.88 & KMa96 & stars \\
DDO187 & 14.4 & ImIV & 97.93 & 24.35 & 154 & RC3 & 2.50 & ATK00 & TRGB \\
DDO190 & 13.2 & IAm & 74.10 & 26.90 & 156 & AT00 & 2.90 & AT00 & TRGB \\
ESO294-G10 & 15.6 & dS0/Im & 254.37 & -5.27 & 117 & J98 & 1.71 & J98 & SBF \\
IC342 & 9.1 & SAB & 10.60 & 0.37 & 34 & RC3 & 2.12 & KT94 & stars \\
IC4182 & 13.0 & SA & 80.22 & 11.60 & 321 & RC3 & 4.49 & FMG01 & ceph \\
IC2574 & 10.8 & SAB & 43.60 & 2.30 & 47 & RC3 & 3.78 & KMa96 & stars \\
KDG52 & 16.5 & I: & 33.52 & -1.93 & 113 & KMa96 & 2.95 & KMa96 & stars \\
KDG61 & 15.6 & dE & 41.54 & 0.33 & -135 & NED & 3.60 & KKD00 & TRGB \\
KDG73 & 14.9 & Im & 44.03 & 4.75 & -132 & NED & 4.04 & KMa96 & stars \\
KK251 & 16.5 & Ir: & 10.79 & 42.12 & 126 & KSH00 & 5.30 & KSH00 & stars \\
KK252 & 17.1 & Sph: & 10.97 & 41.67 & 132 & KSH00 & 5.30 & KSH00 & stars \\
KKR25 & 17.0 & Ir & 56.09 & 40.37 & -135 & NED & 1.86 & KSD01b & TRGB \\
KKR55 & 17.0 & Ir & 8.80 & 41.05 & 23 & NED & 5.40 & KSH00 & stars \\
KKR56 & 17.6 & Ir & 6.94 & 42.25 & -135 & NED & 6.40 & KSH00 & stars \\
KKR59 & 15.7 & Ir & 3.74 & 41.95 & 17 & NED & 4.70 & KSH00 & stars \\
Maffei1 & 11.4 & S0 & 359.29 & 1.44 & 13 & NED & 4.40 & Dv01 & TAGB \\
NGC59 & 13.1 & SA & 273.17 & 3.16 & 362 & J98 & 4.39 & J98 & SBF \\
NGC300 & 8.9 & SA & 259.69 & -9.50 & 142 & RC3 & 2.00 & FMG01 & ceph \\
NGC628 & 9.9 & SA & 314.88 & -5.39 & 656 & RC3 & 7.32 & S96 & stars \\
NGC784 & 12.2 & SBdm: & 328.81 & -6.31 & 198 & NED & 5.00 & DK00 & stars \\
NGC925 & 10.7 & SAB & 335.72 & -9.47 & 553 & RC3 & 9.16 & FMG01 & ceph \\
NGC1560 & 12.2 & SA & 15.85 & 0.79 & -36 & RC3 & 3.73 & KT94 & stars \\
NGC1705 & 12.8 & SA0 & 231.82 & -45.54 & 628 & NED & 5.10 & TSB01 & TRGB \\
NGC2366 & 11.4 & IB & 29.16 & -4.86 & 100 & RC3 & 3.44 & TS95 & ceph \\
NGC2403 & 8.9 & SAB & 30.50 & -8.31 & 131 & RC3 & 3.22 & FMG01 & ceph \\
NGC2683 & 10.6 & SA & 55.87 & -33.42 & 411 & NED & 9.20 & DK00 & stars \\
NGC2903 & 9.7 & SB & 73.53 & -36.44 & 556 & NED & 8.90 & DK00 & stars \\
NGC2976 & 10.8 & SAc & 40.98 & -0.78 & 3 & RC3 & 4.57 & KMa96 & stars \\
NGC3031 & 7.9 & SA & 40.77 & 0.59 & -34 & RC3 & 3.63 & FMG01 & ceph \\
NGC3034 & 9.3 & I0 & 40.38 & 1.06 & 203 & RC3 & 3.89 & S99 & TRGB \\
NGC3077 & 10.6 & I0 & 41.51 & 0.83 & 14 & RC3 & 3.90 & SM01 & TRGB \\
NGC3109 & 10.4 & SB & 138.30 & -45.10 & 404 & RC3 & 1.33 & M99 & ceph \\
NGC3274 & 13.2 & SAB & 77.20 & -21.80 & 537 & RC3 & 8.00 & KMa96 & stars \\
NGC3621 & 10.2 & SA & 145.97 & -28.57 & 727 & RC3 & 6.64 & FMG01 & ceph \\
NGC4144 & 12.1 & SAB & 68.84 & 3.83 & 267 & RC3 & 9.70 & Kd98 & stars \\
NGC4236 & 10.1 & SB & 46.76 & 11.38 & -5 & RC3 & 3.24 & KMa96 & stars \\
NGC4244 & 10.9 & SA & 77.58 & 2.41 & 243 & RC3 & 4.50 & Kd98 & stars \\
NGC4258 & 9.1 & SAB & 68.50 & 5.55 & 470 & H99 & 7.60 & H99,FMG01 & geo,ceph \\
NGC4395 & 10.6 & SA & 82.21 & 2.74 & 320 & RC3 & 4.20 & Kd98 & stars \\
NGC4449 & 10.0 & IBm & 72.09 & 6.18 & 201 & RC3 & 2.90 & Kd98 & stars \\
NGC4523 & 14.4 & SAB & 100.42 & -0.86 & 262 & NED & 6.40 & TGD00 & stars \\
NGC4605 & 10.9 & SB & 55.14 & 12.02 & 143 & RC3 & 5.18 & KMa96 & stars \\
NGC5128 & 7.8 & S0 & 159.98 & -5.25 & 562 & RC3 & 3.63 & So96 & TRGB \\
NGC5204 & 11.7 & SA & 59.10 & 17.85 & 204 & RC3 & 4.10 & DK00 & stars \\
NGC5236 & 8.2 & SAB & 148.25 & 0.99 & 516 & RC3 & 4.50 & KMa96 & stars \\
NGC5238 & 13.9 & SAB & 66.60 & 18.40 & 232 & RC3 & 5.18 & KMa96 & stars \\
NGC5253 & 10.9 & Im & 150.12 & 1.00 & 404 & RC3 & 3.15 & FMG01 & ceph \\
NGC5457 & 8.3 & SAB & 63.30 & 22.61 & 241 & RC3 & 6.70 & FMG01 & ceph \\
NGC5474 & 11.3 & SA & 64.03 & 22.90 & 277 & RC3 & 6.80 & DK00 & stars \\
NGC5477 & 14.4 & SA & 63.40 & 22.90 & 304 & RC3 & 7.73 & KMa96 & stars \\
NGC5585 & 11.2 & SAB & 60.40 & 24.70 & 305 & RC3 & 8.70 & DK00 & stars \\
NGC6789 & 13.8 & Im & 23.27 & 41.59 & -141 & DT99 & 3.60 & DSH01 & TRGB \\
NGC6946 & 9.6 & SAB & 10.03 & 42.00 & 51 & KSH00 & 6.80 & KSH00 & stars \\
ORION & 18.0 & SBd & 345.25 & -62.95 & 365 & KM96 & 6.40 & KM96 & stars \\
UGC288 & 16.0 & Im & 338.00 & 15.38 & 188 & RC3 & 6.73 & GKT97 & stars \\
UGC1104 & 14.2 & Im & 317.00 & -3.70 & 669 & RC3 & 7.55 & S96 & stars \\
UGC1171 & 17.0 & Im & 315.22 & -6.04 & 667 & RC3 & 7.35 & S96 & stars \\
UGC2905 & 15.3 & Im & 330.78 & -35.54 & 292 & NED & 5.83 & GKT97 & stars \\
UGC3755 & 14.1 & Im & 36.82 & -63.36 & 314 & NED & 4.14 & GKT97 & stars \\
UGC3860 & 15.1 & Im & 33.96 & -33.00 & 354 & NED & 7.00 & GKT97 & stars \\
UGC3966 & 13.9 & Im & 37.04 & -33.14 & 361 & NED & 6.85 & GKT97 & stars \\
UGC3974 & 13.6 & IBm & 46.54 & -55.49 & 272 & NED & 4.27 & GKT97 & stars \\
UGC4115 & 15.2 & IAm & 54.21 & -56.22 & 338 & NED & 5.27 & GKT97 & stars \\
UGC4483 & 15.1 & * & 34.69 & -2.65 & 156 & RC3 & 3.20 & DMK01 & TRGB \\
UGC5721 & 13.2 & SAB & 77.24 & -21.81 & 537 & NED & 7.98 & GKT97 & stars \\
UGC6451 & 16.5 & * & 64.19 & -0.79 & 249 & NED & 4.20 & SHG00 & TRGB \\
UGC6456 & 14.5 & P & 36.54 & 11.40 & -93 & RC3 & 4.79 & LT99 & TRGB \\
UGC6541 & * & * & 64.19 & -0.79 & 249 & NED & 3.52 & GKT97 & stars \\
UGC6565 & 12.1 & Irr & 59.57 & 1.79 & 229 & RC3 & 3.52 & GKT97 & stars \\
UGC6572 & 14.3 & ImIII & 67.96 & -2.08 & 229 & NED & 3.47 & GKT97 & stars \\
UGC6817 & 13.4 & Im & 74.93 & -2.12 & 242 & NED & 3.92 & GKT97 & stars \\
UGC7559 & 14.2 & IBm & 78.93 & 4.03 & 218 & RC3 & 3.93 & GKT97 & stars \\
UGC7857 & 14.7 & Sd & 102.31 & 0.63 & 18 & NED & 6.30 & TGD00 & stars \\
UGC8320 & 12.7 & IBm & 71.98 & 14.55 & 195 & RC3 & 3.30 & KMa96 & stars \\
UGC8331 & 14.6 & IAm & 70.37 & 14.93 & 260 & RC3 & 8.20 & Kd98 & stars \\
UGC8508 & 14.4 & IAm & 62.81 & 17.91 & 62 & RC3 & 3.68 & KMa96 & stars \\
UGC9405 & 17.0 & Im & 59.38 & 26.65 & 222 & RC3 & 7.62 & KMa96 & stars \\
UGC11583 & 17.0 & Irr & 10.89 & 42.07 & 127 & KSH00 & 8.20 & KSH00 & stars \\
UGCA86 & 13.5 & Im: & 10.71 & -1.18 & 67 & KMa96 & 1.77 & KT94 & stars \\
UGCA92 & 13.8 & Im: & 11.18 & -6.01 & -99 & KMa96 & 2.21 & KMa96 & stars \\
UGCA105 & 13.9 & Im: & 14.82 & -9.26 & 111 & RC3 & 3.24 & KT94 & stars \\
UGCA281 & 15.2 & Sm & 67.94 & 7.06 & 281 & RC3 & 5.90 & SHG01 & TRGB \\
UGCA290 & 16.0 & * & 77.94 & 6.41 & 445 & NED & 6.70 & CSH00 & TRGB \\
UGCA438 & 13.9 & IB & 258.88 & 9.28 & 62 & L78 & 2.08 & LB99 & TRGB \\
\enddata
\tablecomments{References: A97, \citet{A97}; AT00, \citet{AT00}; ATK00,
\citet{ATK00}; CSH00, \citet{CSH00}; DK00, \citet{DK00}; DMK01, \citet{DMK01};
DSH01, \citet{DSH01}; DT99, \citet{DT99};
Dv01, \citet{Dv01}; 
FMG01, \citet{FMG01}; GKT97, \citet{GKT97}; H98, \citet{H98}; H99, \citet{H99};
J98, \citet{J98}; Kd98, \citet{Kd98};
KKD00, \citet{KKD00}; KM96, \citet{KM96}; KMa96, \citet{KMa96};
KSD01b, \citet{KSD01b}; KSH00, \citet{KSH00};
KT94, \citet{KT94}; L78, \citet{L78}; LB99, \citet{LB99};
LT99, \citet{LT99}; M99, \citet{M99};
NED, the NASA Extragalactic Database; P94, \citet{P94}; RC3,
\citet{RC3};
S96, \citet{S96}; S97, \citet{S97}; S99, \citet{S99}; SHG00, \citet{SHG00};
SHG01, \citet{SHG01};
SM01, \citet{SM01}; So96, \citet{So96};
T95, \citet{T95}; TGD00, \citet{TGD00};
TS95, \citet{TS95}; TSB01, \citet{TSB01}.}
\end{deluxetable}

\clearpage

\subsection{Results}

The various results of the Hubble-tensor calculation are shown in
several following tables.
Calculations were done for four cases: anisotropic expansion using the
whole 98-galaxy sample, anisotropic using the 35 galaxies with more accurate
distances, and isotropic using each sample.   For reference, some
parallel results from the literature are included.

\begin{table}
\caption{Solar Velocity}
\begin{tabular}{lcrr}
Solution & $v$, km ${\rm s}^{-1}$ & L & B \\
98, isotropic & 290 & 10 & 45 \\
98, anisotropic & 310 & 352 & 62 \\
35, isotropic & 350 & 355 & 44  \\
35, anisotropic & 330 & 350 & 49 \\
YTS & 334 & 22.1 & 28.8 \\
KM01 & 325 & 10.9 & 41.3 \\
\end{tabular}
\tablecomments{The solar velocity relative to external galaxies according to
various computations.  The columns are: the calculation; magnitude of velocity
in km ${\rm s}^{-1}$; Supergalactic longitude and latitude, in degrees.
The four calculations performed in the present paper
are listed first, followed by \citet{YTS77} (based only on Local Group galaxies)
then \citet{KM01}, the study most closely comparable to the present one.}
\label{solar}
\end{table}

Considering the reflex solar velocity (${\bf v}_0$, shown in Table~\ref{solar})
first, taking different samples and treating them different ways leads
to a speed varying over a range of 60 km ${\rm s}^{-1}$ and a
direction changing over a dozen degrees.  Comparing these
results with other determinations we find a similar variation.
That of \cite{KM01} uses a somewhat
larger but overlapping sample of galaxies going out to a similar distance.
\citet{YTS77} only used Local Group galaxies, so their result is not
necessarily comparable (though it has been used to correct the radial
velocities of more distant galaxies by, for example, \citet{SB92}).  The
expected amount of variation, given the data, will be treated quantitatively
below.

\begin{table}
\caption{Hubble Tensor Results}
\begin{tabular}{lrrrrr}
Solution & $\sigma$ &$H_{ii}$ & L & B & $\Lambda_{iiii}$ \\
98 Galaxies & & & & & \\ \hline
 $H_{uu}$ & 103 & 83 & 127 & 3 & 3.46 \\
 $H_{vv}$ &  & 51 & 34 & 46 & 1.53 \\
 $H_{ww}$ & & 32 & 40 & -44 & 0.48 \\
 isotropic & 118 & 64 & & & \\
35 Galaxies & & & & & \\ \hline
 $H_{uu}$ & 77 & 138 & 346 & -65 & 1.19 \\
 $H_{vv}$ & & 84 & 104 & -12 & 4.53 \\
$H_{ww}$ & & 35 & 19 & 21 & 1.02 \\
 isotropic & 89 & 70 & & &  \\ \hline
\citet{KM01} & 74 & 82 & 132 & 0 &  \\
        & & 62 & 42 & 0 &  \\
        & & 48 & ... & 90 &  \\
\end{tabular}
\tablecomments{Solutions for the Hubble tensor calculations, plus isotropic
solutions and Hubble tensor components calculated by \citet{KM01}.  Columns
are: the solution (three lines for tensor solutions); rms dispersion, in km ${\rm s}^{-1}$;
Hubble component value, in km ${\rm s}^{-1} {\rm Mpc}^{-1}$; Supergalactic longitude
and latitude in degrees; corresponding (dimensionless) value of the error tensor
(whose use is discussed in the text).}
\label{tensor}
\end{table}

The results for the Hubble tensor (as well as the isotropic solutions)
are displayed in Table~\ref{tensor}.
The directions U, V and W are those of the eigenvectors, in no
specific order (that is, U is not necessarily the closest eigenvector to X, and
U in one solution is not necessarily the closest eigenvector to the U in the
other).

Given the amount of attention which is to be paid to the deviations from the models,
it is important to try to separate the effects of observational errors from real velocities.
For the linear relation between distance and radial velocity, deviations from the
models are related by
\begin{equation}
\delta v = H \delta r
\end{equation}
so the variances are
\begin{equation}
\frac{1}{N} \sum_i \delta v_i^2 = H^2 \frac{1}{N} \sum_i \delta r_i^2
\end{equation}
If we assume that the distance errors are proportional to distance\footnote{Strictly
speaking, distance errors derived from stellar photometry depend in a rather
complicated way on the flux from the target stars and standards and thus on the
telescopes and exposure times used.  In practice,
observations are planned and conducted to a given signal-to-noise ratio and yield
errors as an almost constant fraction of distance.}, and that their distribution
once the distance is factored out is independent of them,
\begin{eqnarray*}
\frac{1}{N} \sum_i \delta v_i^2 & = & H^2 \frac{1}{N} \sum_i r^2_i \delta x^2_i \\
	& = & H^2 \frac{1}{N} \sum_i r_i^2 \sum_i \delta x^2
\end{eqnarray*}
For the galaxies with higher-quality distances, the quoted errors range from 0.1 to 0.2
magnitudes with an average of slightly under 0.15, giving a distance error of 7.2\%.
As noted above, the brightest-star errors are not as well known; but taking 0.4 magnitudes
as an estimate yields an average distance error of 20\%.  Putting these into the above
formula gives contributions of 67 and 21 km s$^{-1}$ of distance errors to the total
rms velocity dispersion for the brightest-star and better distances, respectively; which
leaves 97 km s$^{-1}$ of real motion in the 98-galaxy isotropic solution, 78 km s$^{-1}$
in the 98-galaxy anisotropic solution, and 86 and 74 km s$^{-1}$ for the respective 35-galaxy 
solutions\footnote{Radio HI radial-velocity measurements themselves have quoted accuracies of one
to a few km s$^{-1}$, negligible for purposes of these calculations.  There are a handful
of galaxies with much more inaccurate optical radial velocities, but they have no significant
effect on this study.\\
\indent One might try to set up a sort of inverse method to find the various contributions to
the overall velocity dispersion.
If one assumes that the deviations from the models are made up of an intrinsic
part, say $\sigma_a$ for the anisotropic model and $\sigma_i$ for the isotropic model,
and the part due to distance errors, $\sigma_{35}$ and $\sigma_{98}$, added in
quadrature, we can set up four equations in four unknowns to find the true values of
each.
Unfortunately, the ``intrinsic'' velocity dispersions are dispersions around different
models; different enough that the four equations are inconsistent with each other and
have no common solution.}.

Compared to the differences in solar reflex velocity among the various calculations
it is perhaps reassuring to find among the tensor model
results a stable, high value of 82-84
km ${\rm s}^{-1} {\rm Mpc}^{-1}$ close to the supergalactic plane and not
far from the direction of Virgo.
However, there appears to be no agreement in direction among the other
eigenvectors and very wide variation in eigenvalues; indeed, in the
solution with the most reliable data the Virgo-pointing eigenvector
does not correspond to the largest eigenvalue.  Clearly, the reliablility of these
results must be investigated.

An appropriate way to compare solutions as a whole
is the F-ratio test\footnote{See, for example, \citet{H71}, p.\ 269.  \citet{BS98}, p.\ 629
(section 5.2.3.4) also has a useful
discussion.  In practice calculations were performed using the program
Mathematica.}. One finds their
respective variances (average square of the deviation from the model;
here, the square of the velocity dispersions) and the number
of degrees of freedom in each, and then calculates the probability
that the larger variance could be produced by the model which better fits
the data.  Essentially, while a model with more parameters will always
give a smaller dispersion, one demands that it give a {\em significantly}
smaller dispersion.

For the 98-galaxy sample, the  difference is quite significant: the anisotropic solution
is a better fit at the 90\% level.  For the 35-galaxy sample the probability is lower,
77\%, but we may still say that anisotropy is a better fit to the data.  (The fact that
the more reliable data produce the less certain result, though, is troubling, and will be
explored below.)

Clearly some parts of the solutions are better known than others.  In an attempt 
to separate the uncertainties of the various parameters we expand the dispersion as a 
Taylor series about the solution:
\begin{equation}
\sigma^2 = \sigma^2_0 + \frac{d \sigma^2}{d {\bf H}} \cdot \delta {\bf H}
+ \frac{1}{2} \delta {\bf H} \cdot \frac{d^2 \sigma^2}{d {\bf H}^2} \cdot \delta {\bf H}
\end{equation}
At the solution the linear term vanishes, and since the variance itself is
quadratic in {\bf H} all higher derivatives are identically zero (see
equation~\ref{eq:dispersion}).
Restricting ourselves to the
quadratic term and shifting to summation by repeated indices,
\begin{equation}
\Delta \sigma^2 = \frac{1}{2} \frac{\partial^2 \sigma^2}{\partial H_{ij}
\partial H_{kl}} \Delta H_{ij} \Delta H_{kl}
\end{equation}
We seek a parameter which is dimensionless and thus more easily compared
between different situations.  The most obvious is the fractional change
in dispersion divided by the
fractional change in Hubble tensor, giving a fourth-order tensor:
\begin{eqnarray}
\Lambda_{ijkl}& \equiv &
\frac{\Delta \sigma^2 / \sigma^2}{\Delta H_{ij} \Delta H_{kl} / H_{ij}
H_{kl}}  \nonumber \\
        & = & \frac{1}{2} \frac{\Delta H_{ij} \Delta H_{kl}}
{\sigma^2} \frac{\partial^2 \sigma^2}{\partial H_{ij} \partial H_{kl}}
\end{eqnarray}
The components of this tensor are conveniently computed from the data:
\begin{equation}
\Lambda_{ijkl} = \frac{H_{ij} H_{kl}}{N \sigma^2} \sum_n \frac{x_i x_j x_k x_l}{r^2}
\end{equation}
where $x_i$ is the coordinate of a galaxy in the $i$ direction and $r$ its (total)
distance.
The first fraction shows that a component of this error-tensor is larger as the ratio of
eigenvalues to dispersion is larger, and the sum shows the leverage of more distant
data points in the particular spatial directions.  That is, an eigenvalue of
the error-tensor will be larger and the corresponding eigenvalue of the
Hubble tensor will be more certain as the Hubble component is larger compared with
the variance, and as the data in the direction of the Hubble component are more
distant.

As \citet{LB88} note with their similar construction, a tensor of this sort
is difficult to display or interpret in its entirety.  However, we only
need the values corresponding to specific eigenvalues of specific solutions.
The $\Lambda_{uuuu}$, $\Lambda_{vvvv}$ and $\Lambda_{wwww}$ components,
corresponding to the second-order change in the variance for each of
the eigenvalues of the two anisotropic solutions, are given in
Table~\ref{tensor}.

We continue with the F-ratio test as a way of interpreting these components.
Recall that a change in the ratio of variance by a certain amount
corresponds to a certain probability that one solution is
significantly different from another.  For the number of degrees of freedom
in the 35-galaxy sample, for instance, a ratio of 1.8 means a 95\% probability
that the smaller variance corresponds to the better solution.  The
allowable change in the eigenvalue to remain within this 95\% window is
thus
\begin{equation}
\frac{\Delta H_{ii}}{H_{ii}} = \sqrt{\frac{1.8 -1}{\Lambda_{iiii}}}
\end{equation}
The results of this kind of calculation are shown in Table~\ref{error}.

\begin{table}
\caption{Uncertainty in Tensor Components}
\begin{tabular}{lccc}
Component & $\Delta H/H$, 90\% & $\Delta H/H$, 80\% & $\Delta H/H$, 70\% \\
98 $H_{uu}$ & 0.30 & 0.23 & 0.19 \\
98 $H_{vv}$ & 0.46 & 0.35 & 0.28 \\
98 $H_{ww}$ & 0.82 & 0.63 & 0.50 \\
35 $H_{uu}$ & 0.74 & 0.58 & 0.44 \\
35 $H_{vv}$ & 0.38 & 0.30 & 0.22 \\
35 $H_{ww}$ & 0.80 & 0.63 & 0.47 \\
\end{tabular}
\tablecomments{Uncertainty in Hubble Tensor components, calculated by means
of the error tensor.  For each of the components in the two solutions, the
fractional amount it may change before the solution becomes worse (as measured
by an increase in the dispersion) at various confidence levels is shown.  Thus,
for example, the $H_{ww}$ component in the 98-galaxy solution can change by
half its value before it is 70\% certain that such a change leads
to a worse-fitting solution.}
\label{error}
\end{table}

From the table, the uncertainty in the out-of-plane tensor components is rather large.
If 70\% confidence is required, for instance, the $H_{ww}$ eigenvalues in each
solution are still only known with a 50\% error.  The effect of these uncertainties
on the whole solution may be illustrated by the 35-galaxy $H_{uu}$ eigenvalue.  At
70\% confidence it may vary by a fraction of 0.44, which means it could have a value of 61
km s$^{-1}$ Mpc$^{-1}$, smaller indeed than the middle eigenvalue.  In that case
its direction ceases to be an eigenvector.  As suggested by the great differences
in direction and magnitude among the models, the details of anisotropic flow are
quite uncertain.

An exactly similar derivation gives the error tensor corresponding to solar reflex motion:
\begin{eqnarray*}
\Lambda_{ij} & \equiv & \frac{\Delta \sigma^2 / \sigma^2}{\Delta V_i \Delta V_j / V_i V_j} \\
& = & \frac{ V_i V_j}{N \sigma^2} \sum_n \frac{ x_i x_j}{r^2} \\
\end{eqnarray*}
and in the same way (using the F-ratio test) it can be turned into uncertainties of components
of the reflex motion vector.  To allow comparison with the entries in Table~\ref{solar}, the
70\%-confidence limits of each component (corresponding to about one-sigma error bars) were
combined into limits on the magnitude and direction and listed in Table~\ref{vectorprop}. 
Some of the uncertainty comes from distance errors, which translate into correlated errors in
the components; most is uncorrelated.  Figures are given for both extremes, with the actual
situation being closer to the uncorrelated ideal.  For large values of angle uncertainties
(about $15\arcdeg$ or over) the limits become strongly asymetrical; the figures listed
are averages.

\begin{table}
\caption{Uncertainty in the Solar Reflex Velocity}
\begin{tabular}{lccc}
Solution & Magnitude, km s$^{-1}$ & SG Longitude & SG Latitude \\
98 isotropic & 290 $\pm$ 90, 130 & 10 $\pm$ 15, 12 & 45 $\pm$ 24, 23 \\
98  anisotropic & 310 $\pm$ 90, 110 & 352 $\pm$ 19, 23  & 62 $\pm$ 13, 4 \\
35 isotropic & 350 $\pm$ 80, 120 & 355 $\pm$ 6, 7 & 44 $\pm$ 26, 30 \\
35 anisotropic & 330 $\pm$ 80, 110 & 350 $\pm$ 14, 16 & 49 $\pm$ 24, 26 \\
\end{tabular}
\tablecomments{Uncertainty in the Solar Reflex Velocity, calculated by means of the error
tensor.  The value for each component (magnitude, Supergalactic Longitude, Supergalactic
Latitude) is followed by two numbers: the first corresponds to the case of completely
uncorrelated errors among the Cartesian vector components, the second to completely
correlated errors.  The actual situation is closer to the uncorrelated case.  Large
values for errors in angle are not precise, since nonlinear effects become important
for figures larger than about $15\arcdeg$.}
\label{vectorprop}
\end{table}

For comparison, \citet{KM01} give their uncertainties (standard errors) as $\pm$ 11
km s$^{-1}$ in reflex velocity magnitude and one or two degrees in direction; and
3-5 km s$^{-1}$ Mpc$^{-1}$ in Hubble tensor components.

\subsection{Uncertainty of the Uncertainties}
\label{sec-nogauss}

We have at hand three different ways of estimating the uncertainty in our numbers:
the variation in parameters between calculations; the formal standard error; and the
error tensor.  Unfortunately, they do not agree.

Considering the solar reflex velocity (magnitude and direction) and out-of-plane
Hubble tensor eigenvectors and eigenvalues, variations between the calculations are much larger
than the standard error allows.  They are, however, consistent with estimates based
on the error tensor.  The standard error assumes a rather strict Gaussian distribution
of deviations from the model, and that the sample at hand is an unbiased description of
it.  The error tensor also assumes a Gaussian distribution, but is less strict, in that
it takes into account just how the variance changes with this particular data set
when changing a parameter.  We must conclude, then, that the data at hand do not
satisfy the necessary conditions for the standard error: an unbiased sample of an
underlying model, with a strictly Gaussian distribution of errors around it.  Exactly
how the samples fail is not clear from calculations to this point, but two general
areas are identifiable. 

First, the samples could be chosen with some bias which prevents a good representation of the
kinematics of the Local Volume.  A good fit to the data may then translate to a poor fit to
the real kinematics.  This is indicated by the fact that some parameters show a very strong
dependence on the particular data set used, and that in places where the data sets overlap
(such as in the direction of the 82-84 km s$^{-1}$ Mpc$^{-1}$ Hubble tensor component)
the results are more similar than the error tensor indicates they should be.

Second, the assumed kinematic models could be inappropriate, in that there are systematic
motions not considered.  This would make the peculiar velocity distribution about the models
non-Gaussian.  Deviations which are systematic in space will be investigated in the next
section, followed by other possibilities.

\section{Spatial Deviations from the Hubble Models}

\begin{figure}
\plottwo{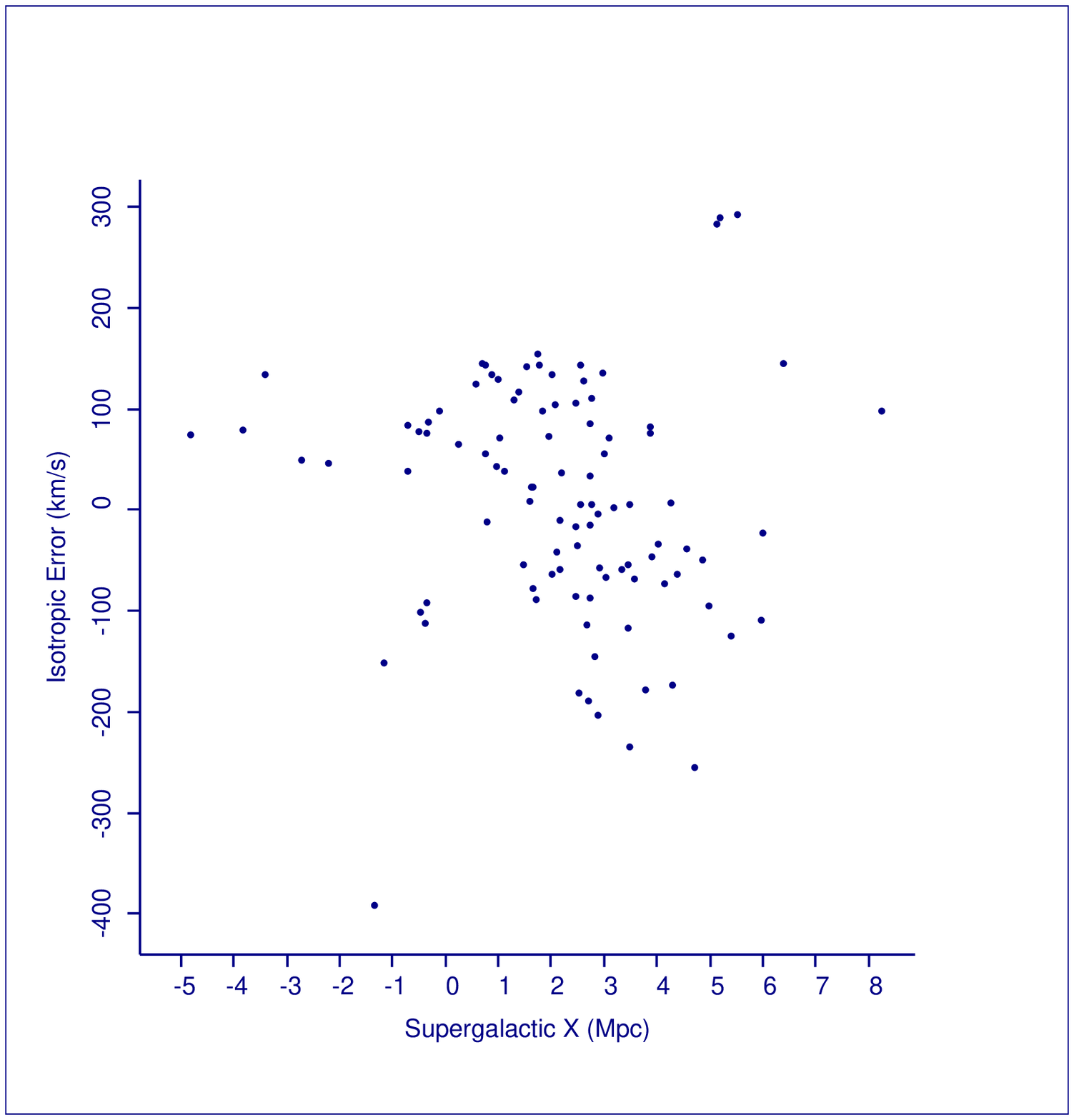}{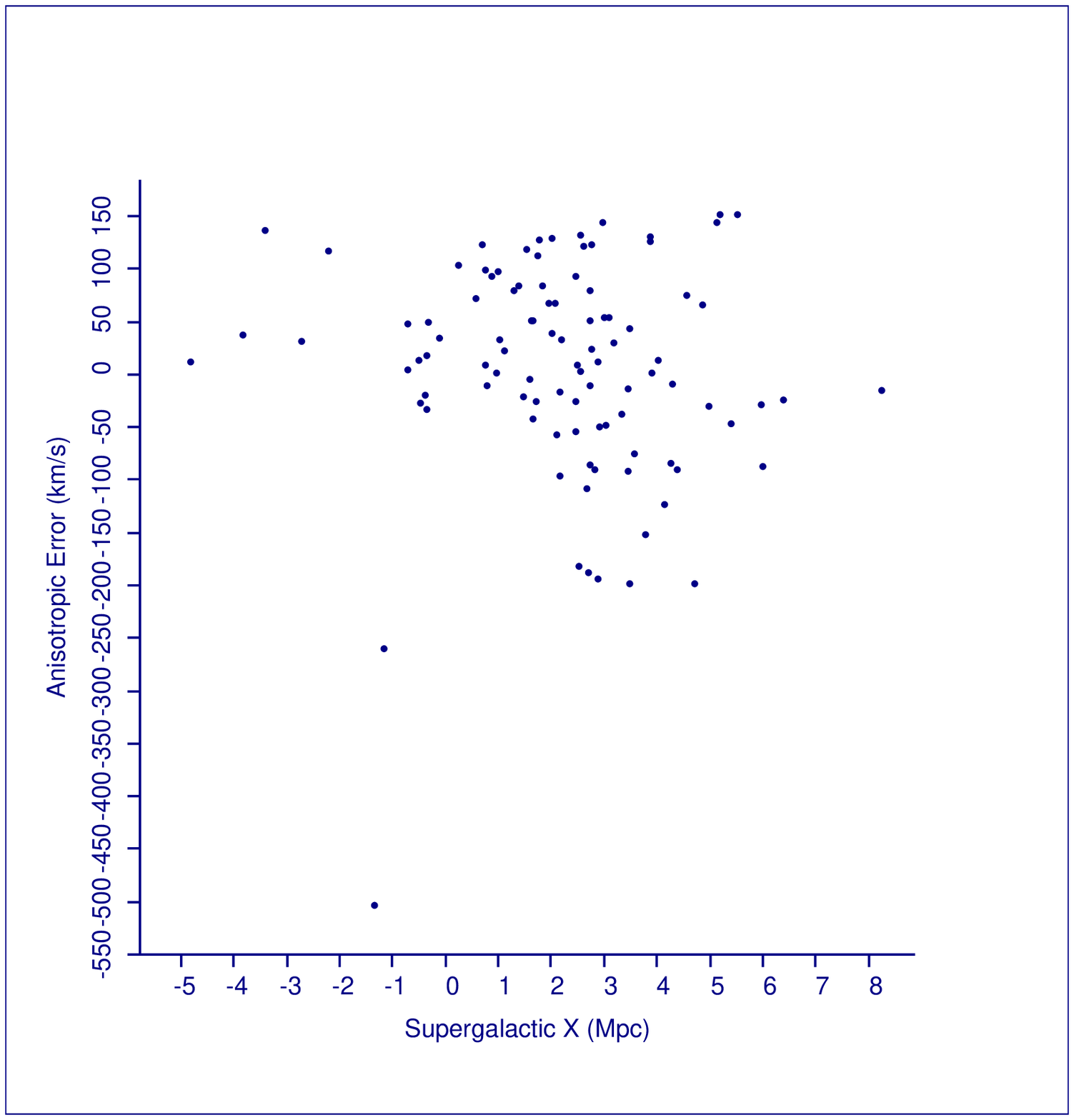}
\caption{Dispersions around the solution against Supergalactic X for the 98 galaxy
isotropic (left) and tensor (right) solutions. \label{fig:99X}}
\end{figure}

\begin{figure}
\plottwo{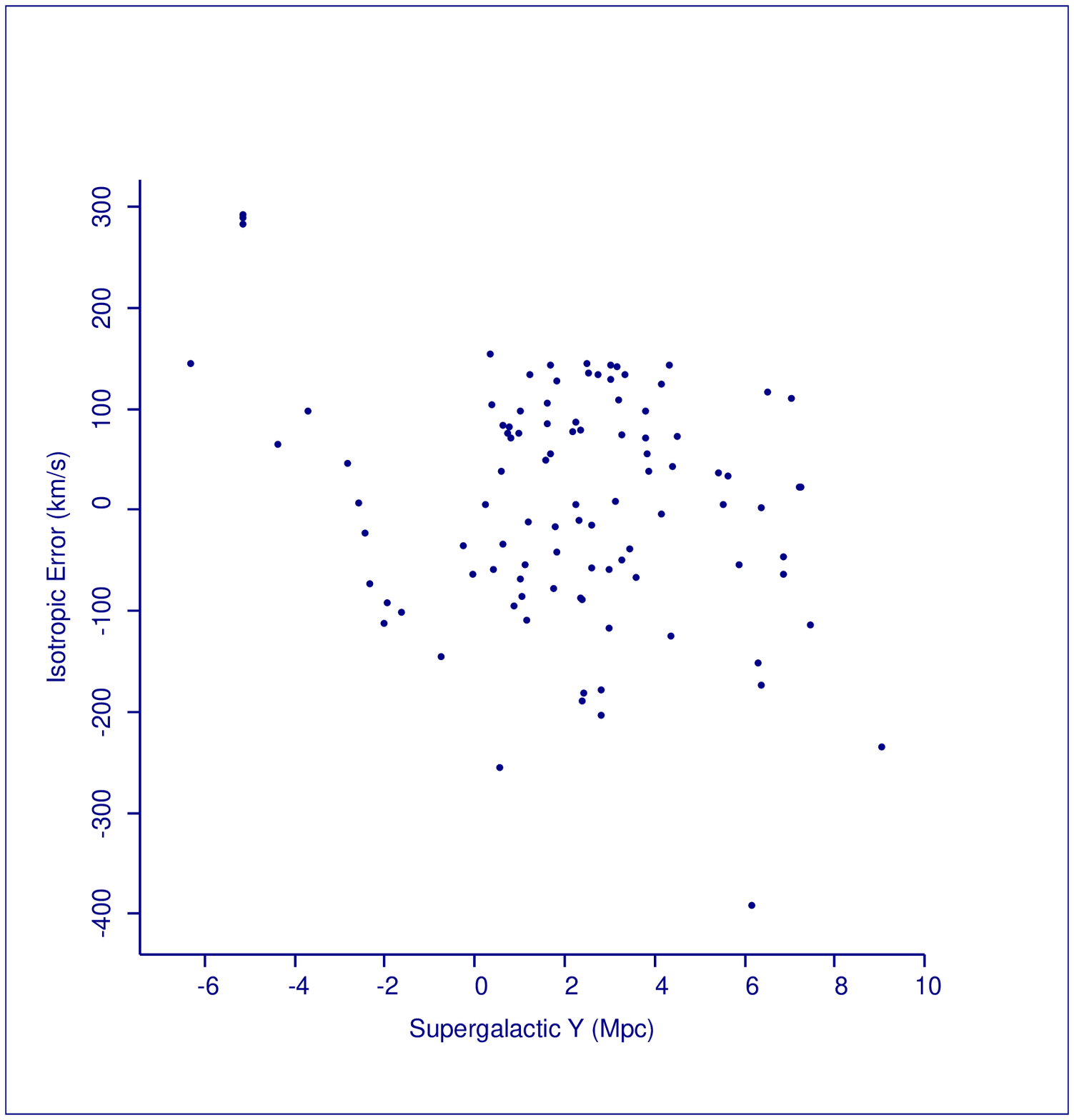}{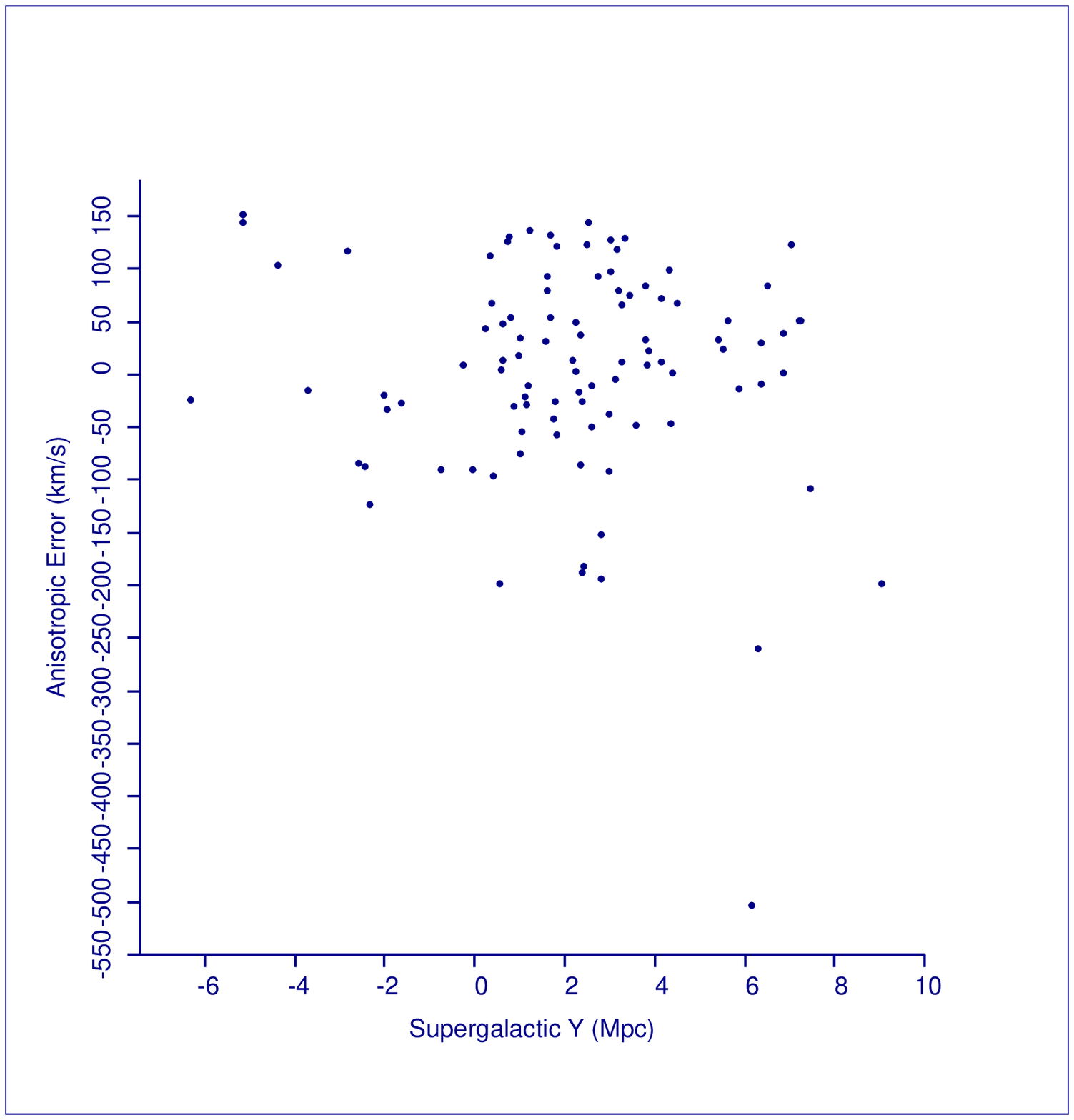}
\caption{Dispersions around the solution against Supergalactic Y for the 98 galaxy
isotropic (left) and tensor (right) solutions. \label{fig:99Y}}
\end{figure}

\begin{figure}
\plottwo{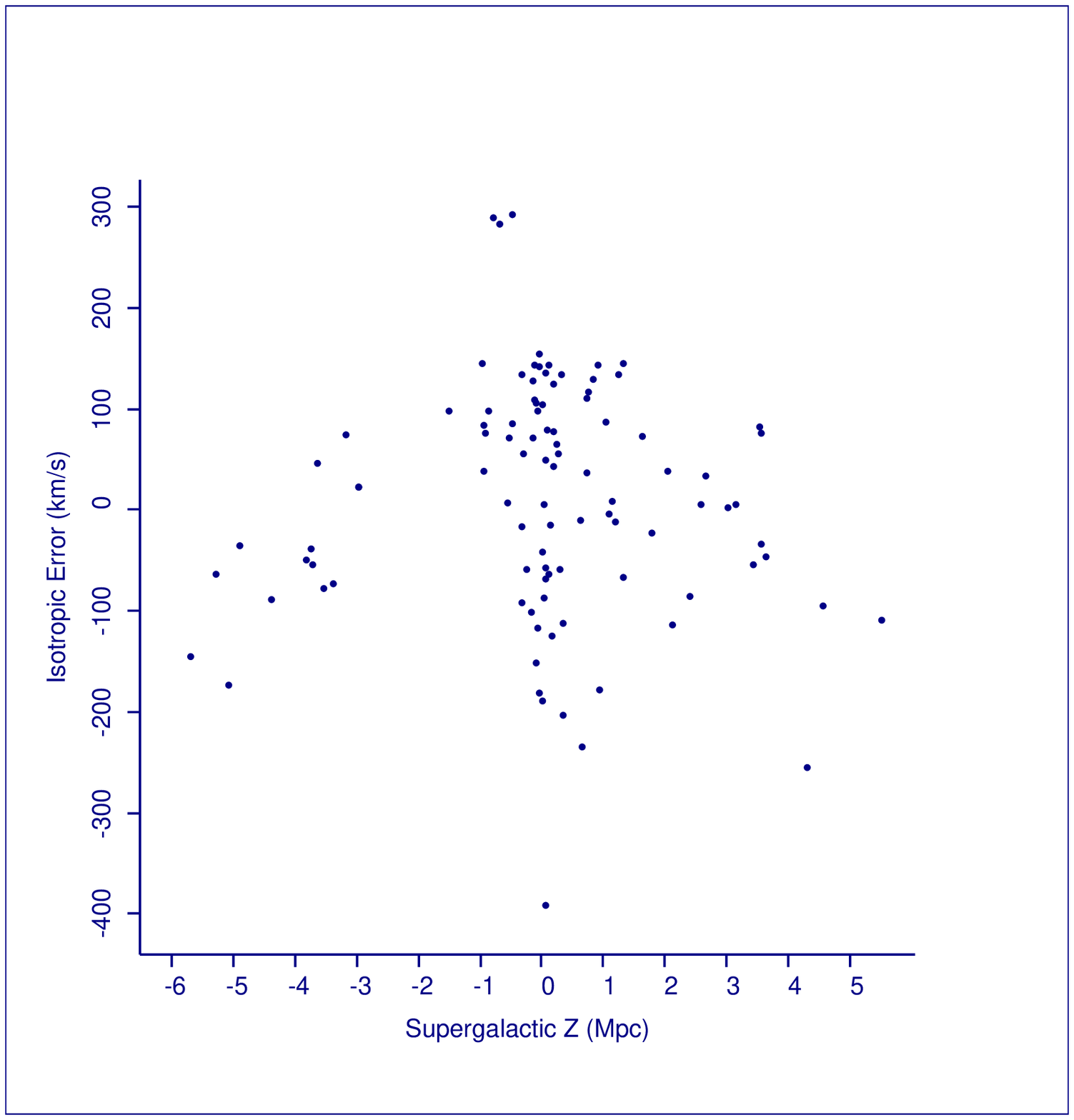}{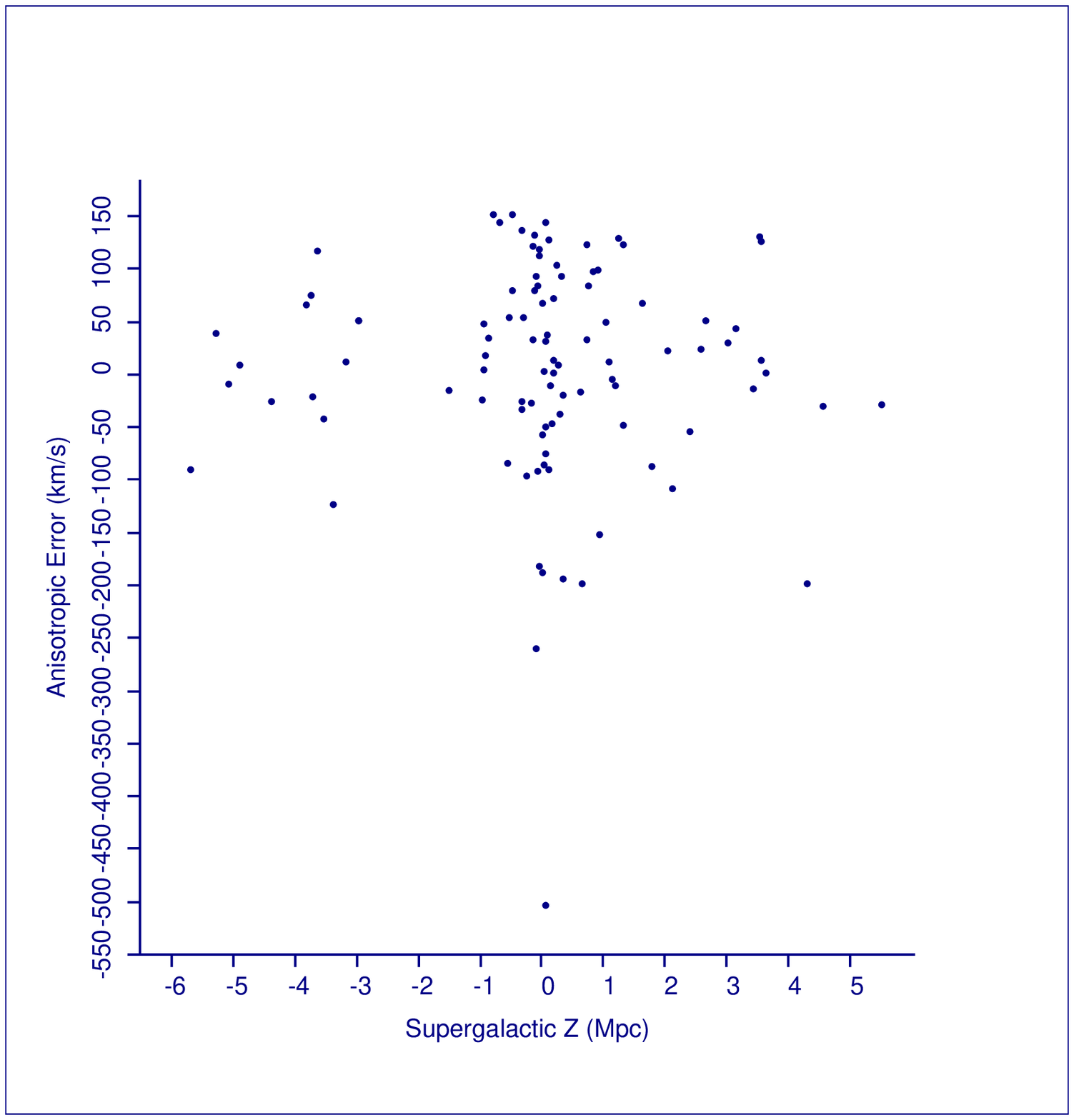}
\caption{Dispersions around the solution against Supergalactic Z for the 98 galaxy
isotropic (left) and tensor (right) solutions.}
\label{99Z}
\end{figure}

\begin{figure}
\plottwo{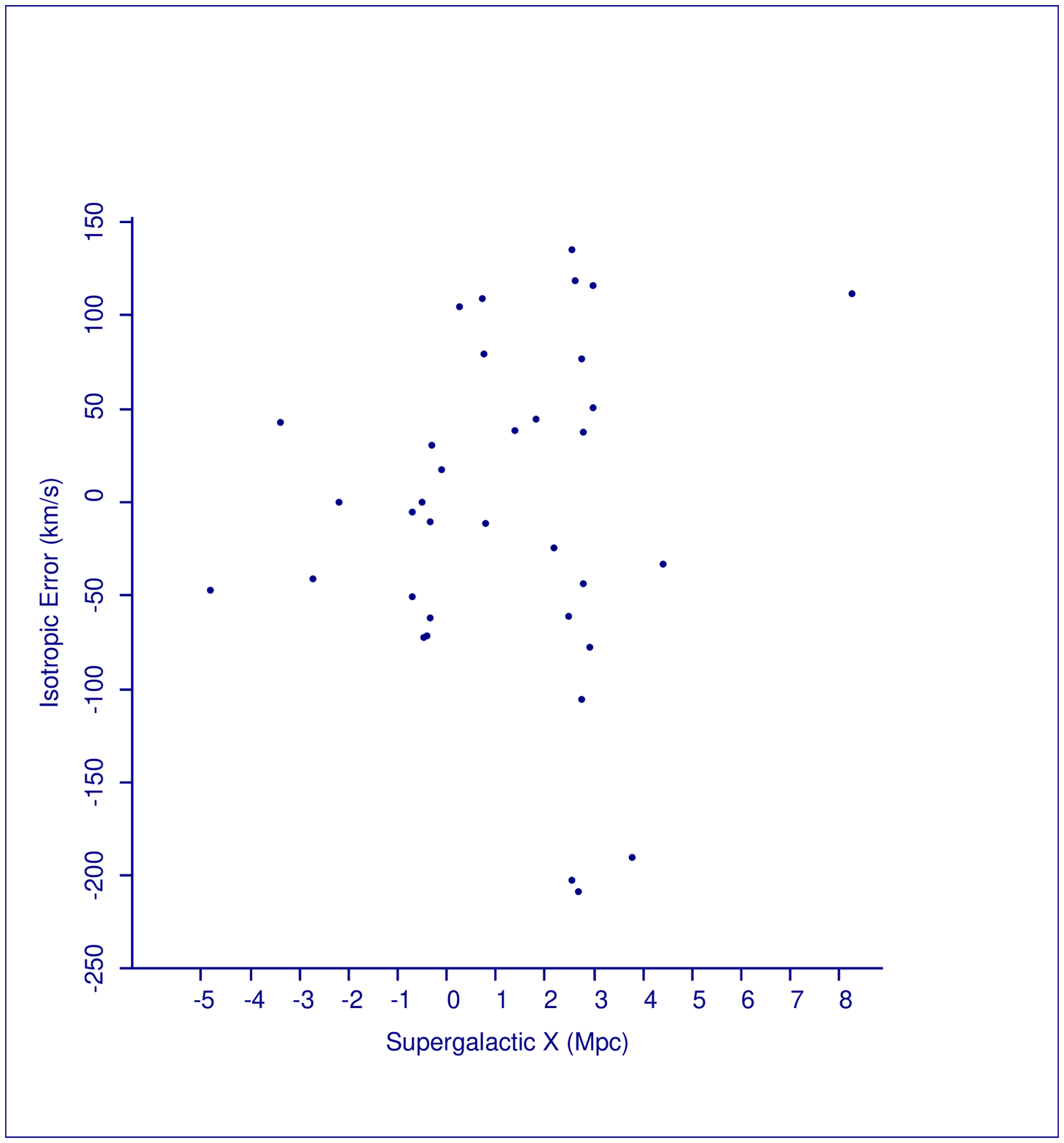}{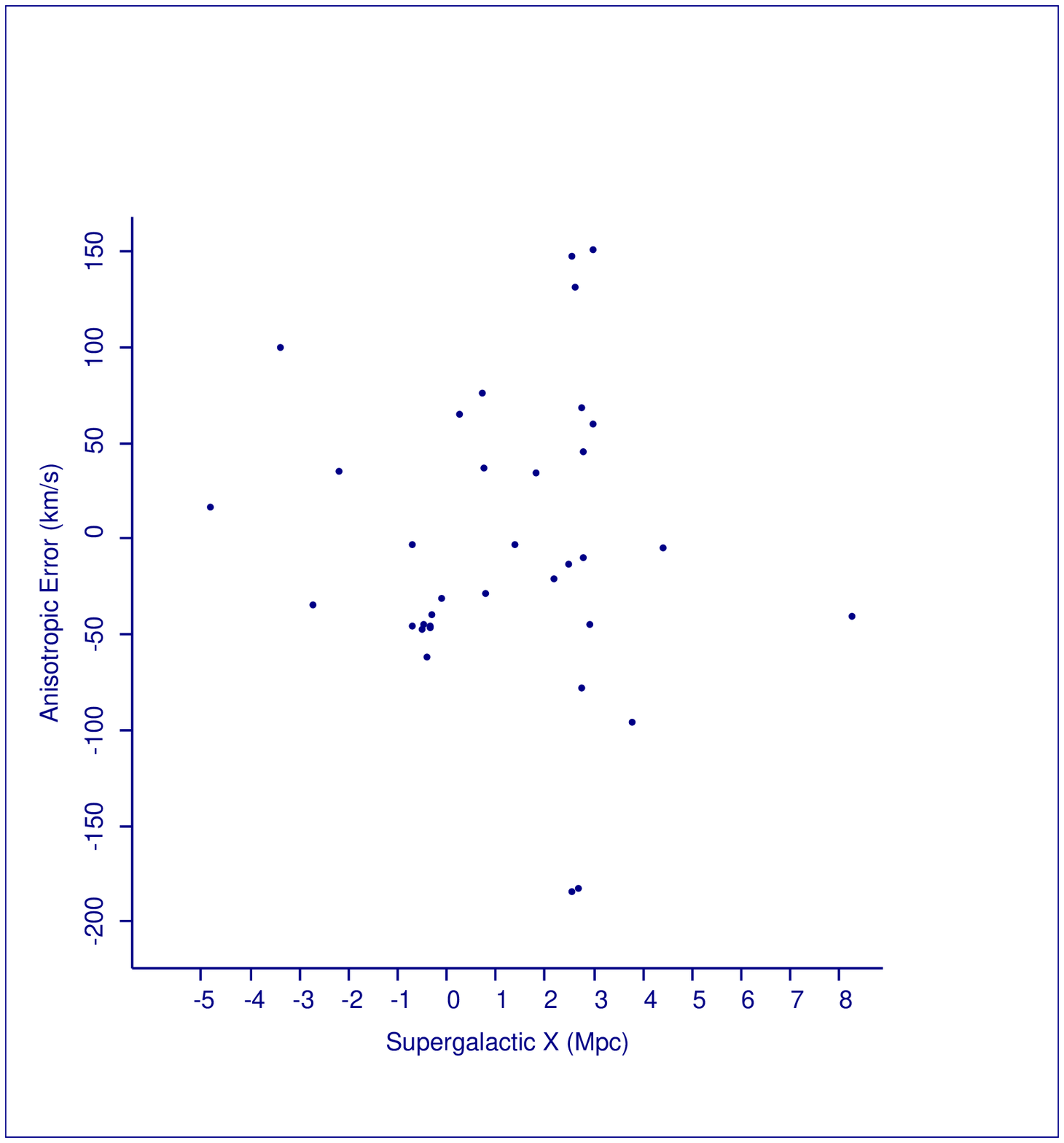}
\caption{Dispersions around the solution against Supergalactic X for the 35 galaxy
isotropic (left) and tensor (right) solutions.}
\label{35X}
\end{figure}

\begin{figure}
\plottwo{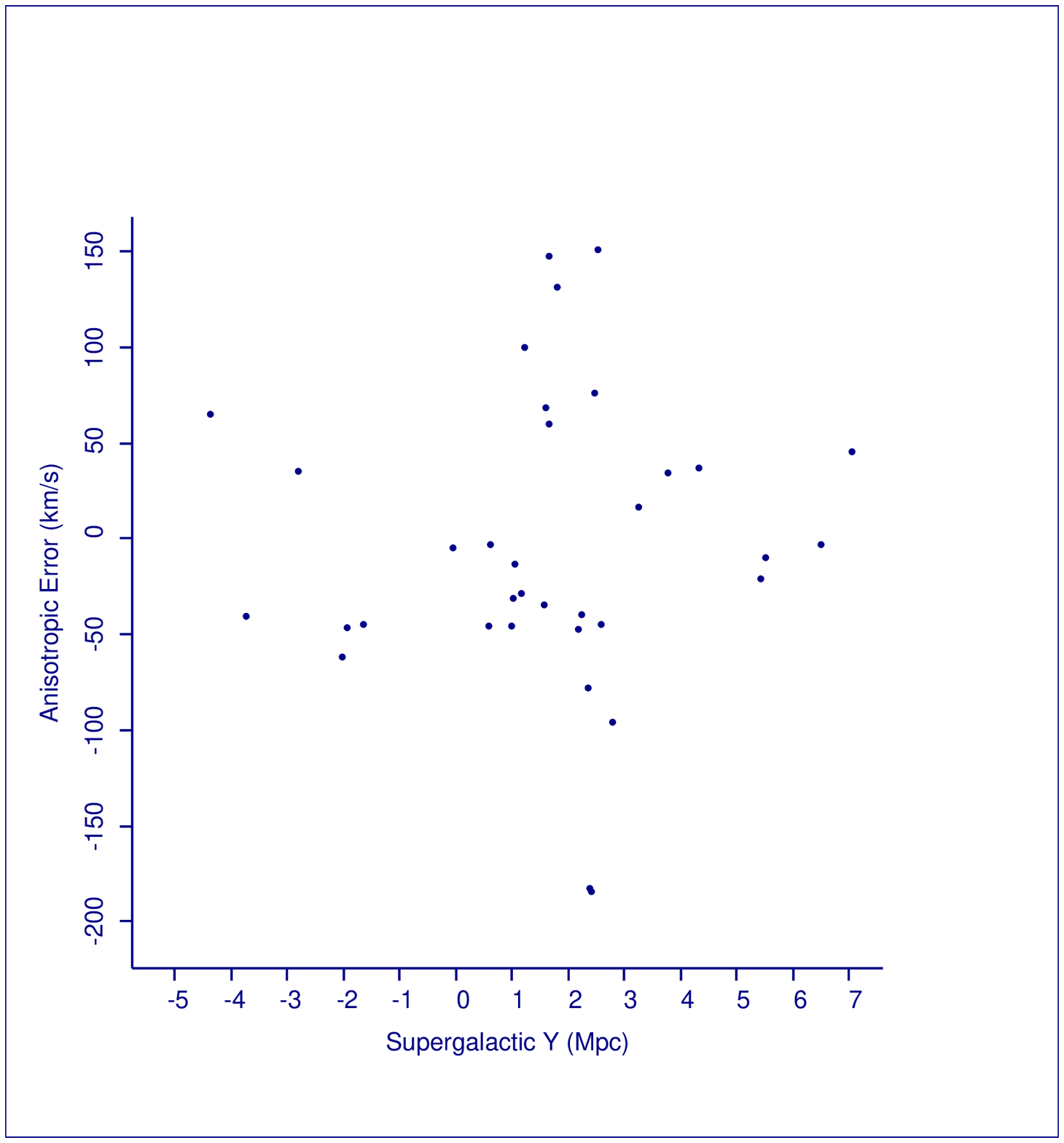}{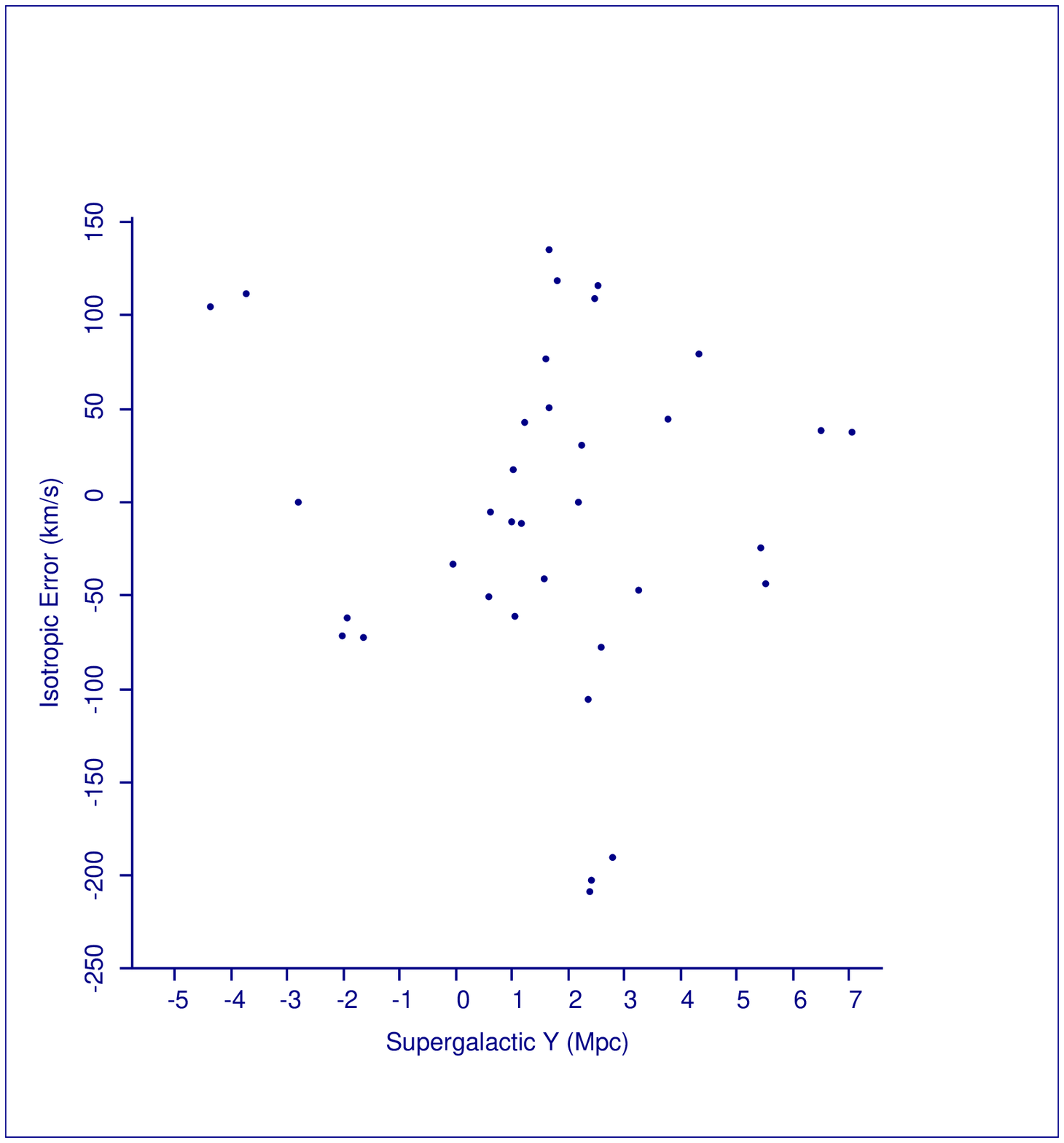}
\caption{Dispersions around the solution against Supergalactic Y for the 35 galaxy
isotropic (left) and tensor (right) solutions.}
\label{35Y}
\end{figure}

\begin{figure}
\plottwo{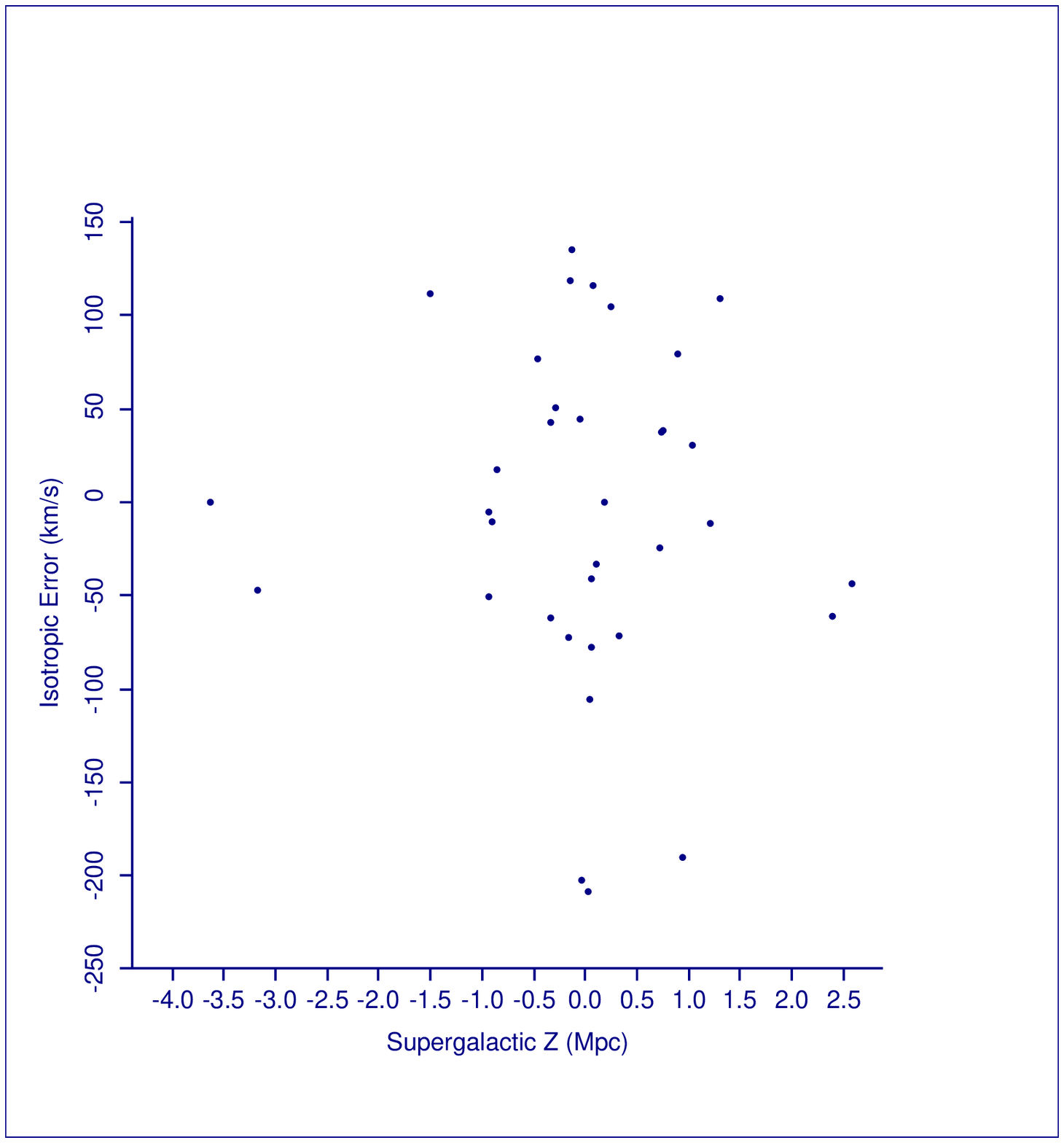}{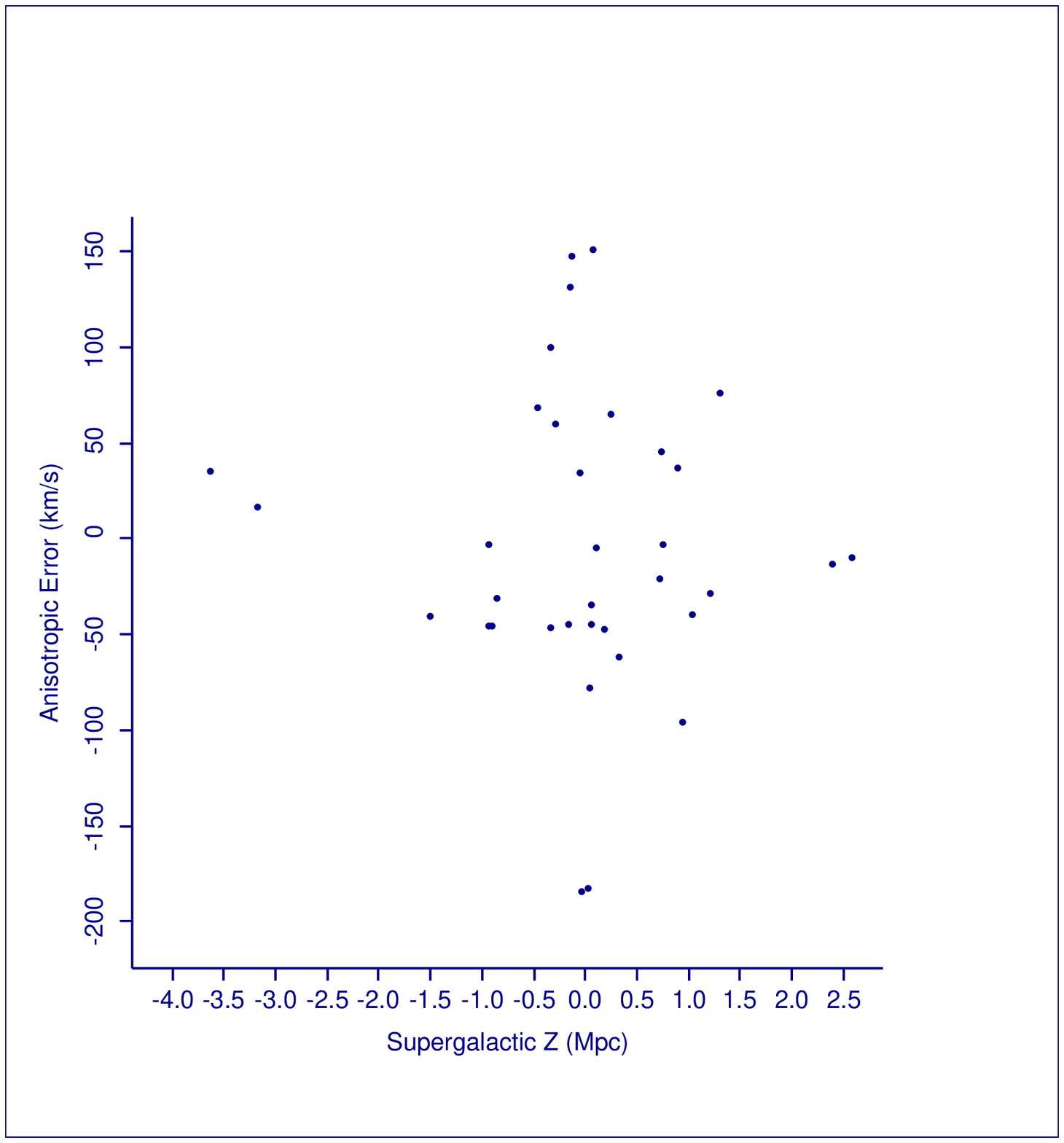}
\caption{Dispersions around the solution against Supergalactic Z for the 35 galaxy
isotropic (left) and tensor (right) solutions.}
\label{35Z}
\end{figure}

\begin{figure}
\plottwo{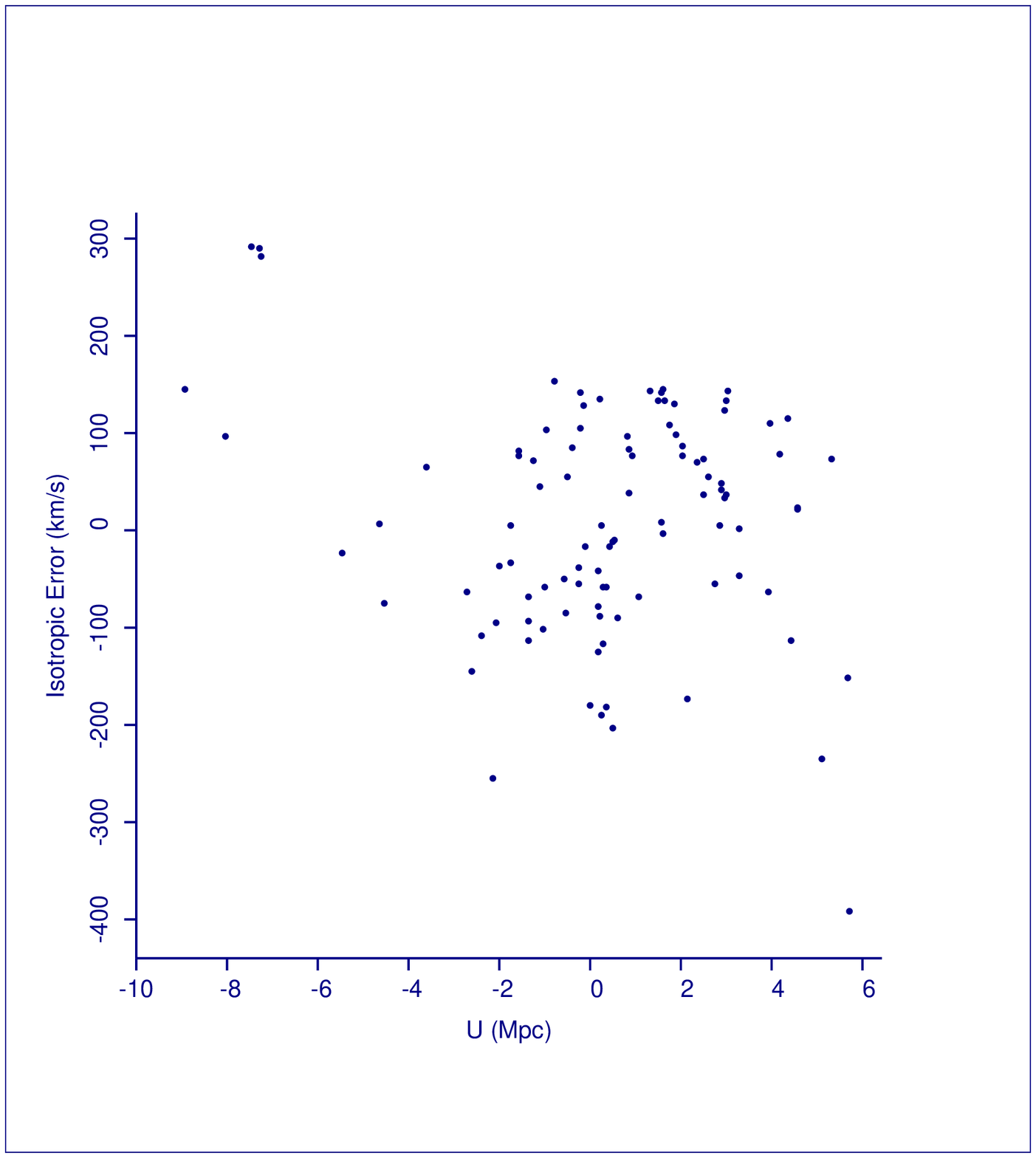}{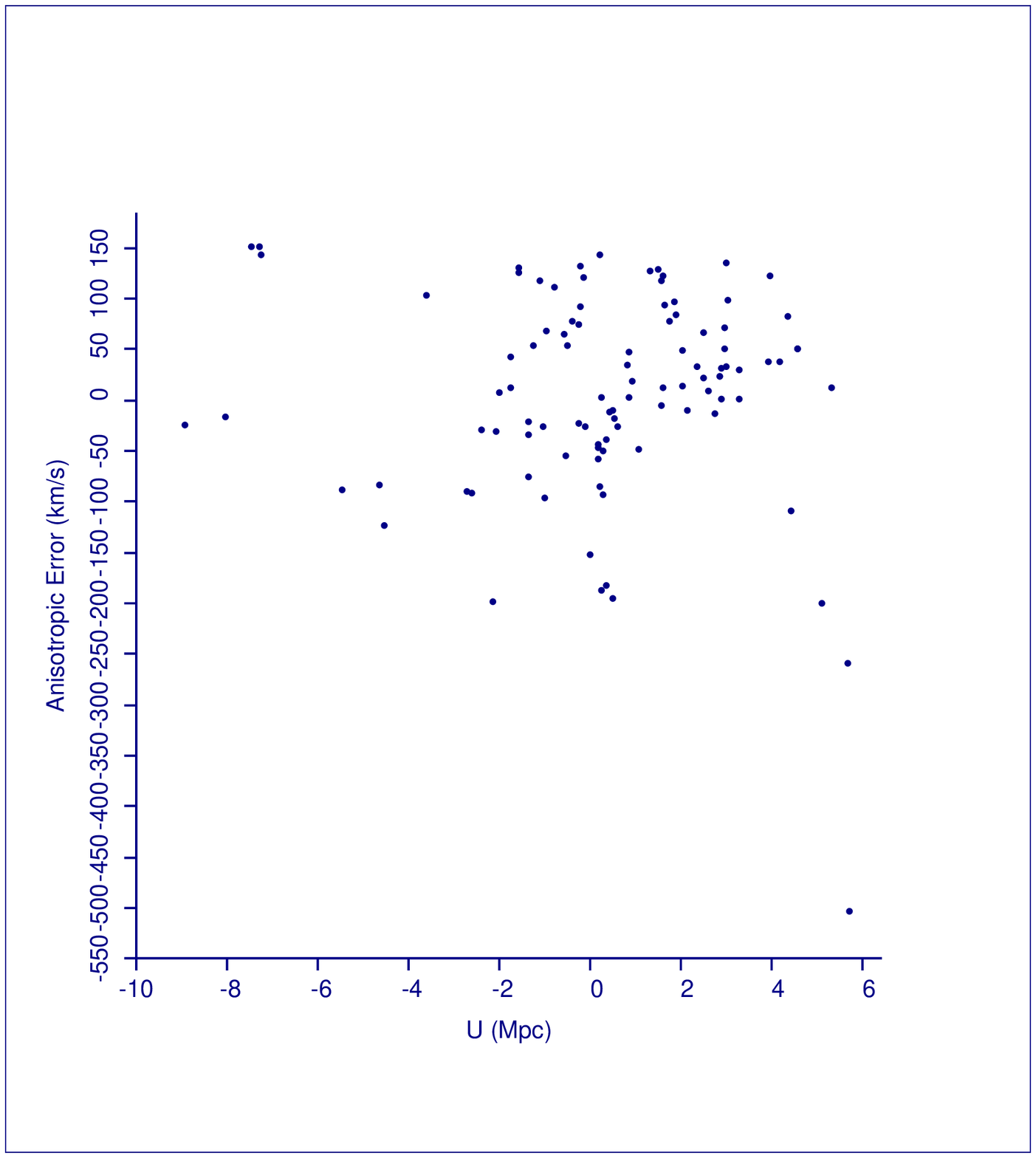}
\caption{Dispersions around the solution against eigenvector U (L = $127^{\rm o}$,
B = $3^{\rm o}$) for the 98 galaxy
isotropic (left) and tensor (right) solutions.}
\label{99U}
\end{figure}

\begin{figure}
\plottwo{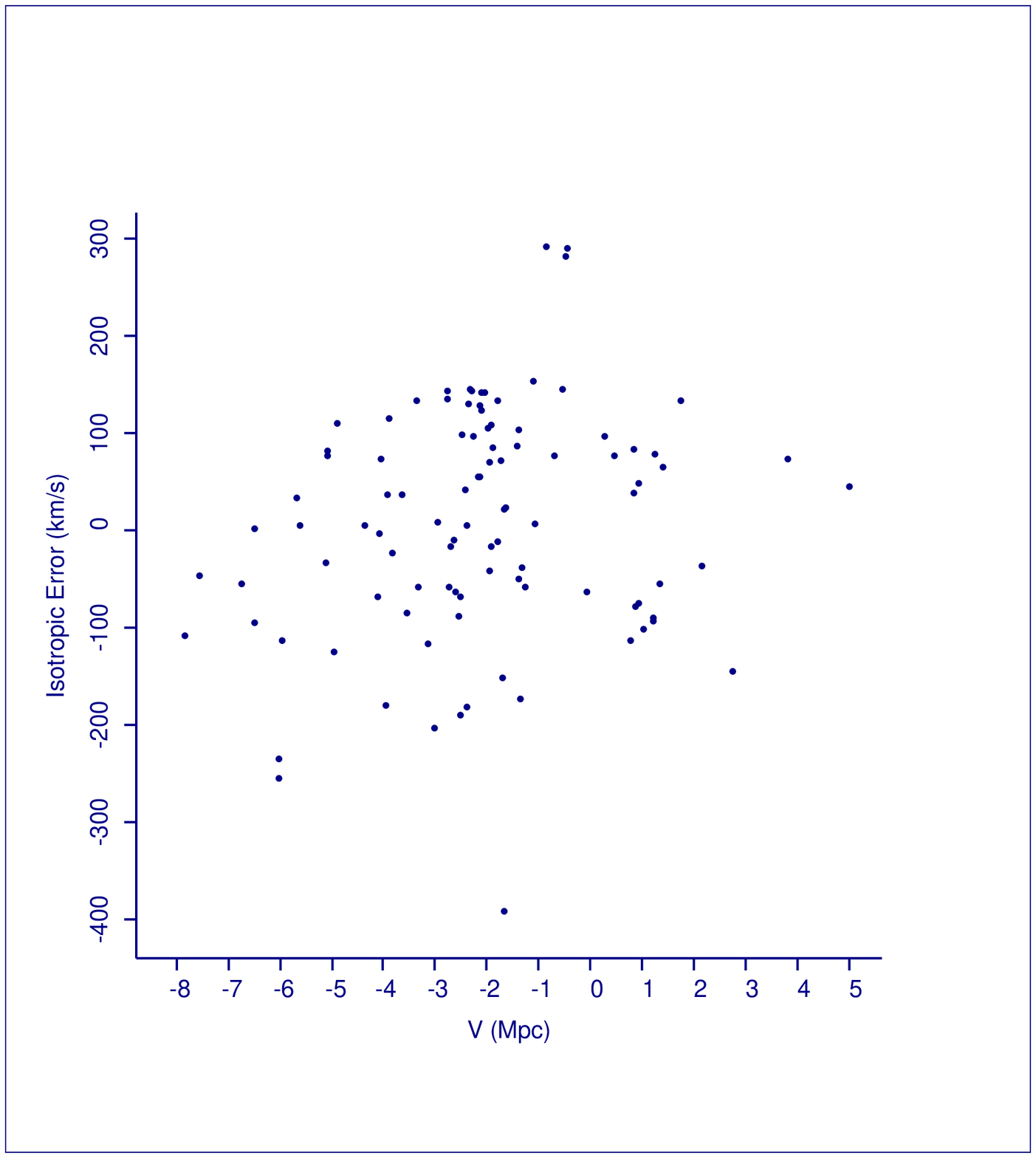}{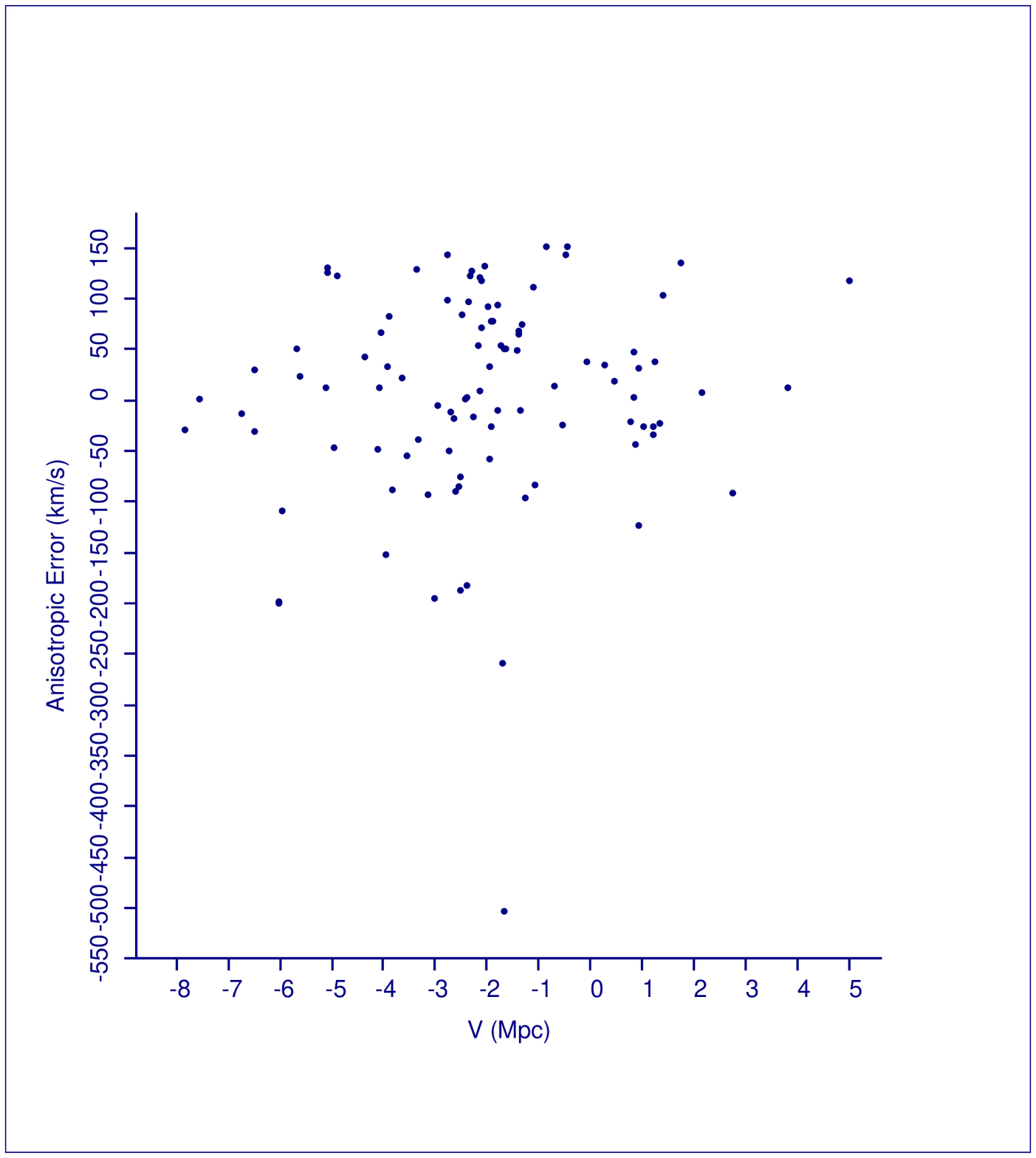}
\caption{Dispersions around the solution against eigenvector V (L = $34^{\rm o}$,
B = $46^{\rm o}$ for the 98 galaxy
isotropic (left) and tensor (right) solutions.}
\label{99V}
\end{figure}

\begin{figure}
\plottwo{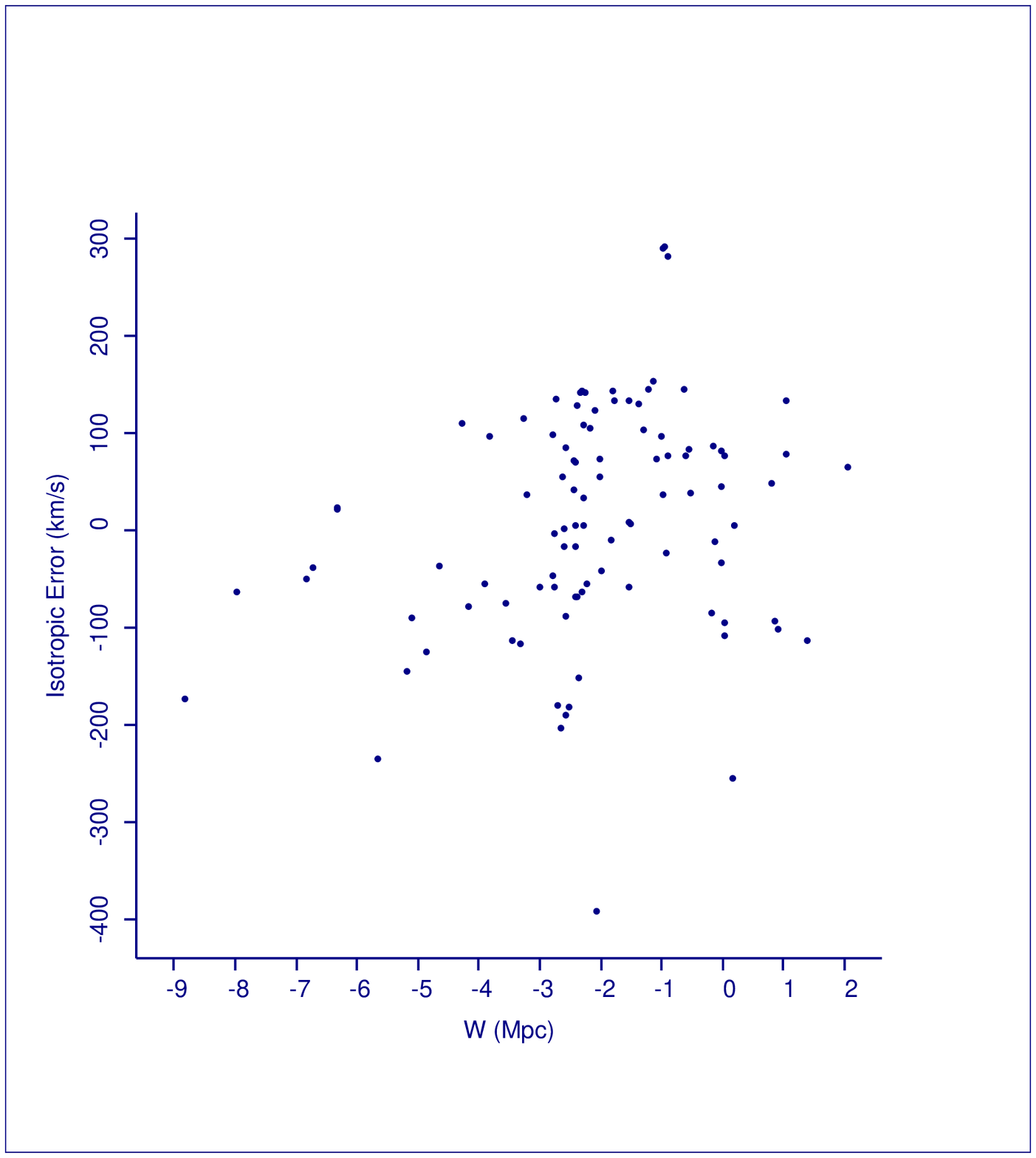}{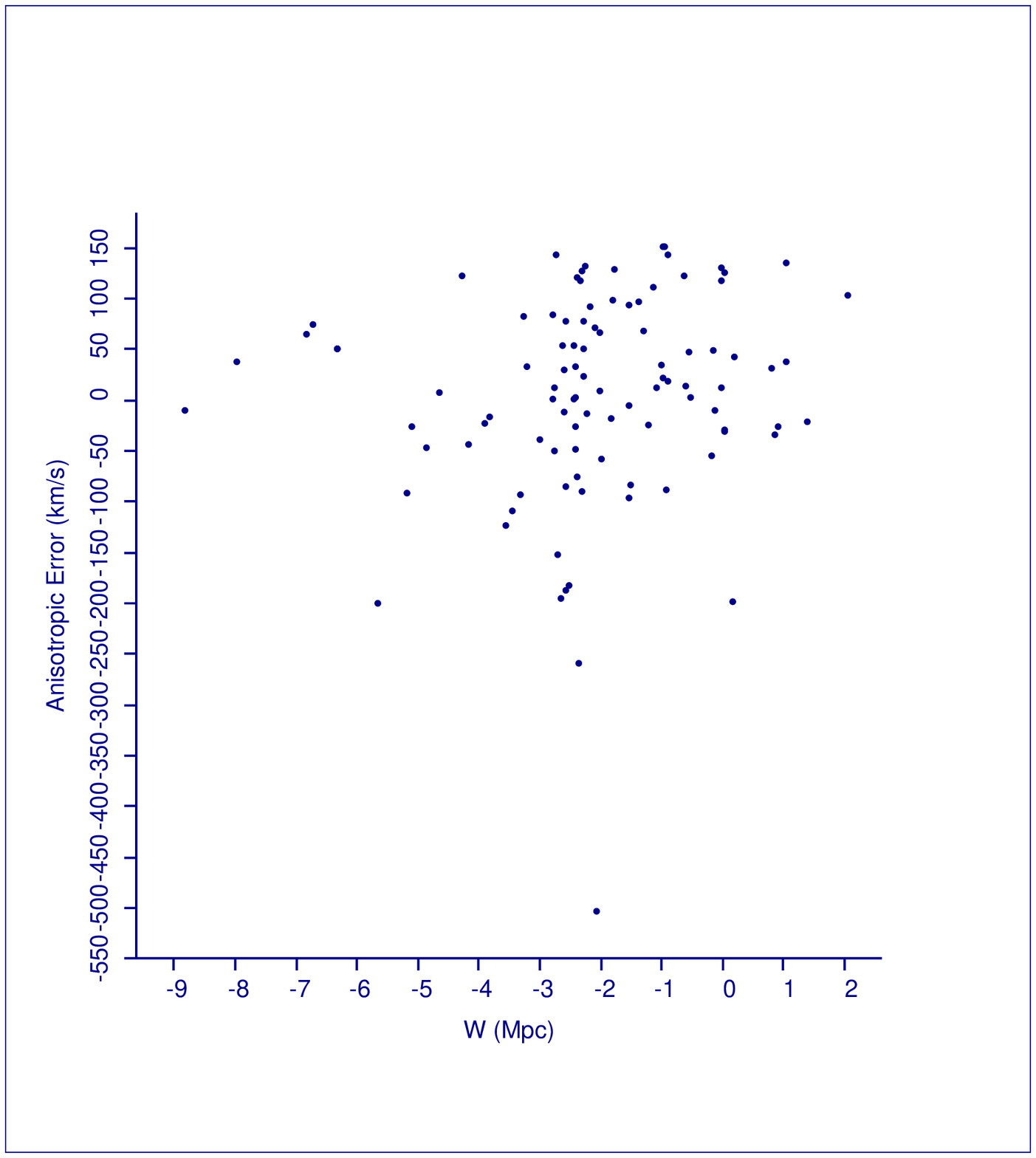}
\caption{Dispersions around the solution against eigenvector W (L = $40^{\rm o}$,
B = $-44^{\rm o}$) for the 98 galaxy
isotropic (left) and tensor (right) solutions.}
\label{99W}
\end{figure}

\begin{figure}
\plottwo{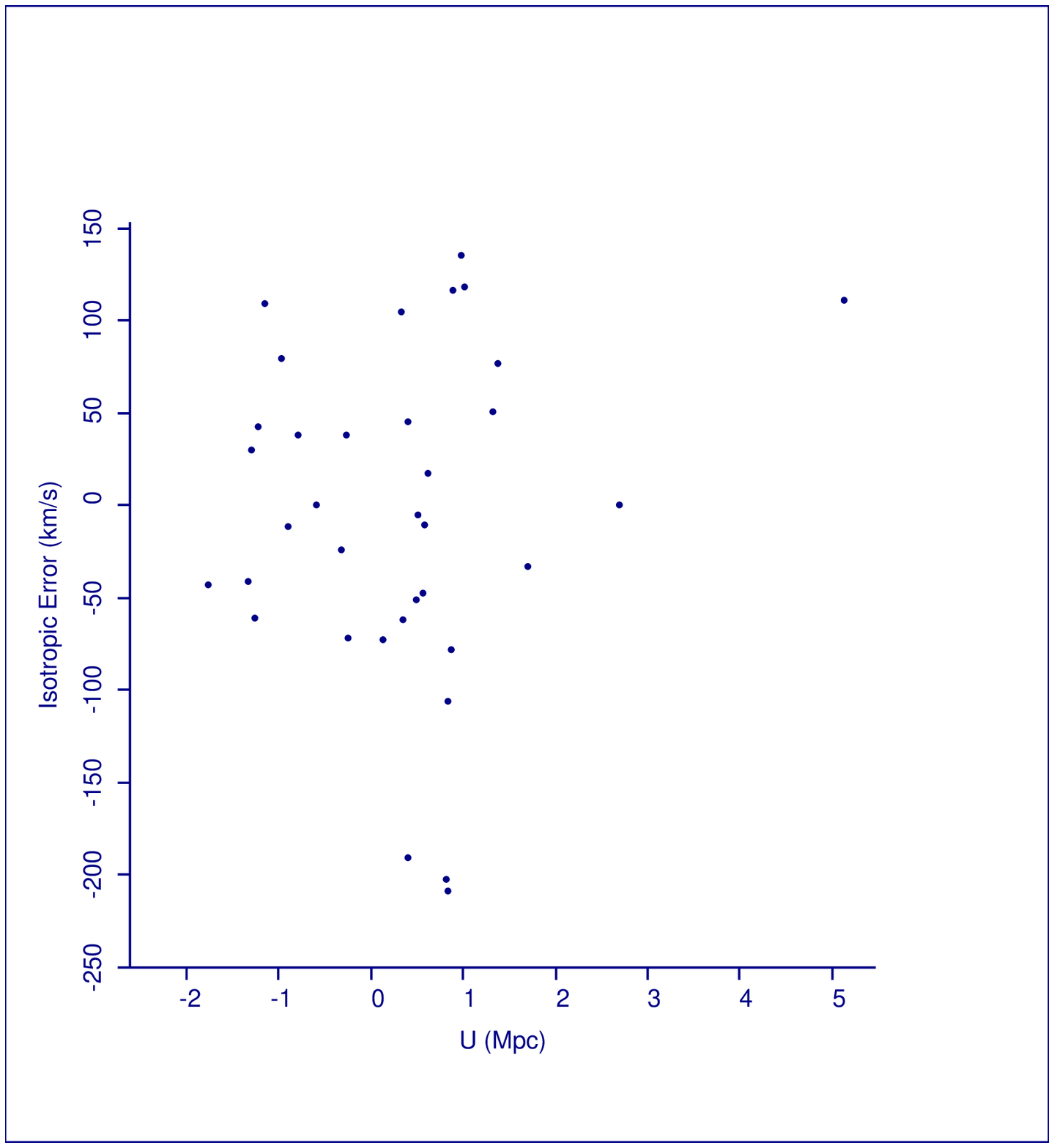}{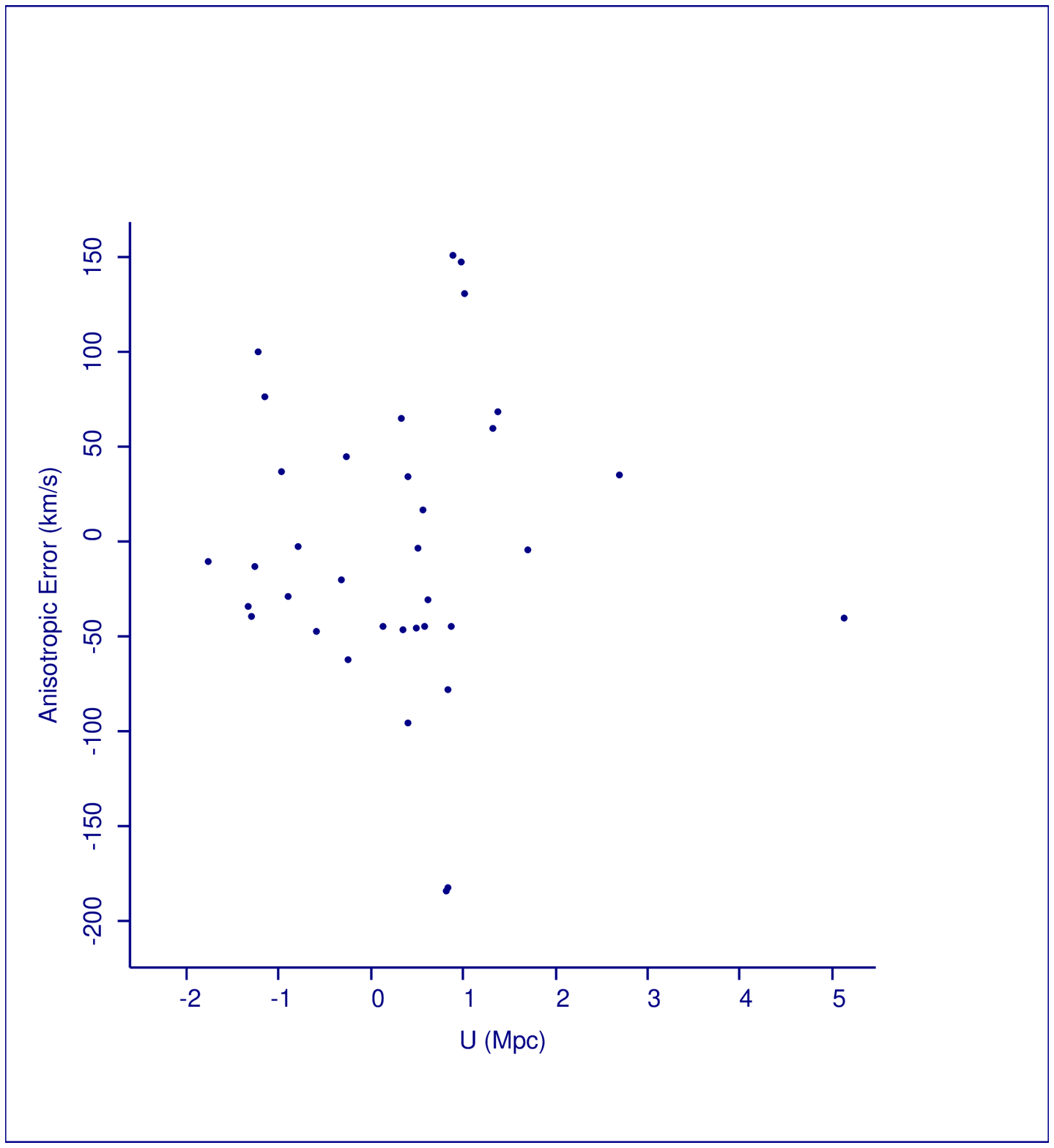}
\caption{Dispersions around the solution against eigenvector U (L = $66^{\rm o}$,
B = $ 65^{\rm o}$) for the 35 galaxy
isotropic (left) and tensor (right) solutions.}
\label{35U}
\end{figure}

\begin{figure}
\plottwo{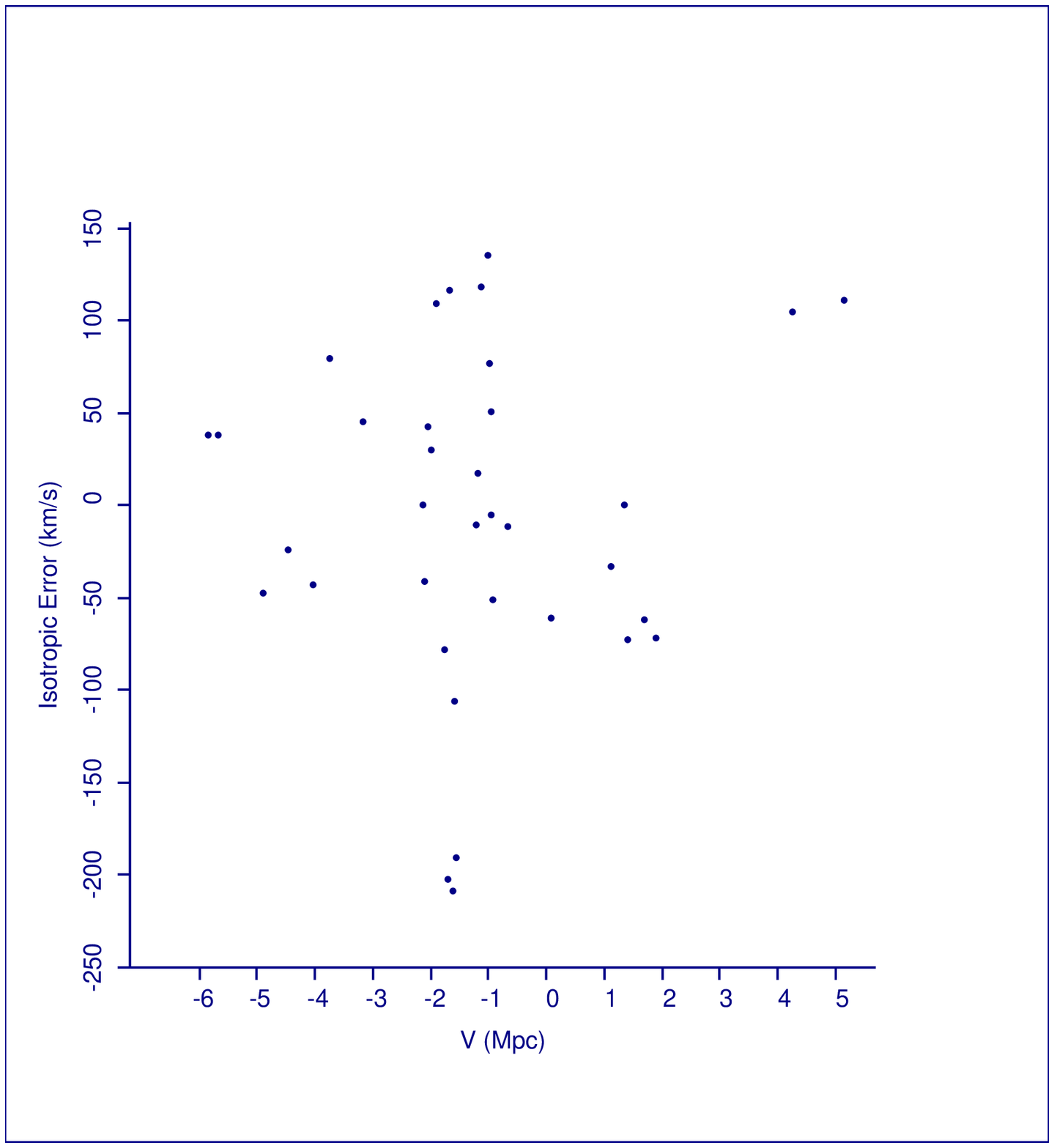}{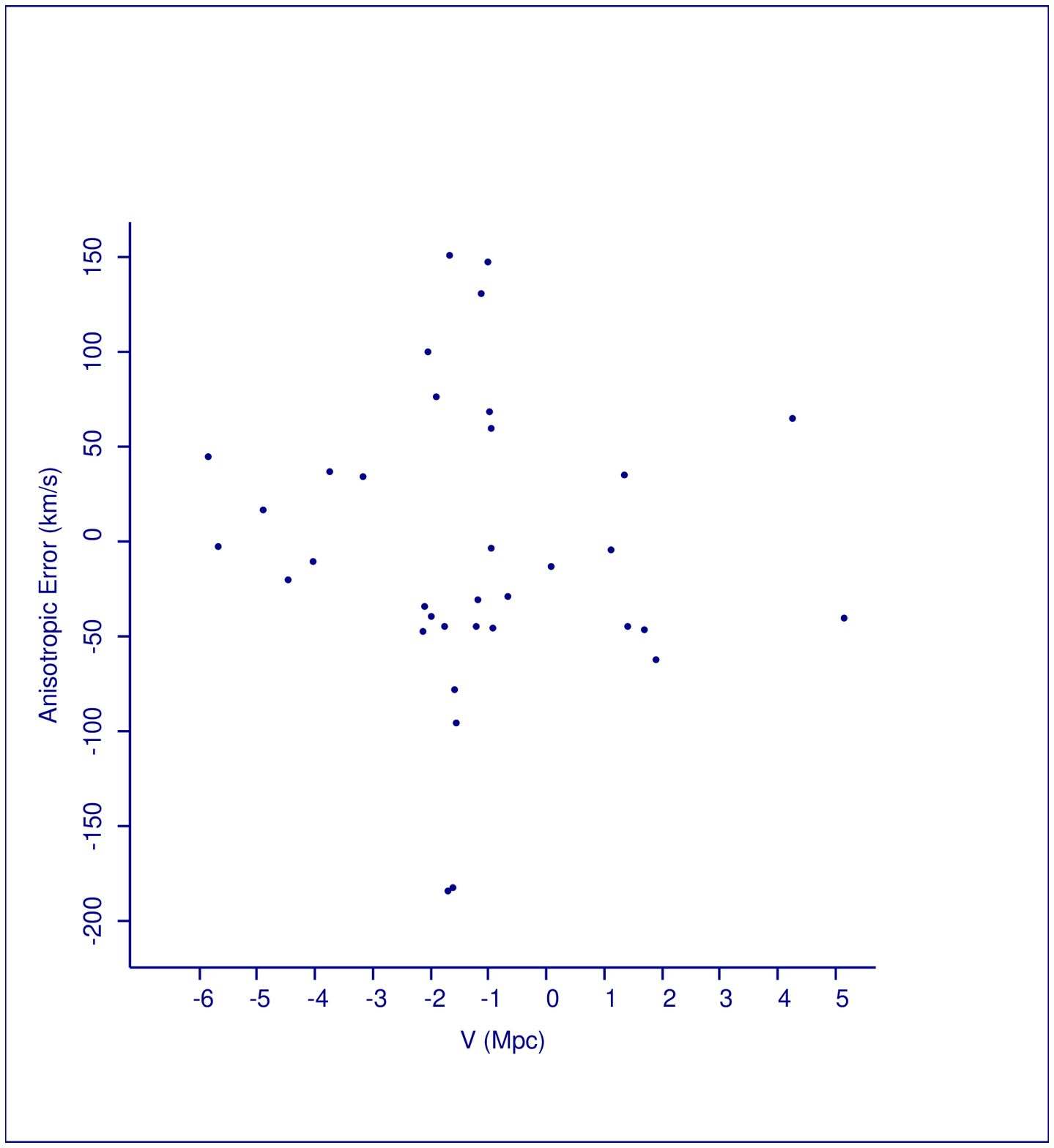}
\caption{Dispersions around the solution against eigenvector V (L = $104^{\rm o}$,
B = $-12^{\rm o}$) for the 35 galaxy
isotropic (left) and tensor (right) solutions.}
\label{35V}
\end{figure}

\begin{figure}
\plottwo{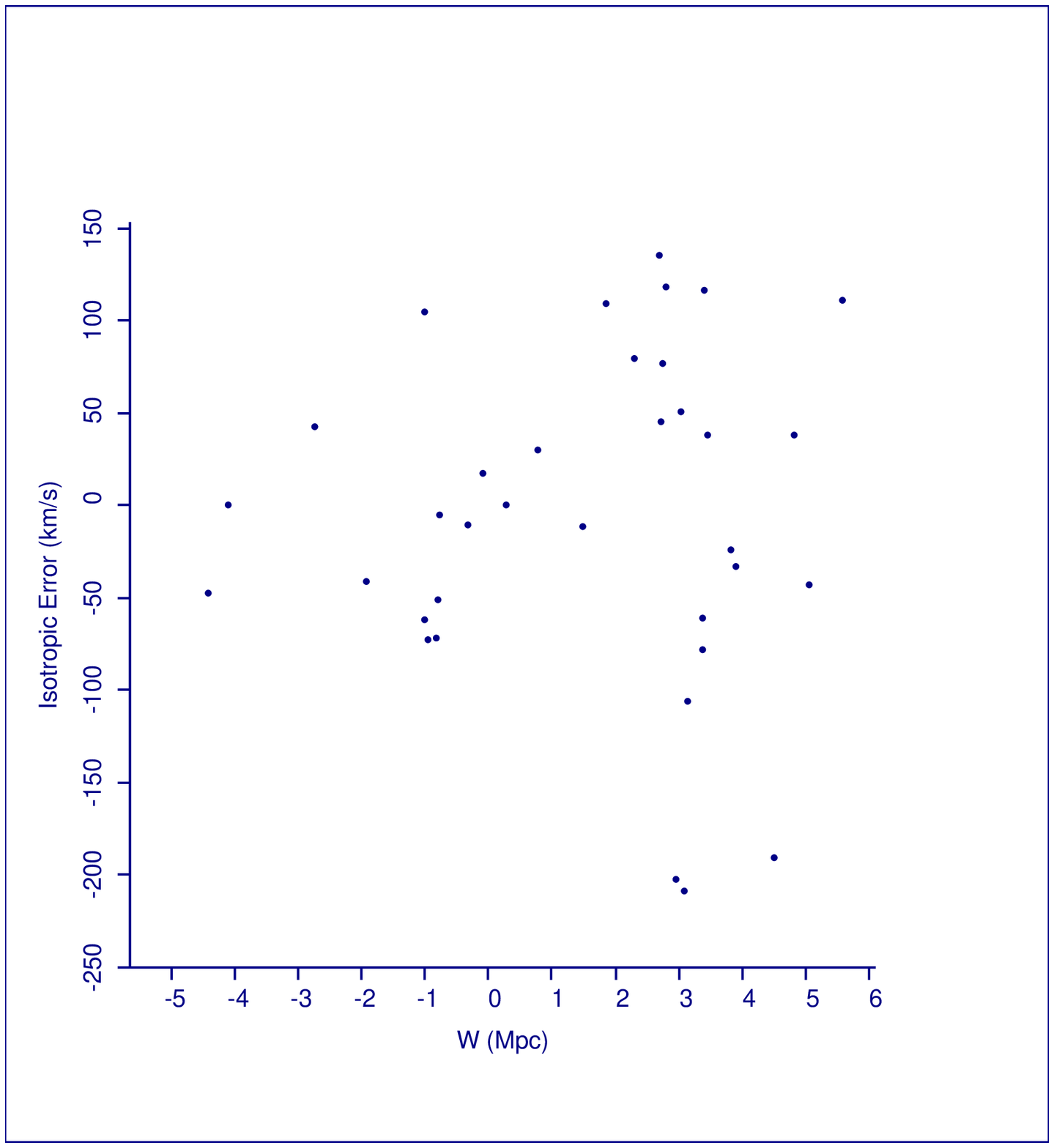}{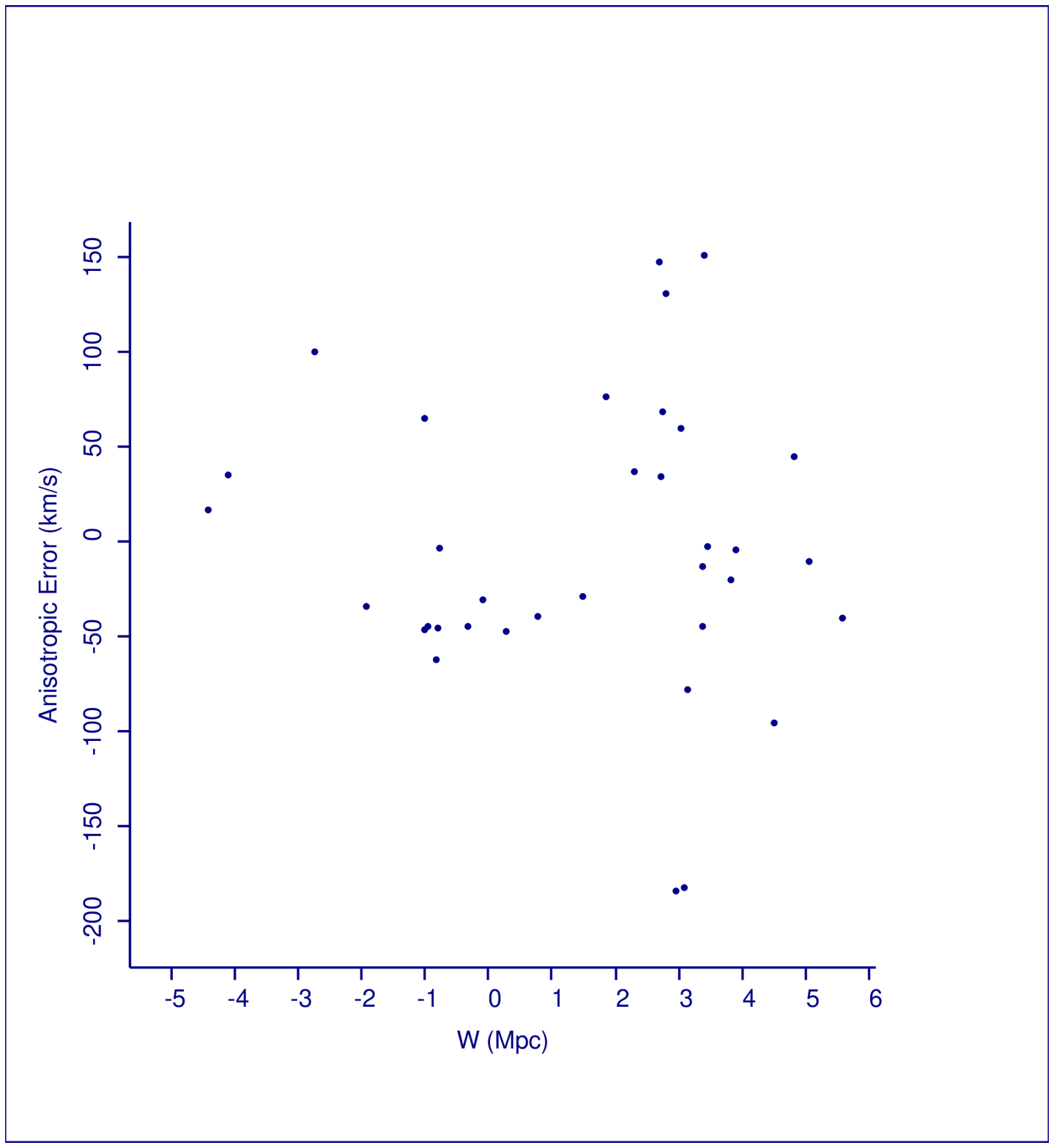}
\caption{Dispersions around the solution against eigenvector W (L = $19^{\rm o}$,
B = $21^{\rm o}$) for the 35 galaxy
isotropic (left) and tensor (right) solutions.}
\label{fig:35W}
\end{figure}

In Figures~\ref{fig:99X} through \ref{fig:35W} are plotted the deviations of each galaxy from
the various solutions (that is, how much its radial velocity differs from what the
model would predict) against the various spatial coordinates.  The first figures use
Supergalactic X, Y and Z; the second half use U, V and W, constructed along the
eigenvectors of the tensor solutions (note that the directions of each U, V and W axis are
{\em different} in the 98-galaxy and 35-galaxy solutions).

\clearpage

\clearpage

The  most important point about these plots is the {\em lack} of any
pattern.  The deviation of any particular galaxy from a model appears to
be quite random in space (with exceptions to be pointed out).

Consider in detail, for example, the Supergalactic X plots.  There is
clearly a lack of galaxies at negative X values, reflecting the uneven distribution
of data.  There is no clear trend of
error with postion, though the total width of the error is reduced somewhat in
the anisotropic solution.  The isolated point with a -400 to -500 km/${\rm sec}^{-1}$ error
is UGC 7857, a dwarf galaxy with only a brightest-star distance, in the general
direction of the Virgo cluster.  The distance may be in error, or the redshift could
be affected by a superimposed star (a problem noted several times in \citet{WHI02}); or,
just possibly, it could represent a very high-velocity tail of the peculiar velocity distribution
(a matter discussed below).  Note also that, in areas with points, the density is
approximately constant.  That is, there is no apparent concentration around any
given value.  This important observation will be expanded below.

The 98-galaxy Supergalactic Z, isotropic plot does show interesting systematic behavior.  
Recall that these are the residual radial velocities after the average expansion of the cloud of
galaxies has been subtracted.   There appears to be a trend for galaxies with ${\rm Z} < -2$ to
line up from lower left to upper right, which would indicate a lower effective Hubble
constant normal to the Supergalactic Plane.  This was noted by \citet{KMa96}
in their similar plot.  This flow is easily interpreted as the expected
lower effective Hubble constant normal to the Plane.

However, it disappears in the plot of higher-quality data (in part, it must be noted,
because the high-quality coverage of high-latitude galaxies is poor).  More important,
it does not show up in the Hubble tensor calculation in the form of a low eigenvalue
in the direction of the Supergalactic Pole, the nearest eigenvector being $45\arcdeg$
away.  It is not, then, a sign of any {\em average} slower Hubble flow out of the Plane.

Going to the coordinate system defined by the eigenvectors of the calculated
tensors, first we note that the 98-galaxy U plots look much like the Supergalactic
Y plots.  This is no surprise, since the respective axes are only a few degrees apart.
There is no clear trend in the other plots.  In particular, the dynamical behavior
discerned above in the 98-galaxy, Supergalactic Z plot (Figure~\ref{99Z}) has faded or
disappeared.  The fact that the calculated eigenvectors might actually conceal information
on the kinematics of the system is an indication that they are not useful in its
description.  This reinforces the conclusion of the previous section: the anisotropic
flow models tell more about a given data set than about the underlying galaxy motions.

The overall lack of any systematic behavior in the spatial plots is very significant.
It means that, at least as far as these data can show, there are no signficant ``bulk
flows'' within the 10 Mpc volume.  Also, any nonlinear effects, such as a second-order
tidal effect from the Virgo cluster, would show as a curve in at least some of the plots;
none are there.  In addition, significant mass concentrations should result
in a localized increase in the peculiar velocity width; nothing of this kind is seen,
although the data are too sparse to rule it out entirely.

\section{The Peculiar Velocity Distribution}

\subsection{The Shape of the Velocity Distribution}

\begin{figure}
\plottwo{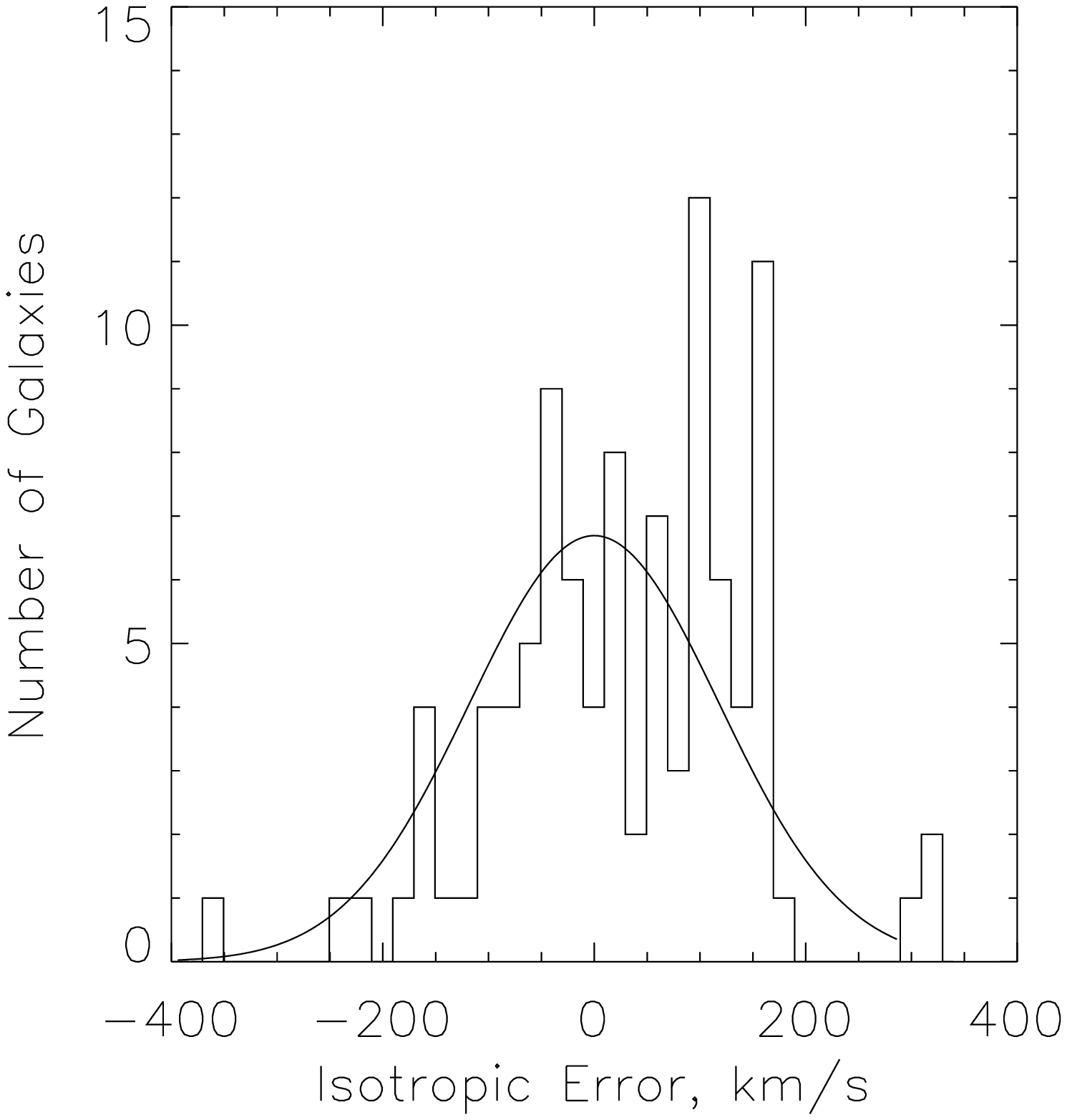}{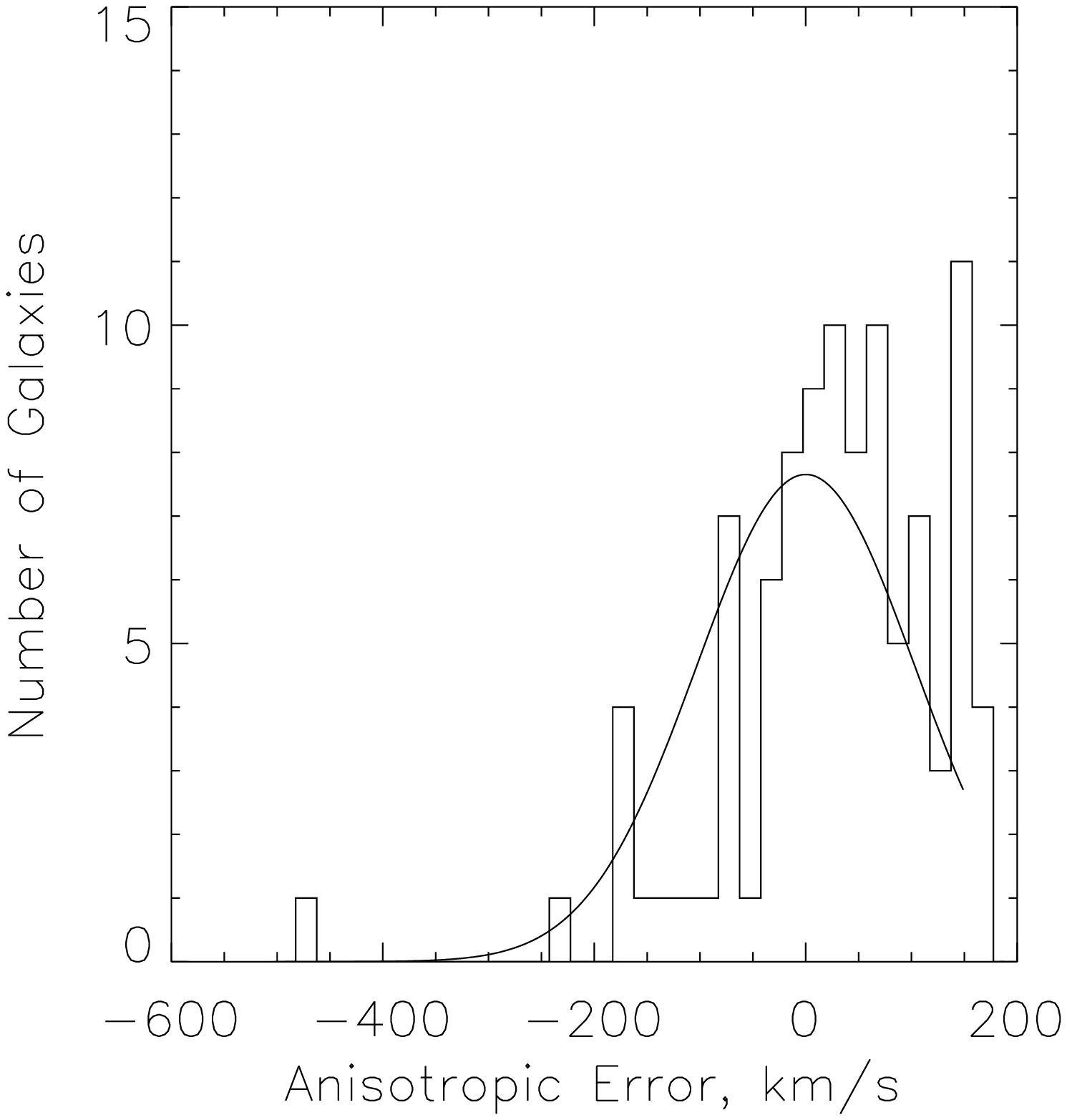}
\caption{Histogram of velocity errors about the isotropic (left) and tensor (right)
solutions for 98 galaxies.  Bins are 20 km ${\rm sec}^{-1}$ wide.  Gaussian distributions
of the same rms width, average, and normalization are shown superimposed.}
\label{hist1}
\end{figure}

\begin{figure}
\plottwo{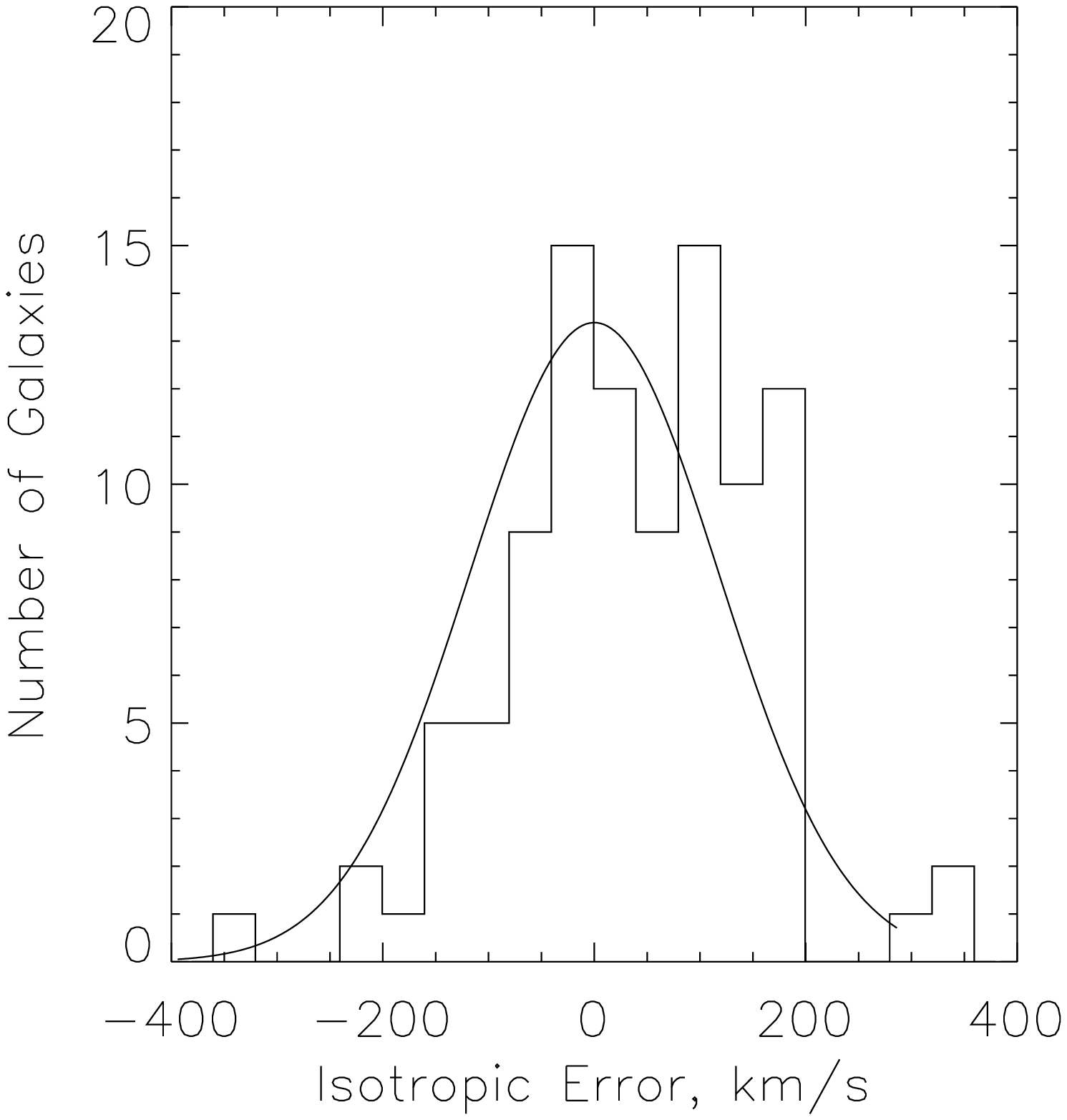}{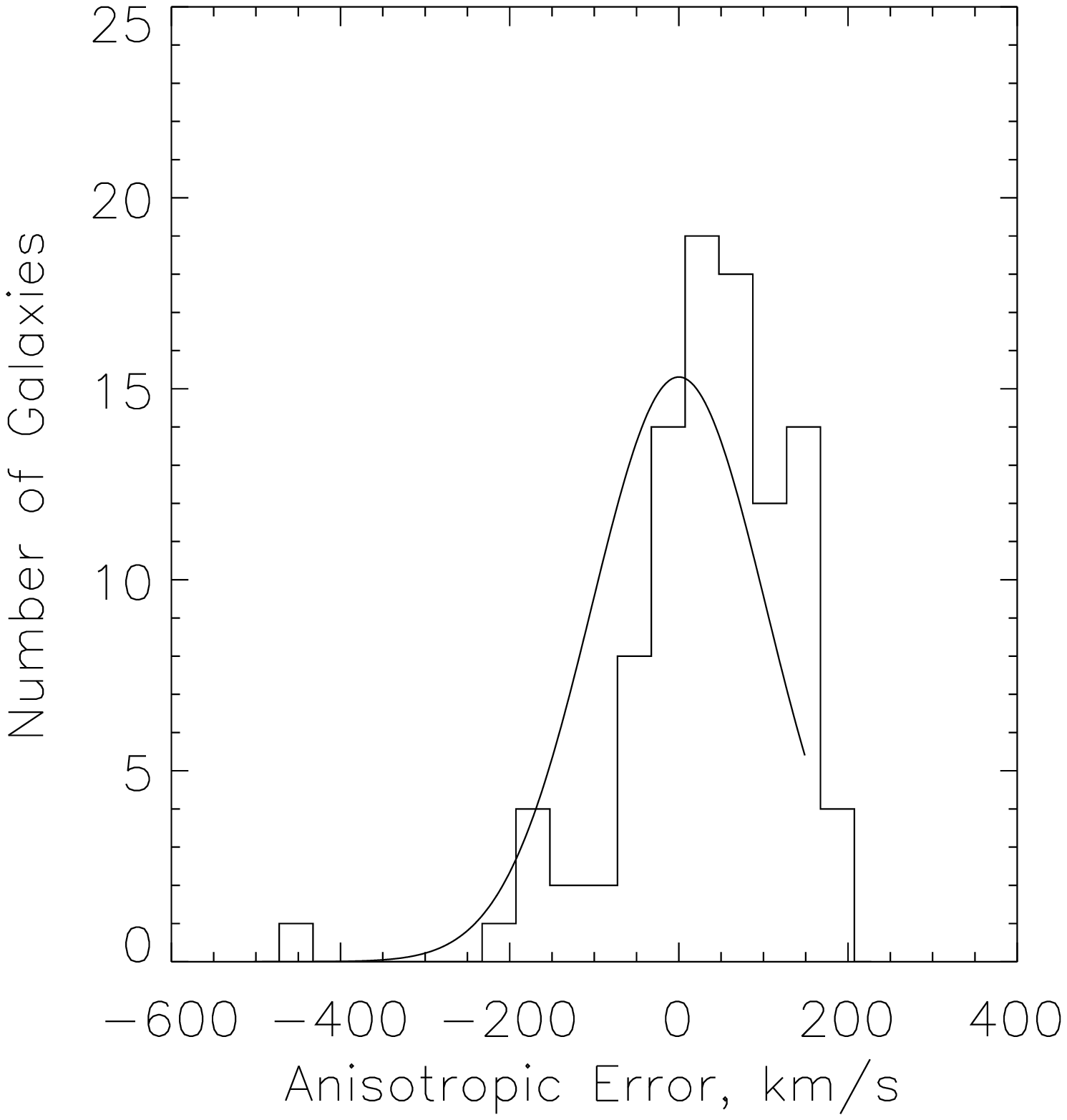}
\caption{Histogram of velocity errors about the isotropic (left) and tensor (right)
solutions for 98 galaxies.  Bins are 40 km ${\rm sec}^{-1}$ wide.  As before, 
corresponding Gaussian distributions are also shown.}
\label{hist2}
\end{figure}

It has been noted that there was no obvious concentration of deviations
from each model near zero; that is, it was roughly as likely to find galaxies far from the
model as close to it.  This is made more quantitative by Figures~\ref{hist1} and
\ref{hist2}, which show histograms of the number of galaxies against the
deviation from model flows\footnote{Only the 98-galaxy solutions are shown, since 
35 galaxies do not provide enough information to produce a useful histogram.  The choice
of bin sizes is a compromise between having enough galaxies in each bin to be significant,
and having enough bins to show a shape.  The 20 km s$^{-1}$ bins are clearly rather
noisy, so 40 km s$^{-1}$ bins are shown for comparison.  Between them, any spurious
signals due to unfortunate binning should be eliminated.}.
In each case a Gaussian distribution with the same average and standard deviation has
been superimposed.

\clearpage

\clearpage

Each histogram is markedly asymmetric, showing a clear slope upward toward positive
errors and a rather abrupt cutoff.  
(Quantitatively, a $\chi$-squared test comparing the four histograms to Gaussian distributions
enables us to rule them out with 98\% to over 99\% confidence.)
This is almost certainly a selection effect.
When making up a list of galaxies out to a certain distance, those with high radial
velocities are generally excluded, the assumption being that most of that velocity is
due to Hubble flow (for instance, \citet{SB92}).  Those with a peculiar velocity slightly
larger than
 200 km ${\rm sec}^{-1}$ in addition to a Hubble flow of 300 km ${\rm sec}^{-1}$
(appropriate to 5 Mpc at the average local Hubble value from the calculations here)
simply will not be included in a sample limited to 500 km ${\rm sec}^{-1}$.  While the
galaxies in the present data samples are known from information other than
radial velocities to be close by, one may expect most of them (at least) to have been chosen for study
initially based on radial velocity.

The lack of high radial velocity galaxies in the sample means that the Hubble constant
and Hubble tensor eigenvalues calculated above are biased low. 
The velocity dispersion given in Table~\ref{tensor} for each solution is also significantly
underestimated, and there must be many known galaxies
not now considered members of the Local Volume which in fact reside there.  (This
statement probably does not apply to the brightest and most important galaxies in the sky,
which have been well-studied and which have other sorts of distance estimates.  But there
should be many of middling brightness and importance which are much closer than
has been assumed, and certainly many dwarfs.)  In addition, statistical estimators based
on Gaussian statistics can no longer be trusted; this reinforces the results in 
section~\ref{sec-nogauss}.

Unfortunately, the actual
values cannot be estimated from the data without a good idea of the quantitative form of the selection
effect, and the quality and quantity of information at hand simply do not allow us to derive it.
Formally, one might have an almost unlimited peculiar velocity distribution, as long as it is
balanced by a high Hubble constant to avoid a large population of galaxies with negative
radial velocities (which would certainly have been noticed, but haven't been observed).

One might try to fit the observed shape of a histogram to an assumed form of the peculiar velocity
distribution, either Gaussian (which assumes a dynamically young system), Maxwellian (which 
assumes a dynamically old system) or the more sophisticated function derived by \citet{SCI2},
assuming that the low-radial velocity end of the  distribution is accurate.  However, note that
if  error bars of $\sqrt{n}$ are added to each bin in Figure~\ref{hist2} (isotropic model),
that is if Poisson statistics are assumed, the
shape is consistent with a flat distribution from -150 to +200 km s$^{-1}$; a monotonic linear
increase from -250 to +200 km s$^{-1}$, abruptly cut off at the upper limit; a Gaussian with
a center about +50 km s$^{-1}$ and its high-velocity end truncated; or a Maxwellian similarly
biased and truncated.  The data at hand simply do not allow accurate curve fitting\footnote{
A $\chi$-squared fit on a flat distribution shows it to be overall a worse fit than the
superimposed Gaussians.  However, the linear ``sawtooth'' mentioned above is much better, 
being rejected at only the 82\% level.  It is not offered seriously as a model, but shows
how asymmetric the distribution really is.}.

If we guess that the distributions shown actually contain most of the peculiar velocity
dispersion (alternatively, that the Hubble constant is not biased very much), then
200 km s$^{-1}$ forms a rough limit to the observed dispersion and the rms
value is something over half that\footnote{These figures include, of course, a dispersion from
observational errors in distances.}.  
A substantially unbiased sample must then include radial velocities out to about
700 km ${\rm sec}^{-1}$ for 5 Mpc and 1000 km ${\rm sec}^{-1}$ for 10 Mpc.

\subsection{Mass, Light and the Coldness of the Local Flow}

Using a sample of nine galaxies extending to about 8 Mpc, \citet{S86} concluded that the
velocity field in the Local Volume was extremely quiet, the velocity dispersion being about
equal to the observational errors in distance, near 60 km ${\rm sec}^{-1}$.  \citet{EBT01}
repeated the calculation with 14 galaxies having Cepheid distances, obtaining a similar dispersion
of 40-60 km ${\rm sec}^{-1}$.  A flow this cold is difficult to explain theoretically, recent
attempts including those of \citet{BCT01} and \citet{AP02}.
However, the fact of a cold flow has been disputed, by \citet{VB79} for instance, and
the present study indicates a dispersion twice that of \citet{EBT01} even ignoring
the sample incompleteness at high radial velocity.  In any investigation of kinematics in the
Local Volume this disagreement requires some explanation.

\begin{figure}
\plottwo{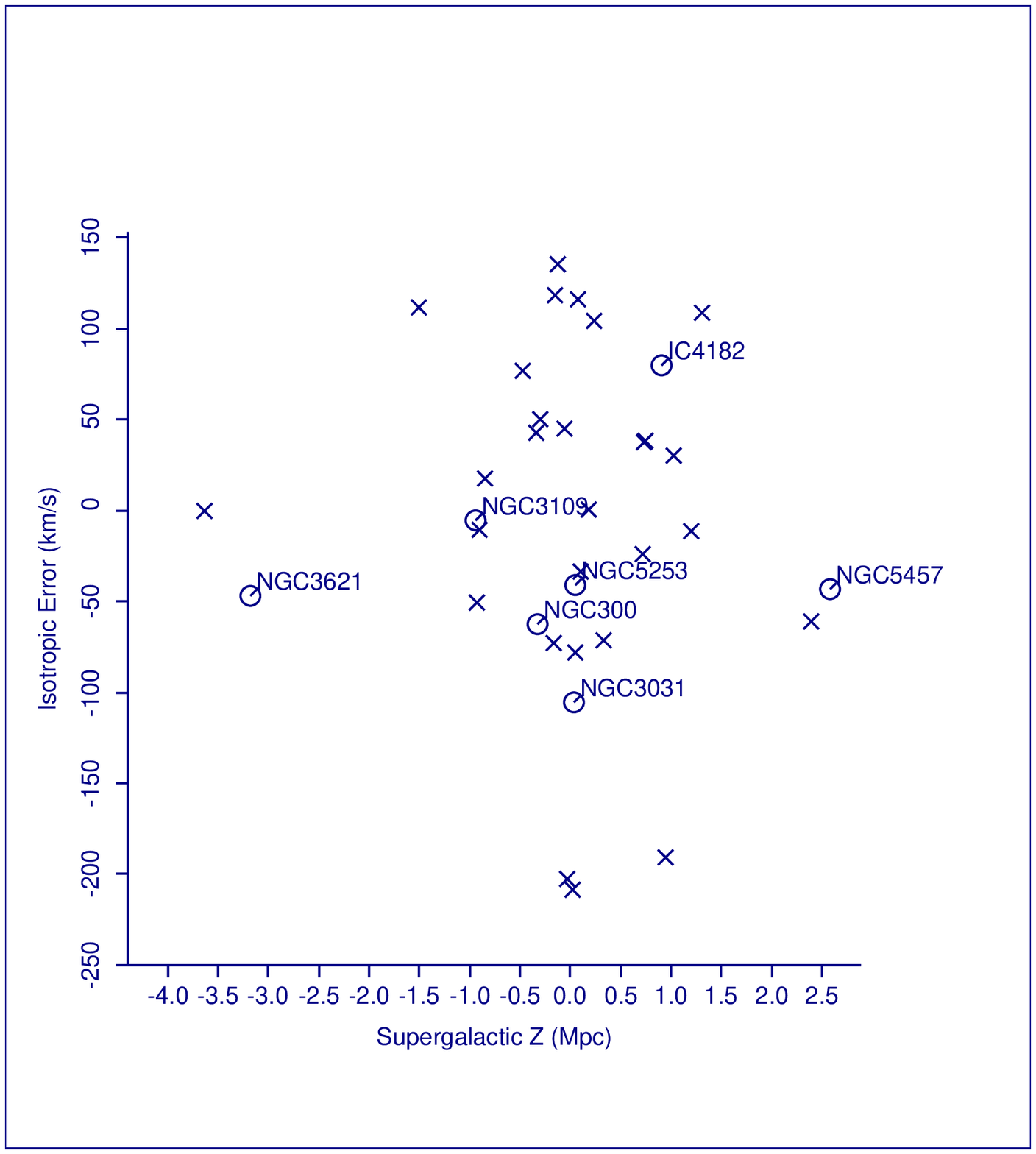}{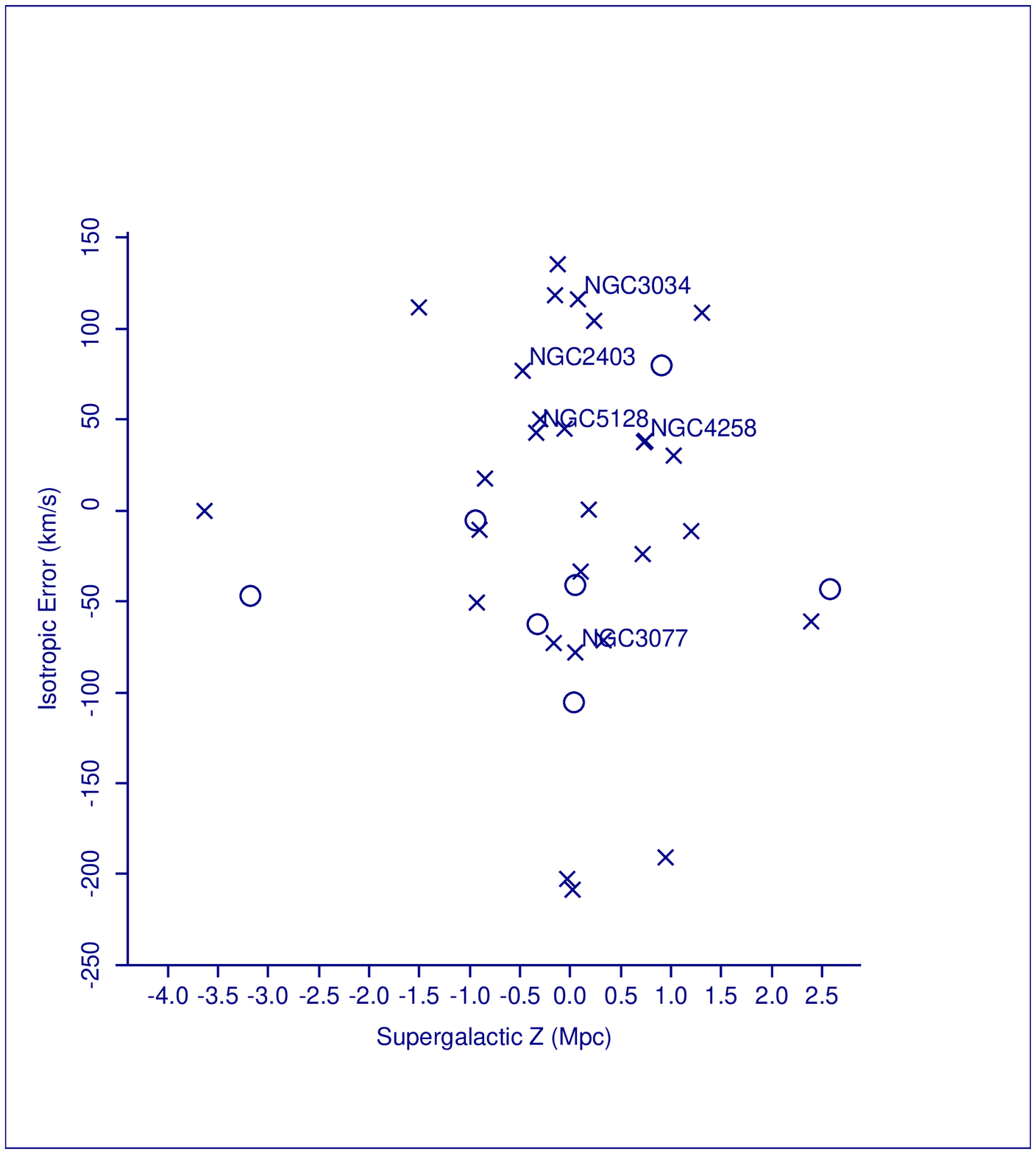}
\caption{Left, non-Local Group
galaxies used by \citet{EBT01} to derive a cold local flow are named as they
appear on the plot of the 35-galaxy solution.  Right, some of the more massive
galaxies not used in their
investigation on the same plot. \label{coldflow}}
\end{figure}

The non-Local Group galaxies used by \citet{EBT01} in their study are shown in Figure~\ref{coldflow},
as they fall in the 35-galaxy anisotropic solution.  They clearly do not explore the full width of the
velocity dispersion.  This is easily explained if the Cepheid galaxies are more massive than
the average, and thus harder to disturb by gravitational interaction (and {\em a priori} plausible,
given that Cepheids are easier to find in massive spirals).  However, in the right hand side of
the same figure several galaxies are singled out which are about as massive as the Cepheid
set, perhaps more so, and show a much greater
dispersion.  It appears that the various cold-flow groups have been misled by small number
statistics\footnote{However, on much larger scales, where motions of
the order of 600 km ${\rm sec}^{-1}$ are found, a dispersion of 100-200 km ${\rm sec}^{-1}$
is still cold.  In this context the cold-flow problem remains, though it changes character
and may not be as intractable theoretically (see, for instance, \citet{WH00}).}.

\clearpage

But this raises a very important point.  We still expect that, if peculiar velocities are
generated by (two- or few-body)
gravitational interaction among galaxies, more massive galaxies should have smaller
peculiar velocities.  Figure~\ref{coldflow} suggests it might not be true; in addition,
\citet{KM01} reported no difference in dispersion
between group and field galaxies, and between giant and dwarf galaxies.  It is a matter worth
investigating in some detail.

Reliable, or even consistent, dynamical estimates of galaxy masses are even 
harder to perform (and thus rarer) 
than distance measurements.  As a surrogate we will use total brightness,
assuming some sort of relation between mass and light to be made more specific later.  
Apparent $B$ magnitudes and extinction
estimates from NED (the \citet{SFD98} extinction figures) were combined with distances 
to produce corrected absolute magnitudes
for 97 of the 98 galaxies in the sample\footnote{UGC 6451 had no listed photometry.}. 
Morphological types were also extracted from NED.
Finally, each galaxy was assigned to a group or to the field, mostly following \citet{SB92},
though further investigation was required for some galaxies not in that paper\footnote{
The assignment of galaxies to groups or to the field is not an exact process, even when
done with more care than has been used here.  Whether a galaxy is gravitationally bound to
a group depends upon the group's total mass, which is poorly known for even the best-studied
groups; and upon distances in three dimensions to all galaxies in the area, which are not
known for most of the galaxies in the Local Volume.  And even were it possible to determine
bound members of all groups, there would still be a population not bound to, but still greatly
affected by, a nearby group.  For the purpose of detecting gross differences in the two
populations, however, the division carried out here should be
sufficient.}.  Apparent
magnitude and type have already been listed in Table~\ref{table:Data}; the derived absolute
magnitude, NED extinction, and group assignments are listed in Table~\ref{table:extinct}.

\begin{deluxetable}{lrrl}
\tablewidth{0pc}
\tablecaption{Absolute Magnitude, Extinction and Groups \label{table:extinct}}
\tablehead{
\colhead{Designation} & \colhead{Mag} & \colhead{Extinction} &
 \colhead{Group} }
\startdata
A0554+0728 & -12.9 & 2.553 & F \\
Antlia & -9.8 & 0.342 & F \\
BK3N & -10.5 & 0.345 & B2 \\
DDO13 & -15.7 & 0.276 & M74 \\
DDO50 & -16.5 & 0.139 & B2 \\
DDO53 & -13.1 & 0.160 & B2 \\
DDO63 & -16.4 & 0.207 & B2 \\
DDO66 & -13.7 & 0.343 & B2 \\
DDO70 & -13.9 & 0.137 & F \\
DDO71 & -10.1 & 0.412 & B2 \\
DDO75 & -14.0 & 0.188 & F \\
DDO82 & -15.0 & 0.188 & B2 \\
DDO155 & -12.2 & 0.113 & F \\
DDO165 & -15.7 & 0.104 & B2 \\
DDO187 & -12.7 & 0.105 & F \\
DDO190 & -14.1 & 0.052 & F \\
ESO294-G10 & -10.6 & 0.024 & B7 \\
IC342 & -19.9 & 2.407 & B1 \\
IC4182 & -15.3 & 0.059 & B5 \\
IC2574 & -17.2 & 0.156 & B2 \\
KDG52 & -10.9 & 0.091 & B2 \\
KDG61 & -12.5 & 0.309 & B2 \\
KDG73 & -13.2 & 0.080 & B2 \\
KK251 & -13.4 & 1.238 & N6946 \\
KK252 & -13.4 & 1.910 & N6946 \\
KKR25 & -9.4 & 0.036 & F \\
KKR55 & -14.6 & 2.941 & N6946 \\
KKR56 & -14.6 & 3.135 & N6946 \\
KKR59 & -16.5 & 3.863 & B2 \\
Maffei1 & -21.9 & 5.046 & B1 \\
NGC59 & -15.2 & 0.088 & F \\
NGC300 & -17.6 & 0.055 & B7 \\
NGC628 & -19.7 & 0.301 & M74 \\
NGC784 & -16.5 & 0.255 & F \\
NGC925 & -19.4 & 0.326 & N1023 \\
NGC1560 & -16.5 & 0.813 & B1 \\
NGC1705 & -15.8 & 0.035 & F \\
NGC2366 & -16.4 & 0.157 & B2 \\
NGC2403 & -18.8 & 0.172 & B2 \\
NGC2683 & -19.3 & 0.142 & F \\
NGC2903 & -20.2 & 0.134 & F \\
NGC2976 & -17.8 & 0.300 & B2 \\
NGC3031 & -20.3 & 0.346 & B2 \\
NGC3034 & -19.3 & 0.685 & B2 \\
NGC3077 & -17.6 & 0.289 & B2 \\
NGC3109 & -15.5 & 0.288 & F \\
NGC3274 & -16.4 & 0.104 & F \\
NGC3621 & -19.3 & 0.346 & F \\
NGC4144 & -17.9 & 0.065 & B4 \\
NGC4236 & -17.6 & 0.063 & B2 \\
NGC4244 & -17.5 & 0.090 & B4 \\
NGC4258 & -20.4 & 0.069 & F \\
NGC4395 & -17.6 & 0.074 & B4 \\
NGC4449 & -17.4 & 0.083 & B4 \\
NGC4523 & -14.8 & 0.166 & F \\
NGC4605 & -17.7 & 0.062 & F \\
NGC5128 & -20.5 & 0.496 & B6 \\
NGC5204 & -16.4 & 0.054 & B3 \\
NGC5236 & -20.4 & 0.284 & B6 \\
NGC5238 & -14.7 & 0.046 & B3 \\
NGC5253 & -16.9 & 0.242 & B6 \\
NGC5457 & -20.9 & 0.037 & B3 \\
NGC5474 & -17.9 & 0.047 & B3 \\
NGC5477 & -15.1 & 0.048 & B3 \\
NGC5585 & -18.6 & 0.067 & B3 \\
NGC6789 & -14.3 & 0.302 & F \\
NGC6946 & -21.0 & 1.475 & N6946 \\
ORION & -14.2 & 3.162 & F \\
UGC288 & -13.5 & 0.331 & F \\
UGC1104 & -15.5 & 0.273 & M74 \\
UGC1171 & -12.6 & 0.252 & M74 \\
UGC2905 & -14.9 & 1.349 & F \\
UGC3755 & -14.4 & 0.384 & F \\
UGC3860 & -14.4 & 0.253 & F \\
UGC3966 & -15.5 & 0.218 & F \\
UGC3974 & -14.7 & 0.145 & F \\
UGC4115 & -13.5 & 0.122 & F \\
UGC4483 & -12.6 &  0.146 & B2 \\
UGC5721 & -16.4 & 0.104 & F \\
UGC6451 & -11.9 & 0.257 & B2 \\
UGC6456 & -14.1 & 0.155 & F \\
UGC6565 & -15.6 & 0.045 & B4 \\
UGC6572 & -13.5 & 0.110 & B4 \\
UGC6817 & -14.7 & 0.113 & B4 \\
UGC7559 & -13.8 & 0.060 & B4 \\
UGC7857 & -14.4 & 0.103 & F \\
UGC8320 & -15.0 & 0.065 & F \\
UGC8331 & -15.0 & 0.039 & F \\
UGC8508 & -13.5 & 0.064 & B3 \\
UGC9405 & -12.5 & 0.051 & B3 \\
UGC11583 & -13.8 & 1.319 & N6946 \\
UGCA86 & -16.8 & 4.061 & B1 \\
UGCA92 & -16.3 & 3.419 & B1 \\
UGCA105 & -15.0 & 1.350 & B1\\
UGCA281 & -13.8 & 0.065 & B5 \\
UGCA290 & -13.2 & 0.060 & B4 \\
UGCA438 & -12.8 & 0.064 & B7 \\
\enddata
\tablecomments{Absolute magnitudes for the galaxies in the sample, derived from NED
photometry and extinctions, and the distances listed in Table~\ref{table:Data}.  The
extinction estimates are shown, as well as the group assignments mostly following
\citet{SB92}. ``F'' denotes a field galaxy; M74, belonging to the M74 group; N1023 and
N6946 one belonging to the NGC 1023 and NGC 6946 groups, respectively.}
\end{deluxetable}

\clearpage

\begin{figure}
\plottwo{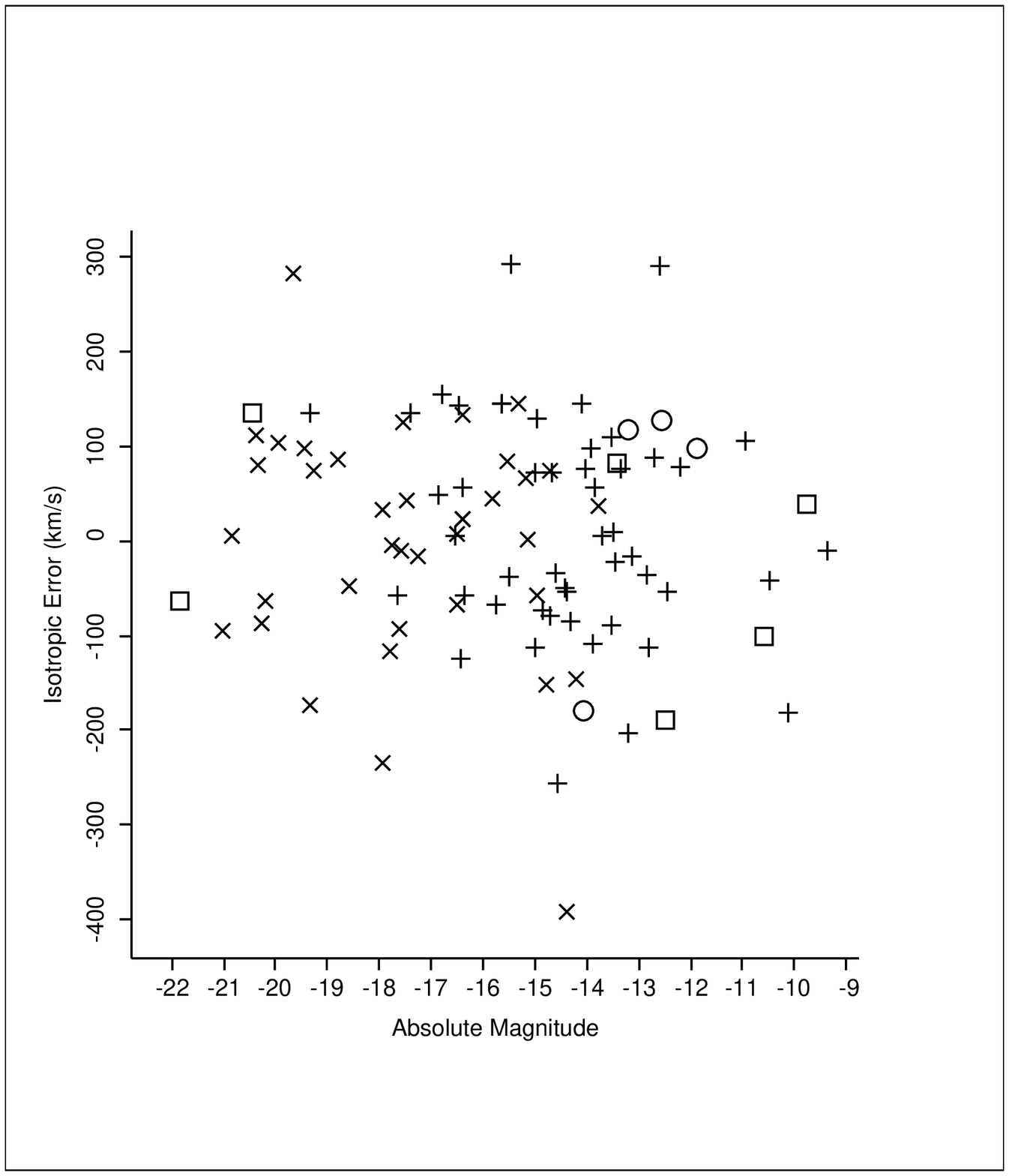}{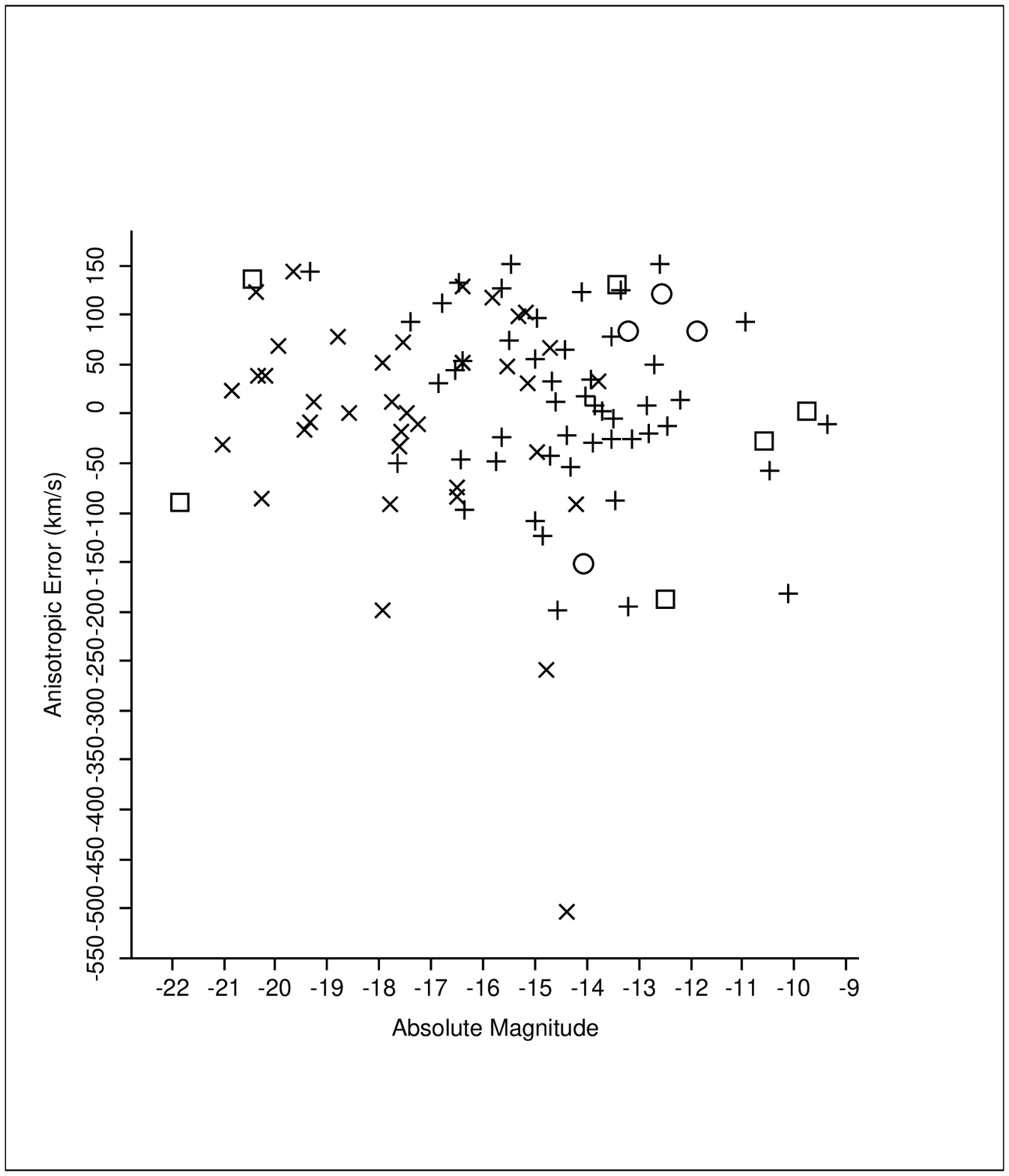}
\caption{Deviation from (left) isotropic and (right) anisotropic Hubble expansion as a function
of absolute magnitude for the 97-galaxy sample, using distances, photometry and extinction
measurements from the literature.  Crosses indicate spiral galaxies; plus, irregular; boxes,
elliptical or lenticular; circles, other, peculiar or unclassified.
\label{mplot98}}
\end{figure}

\begin{figure}
\plottwo{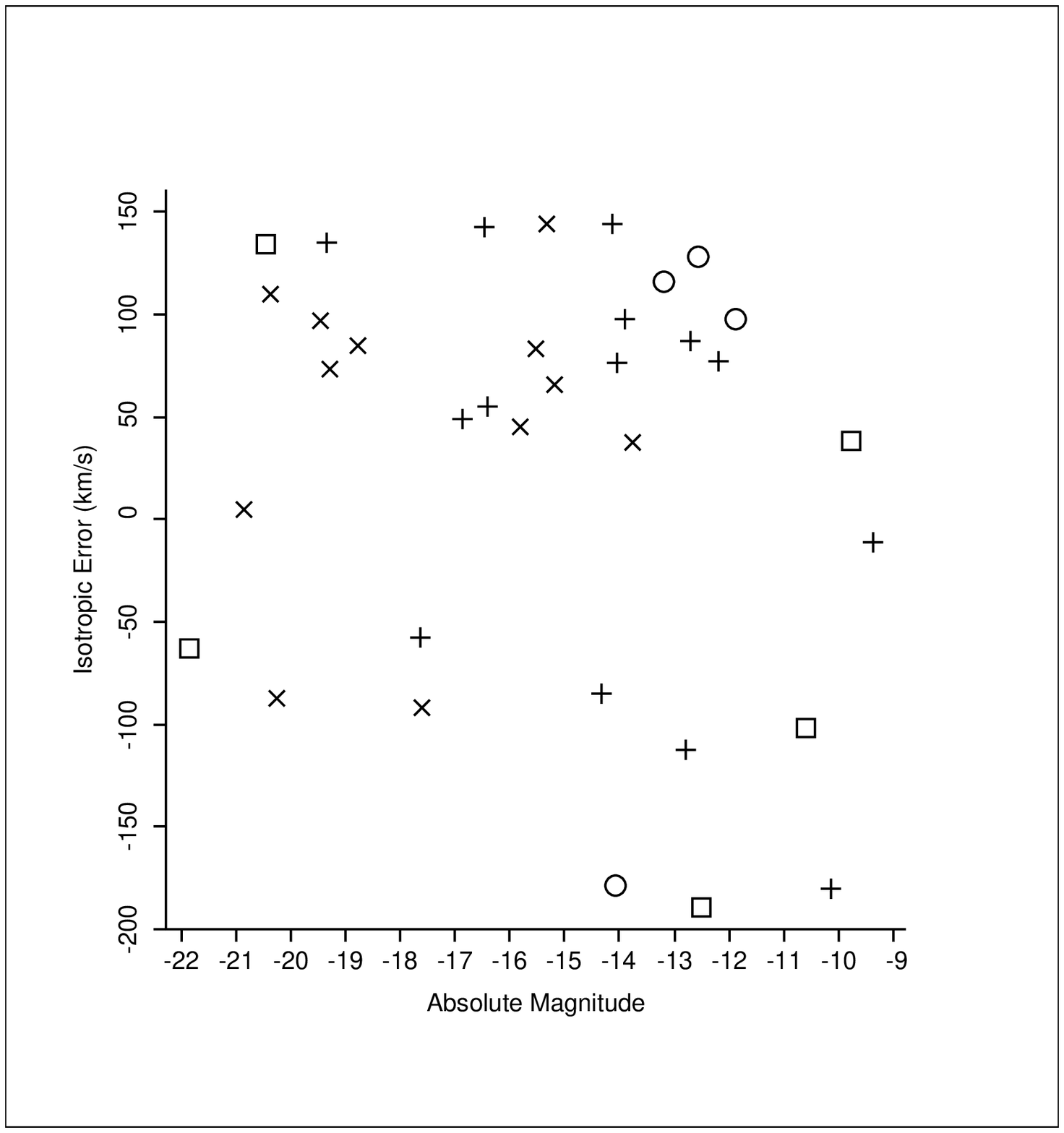}{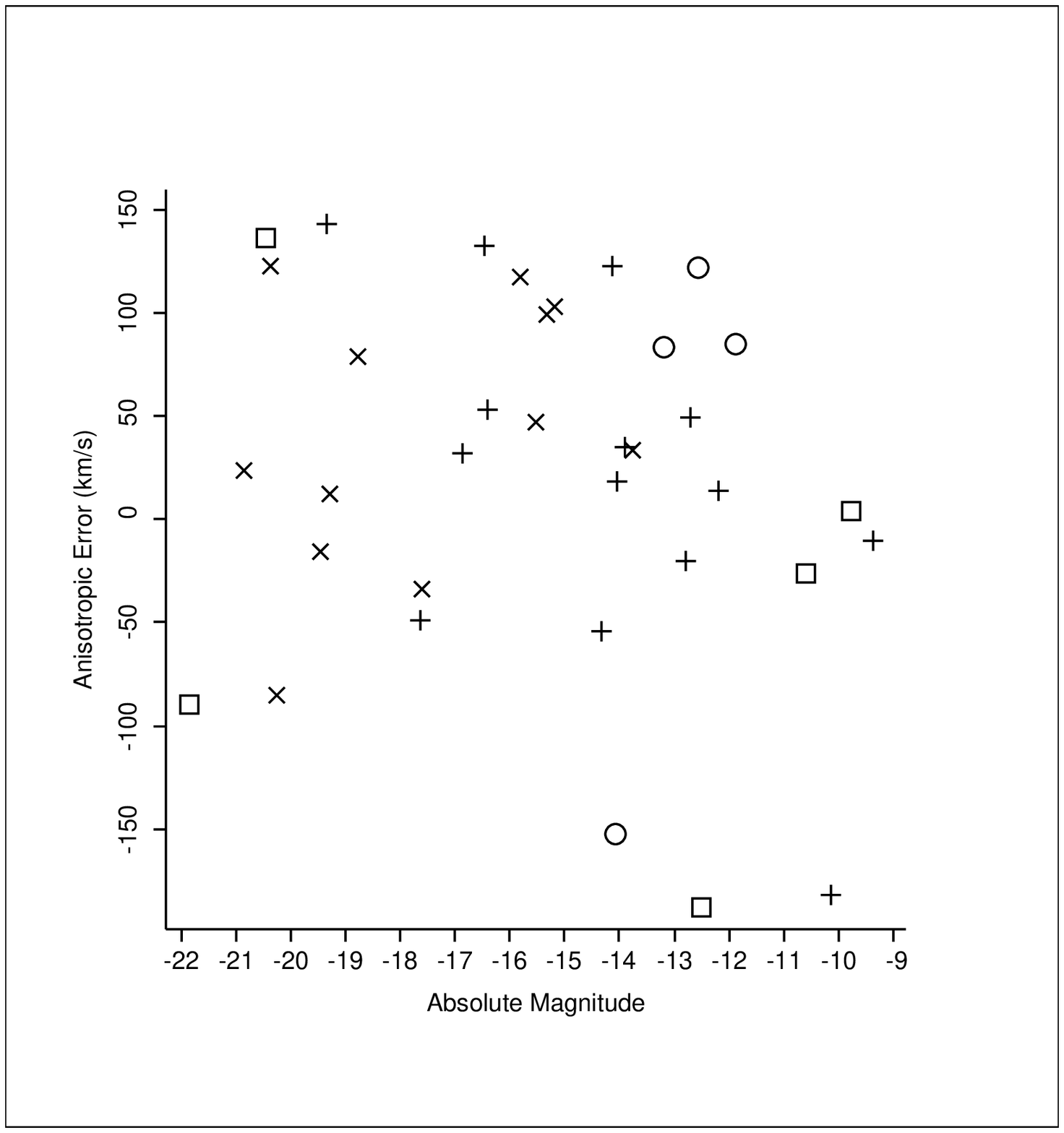}
\caption{Deviation from (left) isotropic and (right) anisotropic Hubble expansion as a function
of absolute magnitude for the 35-galaxy sample, using distances, photometry and extinction
measurements from the literature.  As before, crosses indicate spiral galaxies; plus, irregular; boxes,
elliptical or lenticular; circles, other, peculiar or unclassified.
\label{mplot35}}
\end{figure}

The data are plotted in Figures~\ref{mplot98} and~\ref{mplot35}.   Deviation from the
various flow models is shown as a function of absolute magnitude; in addition, symbols
designate morphological classes of galaxies.

These figures represent a very remarkable result.  Over a range of ten magnitudes in luminosity
there is {\em no} systematic variation in the peculiar velocity dispersion.  Further, there is no
apparent variation with galaxy type.

\begin{figure}
\plottwo{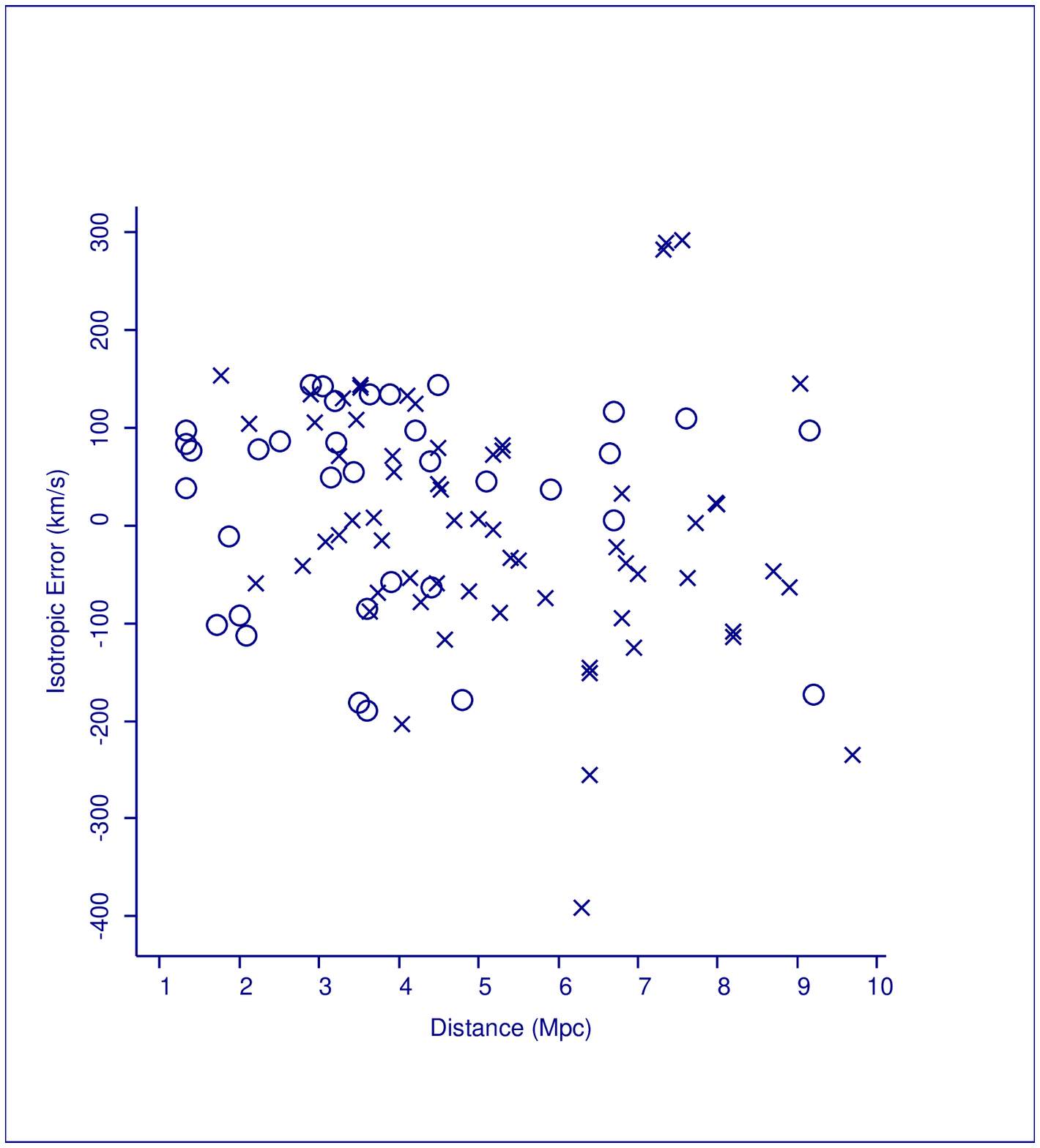}{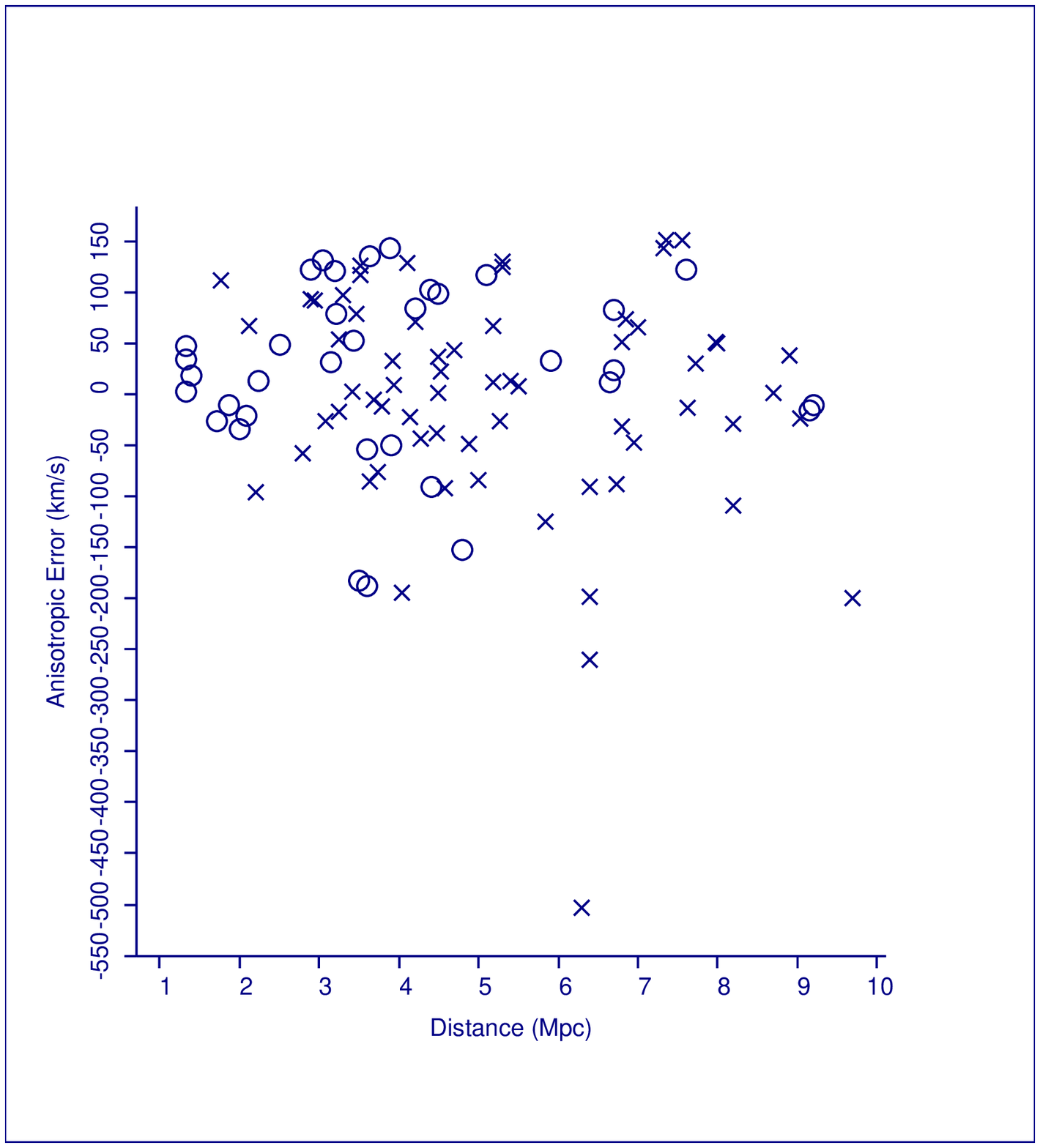}
\caption{Deviation from (left) isotropic and (right) anisotropic Hubble expansion as a function
of distance for the 97-galaxy sample.  The galaxies with higher-quality distances are shown
as open circles.
\label{distplot}}
\end{figure}

Could these plots actually be dominated by the effects of observational errors, rather than
real motions?  That would require Cepheid and TRGB distance uncertainties to be something
like three times their normally estimated size, which seems unlikely.  In addition, a dispersion
due solely to distance errors would be Gaussian in shape, which these are not; would
require a dispersion which increased with distance, which as Figure~\ref{distplot} shows,
is not true\footnote{Since
the number of galaxies in the sample is roughly constant with distance, clear from the plot,
it is a posteriori unnecessary to deal with Malmquist and related biases.}; and would
require that the underlying, physical velocity dispersion be almost zero, a very difficult
thing to manage theoretically (as noted above) and hard to believe in the context of a galaxy
group. 

There is the possibility of errors due to other sources.
The photometry found in NED is admittedly hetrogeneous, and
measurement of the total light from the diffuse and faint outskirts of a galaxy is
notoriously difficult.  But a comparison of various photometric measurements in NED (where
galaxies have been measured more than once) shows a remarkable agreement, figures 
seldom diverging by as much as 0.2 magnitude.  The exceptions are those visually estimated
from photographic plates, which have claimed accuracies of about $\pm$ 0.5 magnitude.
Being conservative and allowing points in Figures~\ref{mplot98} and \ref{mplot35} to
move sideways a full magnitude, and postulating some unknown process which preferentially
biased high the photometry for galaxies with high peculiar velocities, the shape of the
plots remains.

Extinction is another fruitful source of error.  But only two of the points in Figure~\ref{mplot35}
have as much as half a magnitude of extinction.  Though both are outliers, removal still leaves
the shape intact.

It appears, then, that the constancy of the peculiar velocity dispersion with absolute magnitude
is a real effect.  What would we actually expect?

Naively, for any two-body interaction, the velocities
of galaxies should be changed in inverse proportion to their masses; for a completely relaxed
system, with kinetic energy equally partitioned, the random velocities should be inversely
proportional to the square root of the masses.  Suppose that the mass-to-light ratio lowers
by a factor of 100 between absolute magnitude -10 and -20 (for which extreme value there is
no evidence).  Then there should still be a change in dispersion by a factor of
ten; while apparently there is none at all.

Searching for more sophisticated predictions, one quickly encounters one major problem with 
concentrating on the Local
Volume: it is much smaller than the regions dealt with in studies of structure formation.
In \citet{JF98}, Figure 10, for instance (which shows predicted velocity dispersions for
various length scales), a volume of 10 Mpc radius is beyond the small-scale edge 
of the curves drawn.  One can extrapolate them, but the result is quite uncertain; they might
predict a velocity dispersion of anywhere from 50 to 500 km s$^{-1}$.

Of more use for the question at hand,
\citet{IIS92} compared the analytic expression for velocity dispersion in \citet{SCI2} with a
series of n-body calculations.  They found a rather small difference in rms velocity dispersion
with mass: 169 compared to 179 (in scaled units) for sets of galaxies differing in mass by
a factor of 100.  However, they note that in their work the tendency for smaller galaxies
to be affected more by gravitational interactions is offset by the tendency for larger galaxies
to be found in rich clusters.  Within the Local Volume there are no rich clusters.  For such
an effect to operate here, one would have to have groups made up of similarly-sized galaxies:
dwarf groups and giant groups.  This is certainly not true, each group in the Volume having
galaxies with a wide distribution of size and brightness.

\begin{figure}
\plottwo{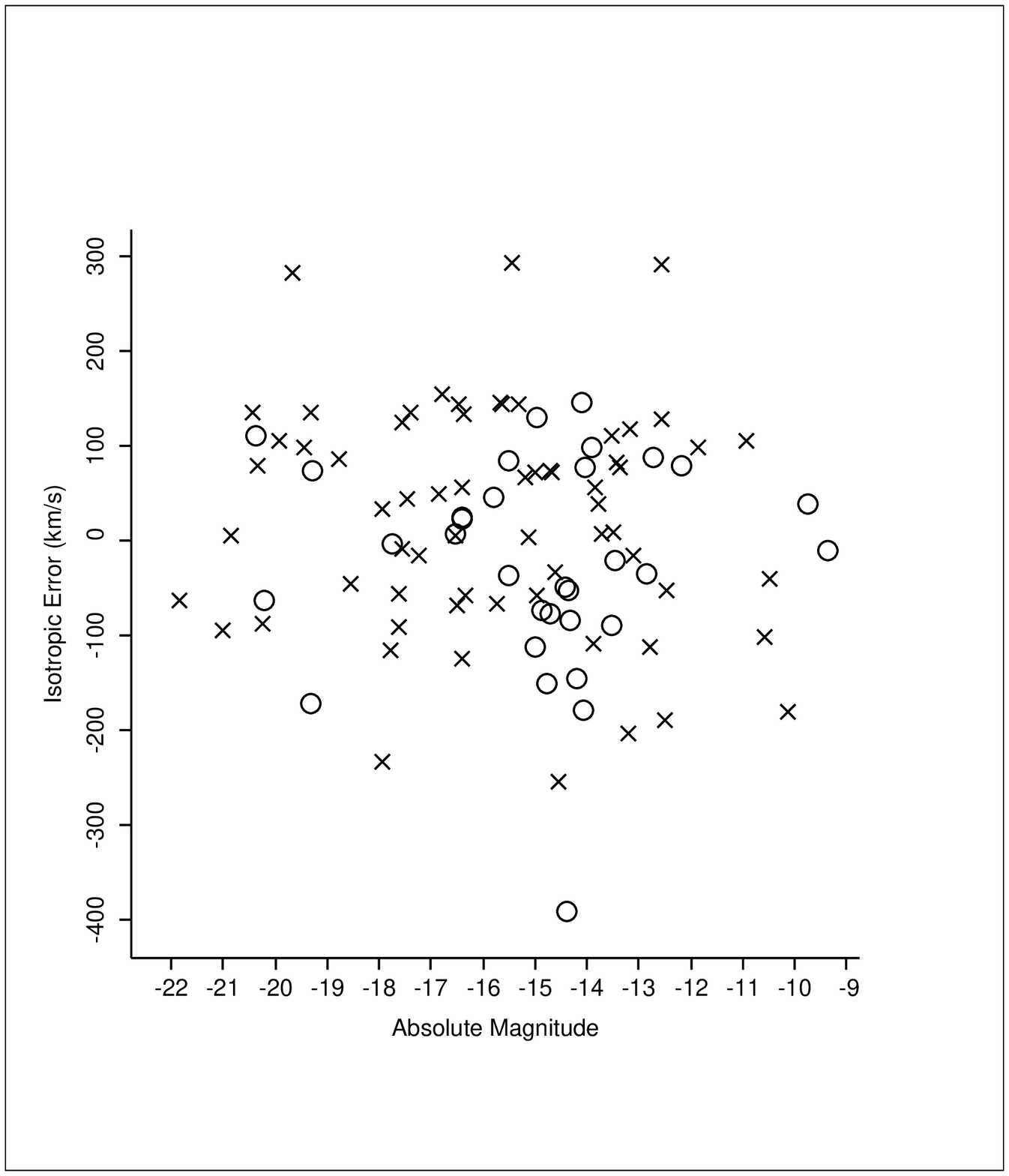}{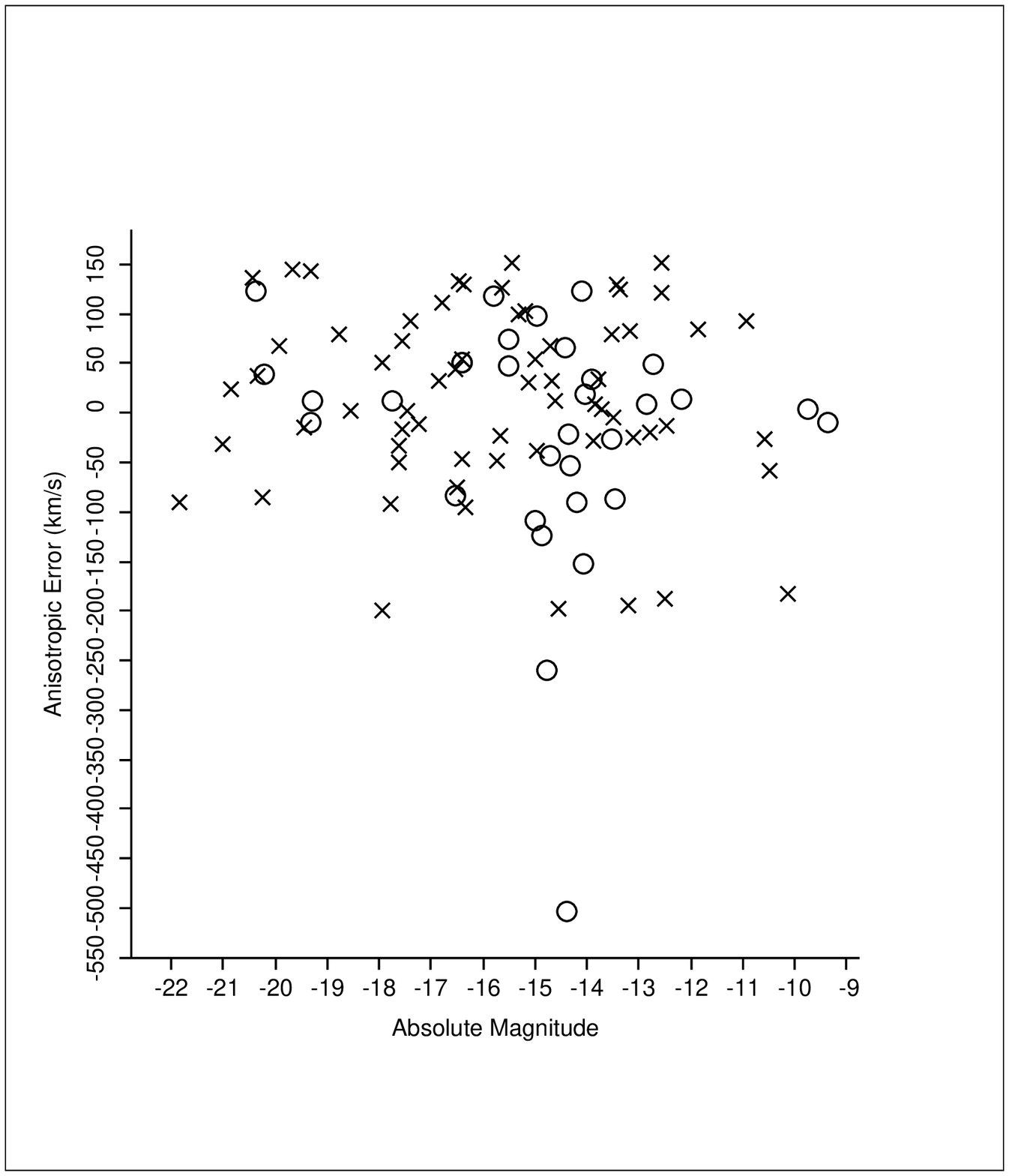}
\caption{Deviation from (left) isotropic and (right) anisotropic Hubble expansion as a function
of absolute magnitude for the 97-galaxy sample, using distances, photometry and extinction
measurements from the literature.  Crosses indicate galaxies in groups, circles field galaxies.
\label{group97}}
\end{figure}

\begin{figure}
\plottwo{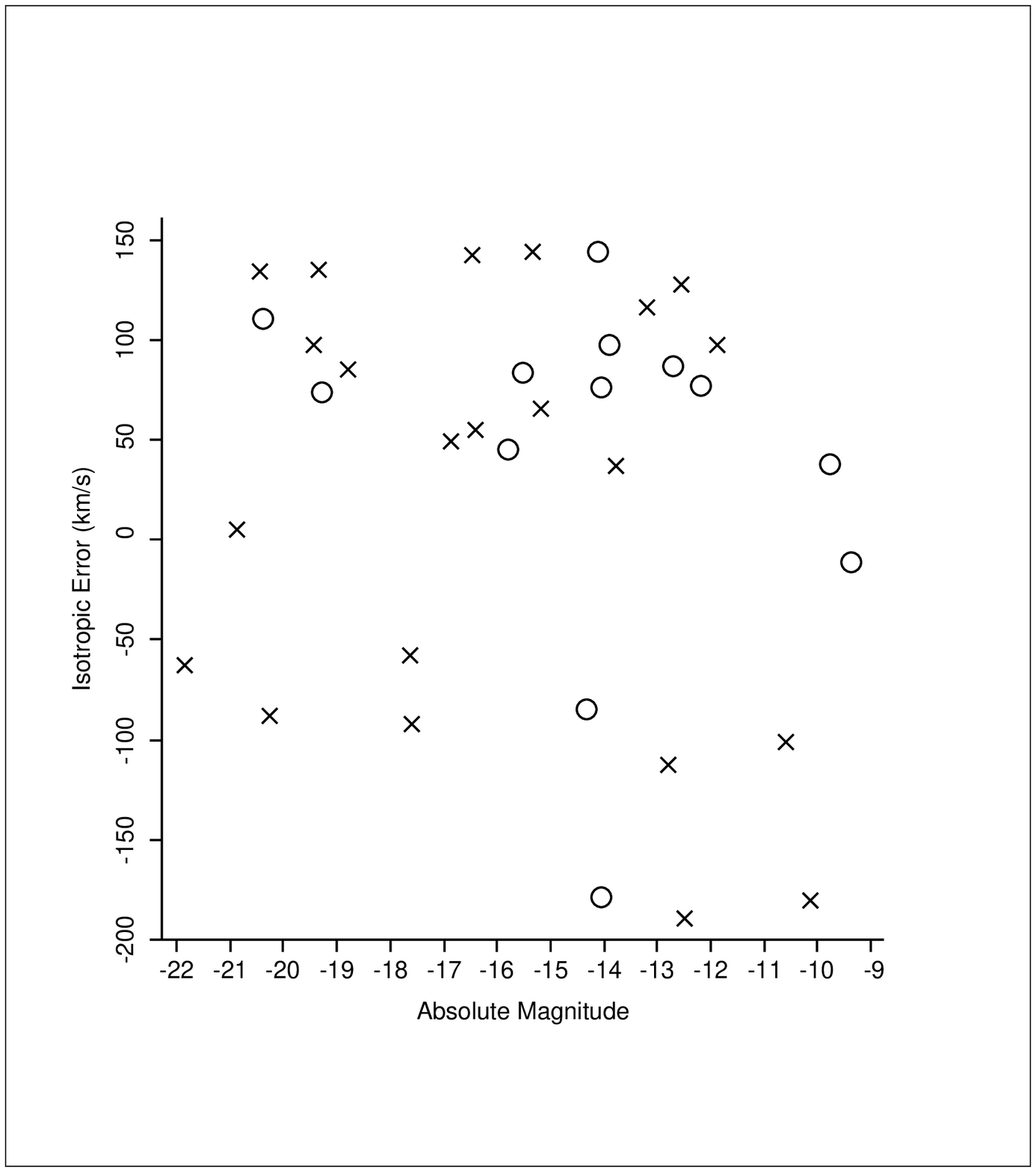}{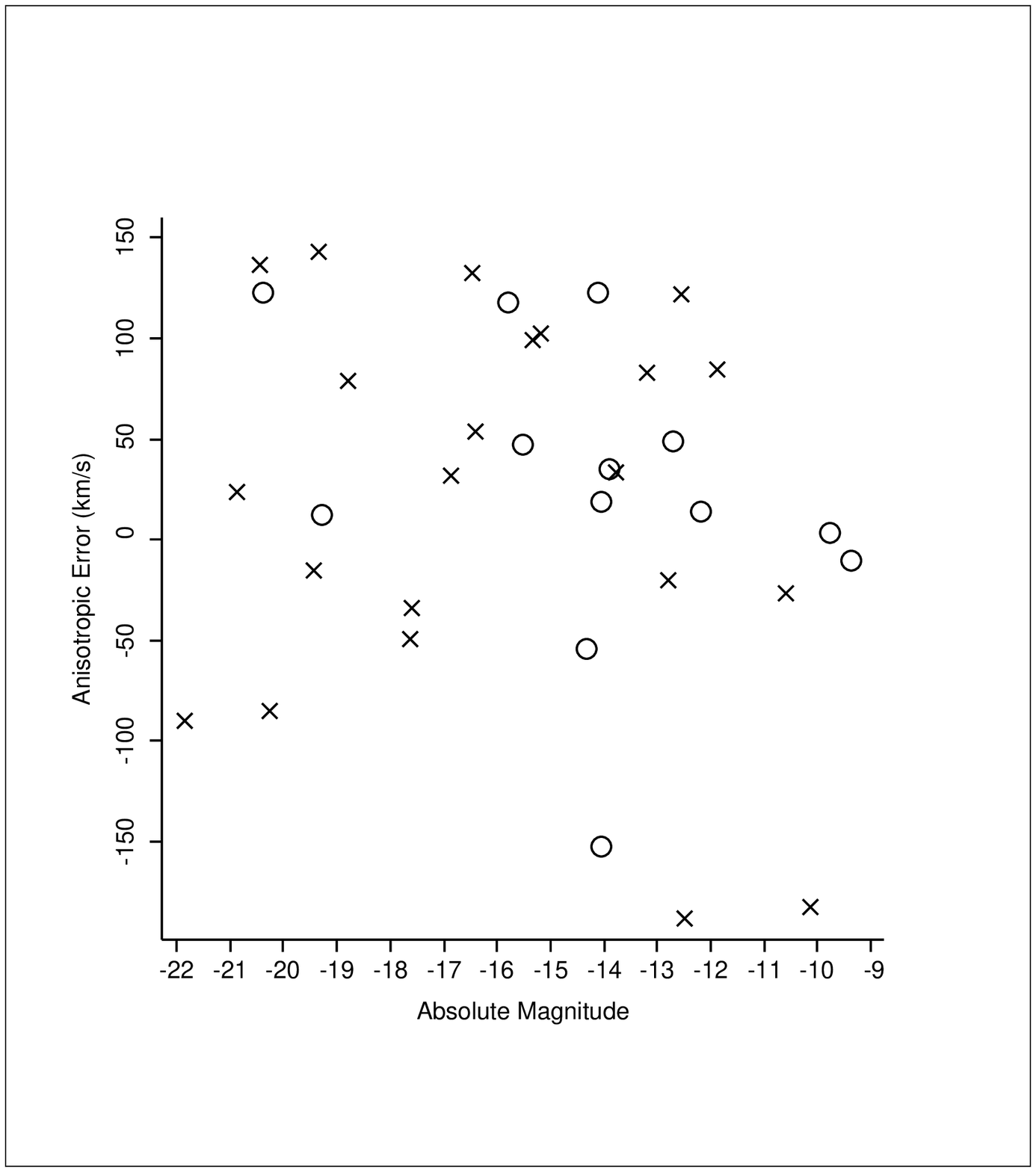}
\caption{Deviation from (left) isotropic and (right) anisotropic Hubble expansion as a function
of absolute magnitude for the 35-galaxy sample, using distances, photometry and extinction
measurements from the literature.  Crosses indicate galaxies in groups, circles field galaxies.
\label{group35}}
\end{figure}

This brings us to the question of field and group galaxies.  \citet{IIS92} predict that field 
galaxies should have much smaller peculiar velocities than cluster galaxies.  This is a quite
plausible prediction: field galaxies should have little or no interactions which take them
away from the Hubble flow, while group (and cluster) galaxies should be at least in the
process of virializing.   The situation with the present sample is shown in Figures~\ref{group97}
and \ref{group35}.

There is no tendency for brighter galaxies to occur in groups, and no obvious tendency for group
galaxies to have a larger velocity dispersion.  To compare the situations quantitatively, we
again use the dispersions about the models and employ (cautiously) 
the F-ratio test.  For the 98-galaxy sample,
isotropic model, the group dispersion is 120 km s$^{-1}$ against a field dispersion of 113
km s$^{-1}$.  The F-ratio test gives a 64\% significance to this; the difference is probably
not significant.  For the 98-galaxy anisotropic model the group dispersion is 93 km s$^{-1}$
against 122 km s$^{-1}$ for the field, giving a 95\% significance---but with the {\em field}
galaxies having a higher dispersion.

For the 35-galaxy isotropic model the field and group dispersions are 68 and 98 km s$^{-1}$,
respectively, for a significance of 87\%; the anisotropic model has 47 and 86 km s$^{-1}$ for
82\%.  Here it appears that we have a true group versus field effect, and acting in the right
direction.  But these figures for significance are, as noted, to be treated cautiously.  Also, with
only 13 galaxies in the field category, we are hardly better off than \citet{EBT01} from the
standpoint of small-number statistics.  I argue that an effect which appears only with
a small sample of data (albeit one with better quality) and goes away (or reverses) with
a larger sample is not real, especially in light of the quantitative unreliability of standard
statistical measures when dealing with these samples. 

The predictions of \citet{IIS92} have been found to agree closely with observations on scales
up to 50 $h^{-1}_{100}$ Mpc \citep{RS96}, as well as n-body calculations, so they may be taken
to be well-established.  Other techniques which match structure with motions on large scales
also seem to work well (for example, those described in \citet{CW00}).  On smaller scales,
dynamical modeling of galaxy clusters and individual galaxies has become quite sophisticated
and successful.  In between, it appears, something different is happening.

\subsection{Dynamics of the Peculiar Velocity Dispersion}

A peculiar velocity distribution which ignores completely such things as absolute magnitude
and the presence or absence of galaxy groups requires some explanation.  The picture
of motions generated by gravity, with that gravity field related to observable (luminous) matter
in some simple way, does not appear to work in the Local Volume.  There are a number of possible
expanations:

1.  Light does not trace mass, in that each observable galaxy is contained in a dark halo of about
the same mass.  Of course this contradicts otherwise successful simulations of large scale structure
(for example, \citet{SWT01}; \citet{MLS02}).
It is hard to see how the extreme variations in amounts of luminous matter (a factor of
something like 10,000 over a ten-magnitude range) could come about within similar-sized
halos.  It is also hard to see how such a uniform field of dark matter halos themselves could
come about.  Alternatively, there could be a range of masses in dark matter halos, but no
relation between their masses and the luminous matter within them; again, it is difficult
to produce such a variation.

2.  Light does not trace mass, in that the peculiar velocities of observable galaxies are produced
by interaction with totally dark objects, which are so much more massive than galaxy halos that
the latter are all equally affected by them.  This has the attraction that such dark objects could
be clustered similarly to galaxies on larger scales (allowing them to define the Supergalactic
Plane, for instance) and still not interfere
with the internal dynamics of galaxy groups.  It is difficult to understand, however, how the
most massive dark matter objects could manage to avoid accreting any luminous matter at all.

3.  The observed velocities of galaxies are not those of their dark matter halos; the luminous
matter is ``sloshing around'' inside the dark potential well.  This is probably a more
attractive idea than the previous two, though a mechanism for such internal motion is lacking.

4.  Peculiar velocities on this scale are not produced by gravitational interaction, but perhaps by
some remnant of ``primeval turbulence.''  This is an old idea in the context of the formation
of galaxies.  Heisenberg (in \citet{H49}) required some original motion in order for matter to clump 
together in galaxies; but this kind of motion is probably unimportant on galactic scales \citep{P93}
(page 541).

5.  We might have had bad luck with data, an unfortunate set of observations giving a misleading result
of the sort that \citet{EBT01} encountered.  Of course this is harder to arrange with a much larger
sample, and in particular it is hard to understand how the very even distribution of
peculiar velocity with absolute magnitude could have been produced by any reasonable
selection of data.  It has indeed been argued above that the data sets at hand do not
sample the true underlying dynamics well; but that is in the context of a particular
model (isotropic or simple anisotropic expansion).

6.  That idea leads on to the possibility that we are using the wrong kinematic model.  There
is indeed a general expansion of the Local Volume; but perhaps there is some different motion.
This is a most attractive idea in that no current notions of cosmology or physics
need necessarily be discarded or even greatly revised; however, it is not obvious
to how to pursue it.  The fact that no clear trend shows up in the plots of deviation versus
the various spatial directions shows that no simple modification of the present models
(including, for instance, a nonlinear effect of Virgocentric flow) is likely to be of use. 

\section{Summary}

An examination of the kinematics of galaxies within 10 Mpc of the Milky Way has thrown up
some surprises and one deep puzzle.  An overall anisotropic expansion, expected from the
distribution of galaxies in the Local Volume, can be calculated; but the uncertainty of its
details and its strong dependence on the particular data set chosen indicate that it is not
a useful description.   

A mismatch between (on one hand) the calculated standard error of various model parameters 
and (on the other) the change of those parameters between data sets indicates that the
present data do not form a good sample of the underlying kinematics, at least in the context
of the models at hand.  At least part of this is due to a selection bias against nearby
galaxies with high radial velocities.

There is {\em no} variation in the width of the peculiar velocity dispersion with absolute magnitude
over a range of ten magnitudes.  Neither is there any apparent relationship with galaxy
type, or between field and cluster galaxies.  This is difficult to understand, as the tendencies
which even out peculiar velocities among masses on larger scales are not at work.  Mass may have
no relation to light on these scales; or peculiar velocities might be produced by other than
gravitational interactions among galaxies; or the set of data at hand might somehow be terribly
misleading; or the reference kinematic model might be wrong.

To clarify the kinematics of this volume, and to shed light on (much less clear up) the puzzle,
much additional data will be necessary.  In particular, to define the shape of the peculiar
velocity histogram many more galaxies need to be added.  Unfortunately, accurate distances to objects
between 2 and 10 Mpc require careful observations with the larger telescopes, and distances
to a useful fraction of
hundreds of galaxies will take much time to compile.  It will be worthwhile to target, initially,
bright and massive galaxies;  those at high supergalactic latitude; those in the field; and
those of high radial velocity but possibly still nearby.

\acknowledgements

It is a pleasure for the author to thank Donald Lynden-Bell for bringing the Hubble Tensor to
his attention and for many helpful conversations.  His debt to the indefatigable data collection
efforts of Igor Karachentsev and his various coworkers should be evident in Table~\ref{table:Data}.
The comments of an anonymous referee have very much clarified and improved this paper.
This work has been supported in part by
the Institute of Astronomy, University of Cambridge and in part by the Physics Department,
U. S. Naval Academy.

\end{document}